\documentclass[acmsmall]{acmart}
\acmJournal{PACMPL}
\acmVolume{1}
\acmNumber{CONF} % CONF = POPL or ICFP or OOPSLA
\acmArticle{1}
\acmYear{2022}
\acmMonth{1}
\acmDOI{} % \acmDOI{10.1145/nnnnnnn.nnnnnnn}
\startPage{1}

\setcopyright{none}
\usepackage{booktabs}   %% For formal tables:
\usepackage{subcaption} %% For complex figures with subfigures/subcaptions
\usepackage{wrapfig}
\usepackage{booktabs}   %% For formal tables:
\usepackage{subcaption} %% For complex figures with subfigures/subcaptions
\usepackage{listings} 
\usepackage{semantic}
\usepackage{tikz,pgf}
\usepackage{stmaryrd}
\usepackage{xcolor}
\usepackage{markdown}
\setpremisesspace{1cm}

\usepackage{authorcomments}
\usepackage{qtree}
\usepackage{dirtree}
\newcommand{\name}{\textsf{Cobalt}\,}
\usepackage{stmaryrd}
\newcommand{\den}[1]{\mbox{$\llbracket$\,#1\,$\rrbracket$}}

\newcommand{\eff}{\mbox{$\mathit{eff}$}}
\newcommand{\ep}{\mbox{$\mathsf{e}_p$}}
\newcommand{\eip}{\mbox{$\mathsf{e}_{ip}$}}

\definecolor{mygreen}{rgb}{0,0.6,0}
\definecolor{mygray}{rgb}{0.5,0.5,0.5}
\definecolor{mymauve}{rgb}{0.58,0,0.82}

\tikzoption{base font}{\def\tikz@base@textfont{#1}}
\tikzset{
  base font=\sffamily,
}
\usepackage{amsmath}
\usepackage{supertabular}
\usepackage{geometry}
\usepackage{ifthen}
\usepackage{alltt}%hack
\usepackage{color}

\usepackage{multirow}
\usepackage{multicol}
\usepackage{semantic}
\usepackage{algpseudocode}
\usepackage[noend,noline]{algorithm2e}
\usepackage{pgfplots, pgfplotstable}
\pgfplotsset{compat=1.13}
\SetKwInput{KwData}{Input}
\SetKwInput{KwResult}{Output}

\begin{document}

%% Title information
\title[Effectful Component-Based Synthesis]
      {Specification-Guided Component-Based Synthesis from Effectful Libraries }         %% [Short Title] is optional;
                                        %% when present, will be used in
                                        %% header instead of Full Title.
%\titlenote{with title note}             %% \titlenote is optional;
                                        %% can be repeated if necessary;
                                        %% contents suppressed with 'anonymous'
%\subtitle{Subtitle}                     %% \subtitle is optional
%\subtitlenote{with subtitle note}       %% \subtitlenote is optional;
                                        %% can be repeated if necessary;
                                        %% contents suppressed with 'anonymous'

%% Author information
%% Contents and number of authors suppressed with 'anonymous'.
%% Each author should be introduced by \author, followed by
%% \authornote (optional), \orcid (optional), \affiliation, and
%% \email.
%% An author may have multiple affiliations and/or emails; repeat the
%% appropriate command.
%% Many elements are not rendered, but should be provided for metadata
%% extraction tools.

%% Author with single affiliation.
\author{Ashish Mishra}
%\authornote{}          %% \authornote is optional;
                                        %% can be repeated if necessary
%\orcid{nnnn-nnnn-nnnn-nnnn}             %% \orcid is optional
\affiliation{
  \position{}
  \department{Department of Computer Science}              %% \department is recommended
  \institution{Purdue University}            %% \institution is required
  \country{USA}                    %% \country is recommended
}
\email{mishr115@purdue.edu}          %% \email is recommended

%% Author with two affiliations and emails.
\author{Suresh Jagannathan}
%\authornote{}          %% \authornote is optional;
                                        %% can be repeated if necessary
%\orcid{nnnn-nnnn-nnnn-nnnn}             %% \orcid is optional
\affiliation{
  \position{}
  \department{Department of Computer Science}             %% \department is recommended
  \institution{Purdue University}           %% \institution is required
  \country{USA}                   %% \country is recommended
}
\email{suresh@cs.purdue.edu}         %% \email is recommended

%% Abstract
%% Note: \begin{abstract}...\end{abstract} environment must come
%% before \maketitle command
\begin{abstract}
  Component-based synthesis seeks to build programs using the APIs
  provided by a set of libraries.  Oftentimes, these APIs have
  effects, which make it challenging to reason about the correctness
  of potential synthesis candidates.  This is because changes to
  global state made by effectful library procedures affect how they
  may be composed together, yielding an intractably large search space
  that can confound typical enumerative synthesis techniques.  If the
  nature of these effects are exposed as part of their specification,
  however, deductive synthesis approaches can be used to help guide
  the search for components.  In this paper, we present a new
  specification-guided synthesis procedure that uses Hoare-style pre-
  and post-conditions to express fine-grained effects of potential
  library component candidates to drive a bi-directional synthesis
  search strategy.  The procedure alternates between a forward search
  process that seeks to build larger terms given an existing context
  but which is otherwise unaware of the actual goal, alongside a
  backward search mechanism that seeks terms consistent with the
  desired goal but which is otherwise unaware of the context from
  which these terms must be synthesized.  To further improve
  efficiency and scalability, we integrate a conflict-driven learning
  procedure into the synthesis algorithm that provides a semantic
  characterization of previously encountered unsuccessful search paths
  that is used to prune the space of possible candidates as synthesis
  proceeds.  We have implemented our ideas in a tool called \name\ and
  demonstrate its effectiveness on a number of challenging synthesis
  problems defined over OCaml libraries equipped with effectful
  specifications.
\end{abstract}

%% 2012 ACM Computing Classification System (CSS) concepts
%% Generate at 'http://dl.acm.org/ccs/ccs.cfm'.

%% End of generated code

%% Keywords
%% comma separated list
  %% \keywords are mandatory in final camera-ready submission

%% \maketitle
%% Note: \maketitle command must come after title commands, author
%% commands, abstract environment, Computing Classification System
%% environment and commands, and keywords command.
\maketitle
\section{Introduction}
\label{sec:introduction}

Many useful programming tasks can be efficiently expressed by
intelligently composing the elements found in a library of available
APIs (or components).  Program synthesis queries, in particular, can
benefit from the ability to use library function calls in synthesizing
terms.  This observation has led, in recent years, to the development
of several useful component-based synthesis
tools~\cite{sypet,tygus,hoogle,rbsyn,frangel}. These methods typically
use examples and/or type annotations to guide the synthesis procedure
in an enumerative fashion.

However, APIs often have effects that must be taken into account when
reasoning about their composition.  Library implementations of
imperative data structures, databases, or parsers are canonical
examples where it is natural to have effectful APIs.  In these
domains, the effects performed by APIs can impose non-trivial
constraints on the choice of synthesis candidates, and the order in
which they must be sequenced. 

%% To illustrate, consider a database
%% library that provides an interface to access and update underlying
%% tables (say, {\sf Newsletters}) using functions (say, \textsf{member},
%% \textsf{subscribe}, and \textsf{unsubscribe}).  Suppose both
%% \textsf{subscribe} and \textsf{unsubscribe} have associated
%% preconditions on the database state that only allow users to
%% subscribe to currently unsubscribed newsletters, to only unsubscribe
%% from currently subscribed ones.  A simple synthesis query on this
%% library might ask for a program that transfers the subscription of a
%% user from newsletter $N_1$ to $N_2$.  A correctly synthesized result,
%% however, cannot simply call \textsf{unsubscribe} on $N_1$ and
%% \textsf{subscribe} on $N_2$ without first performing the appropriate
%% subscription status checks, perhaps using \textsf{member} for this
%% purpose.  This is because actions to add or remove users from the
%% database are effectful and impose constraints over the database state
%% that restrict how and when they can be performed.

Ordinarily, the types associated with such functions would not expose
these kinds of effects.  A typical type for an update operation on the
state maintained by a database library instance, for example, might
simply recognize that it performs a write effect by declaring its
return type to be \textsf{unit}, without providing specific details
about how the database actually changes.  Consequently,
state-of-the-art purely type-directed component-based synthesis
approaches~\cite{sypet,tygus} when applied to these kinds of libraries
could easily synthesize unsound programs because necessary protocol
constraints are not reflected in library function types.  Simply
embellishing a library with coarse read/write effect
annotations~\cite{rbsyn} is also unlikely to be effective for problems
like these because the lack of fine-grained effect tracking would
require the synthesis procedure to explore an intractably large space
of possible candidate compositions, since every successful call to an
effectful operation potentially alters underlying state.

However, advances in mechanized proof assistants and automated theorem
provers have made it increasingly worthwhile for library developers to
provide detailed specifications that can be used as part of a
verification task.  For example, F*~\cite{fstar} and
\textsf{VOCal}~\cite{vocal} are significant efforts aimed at
developing mechanically-verified effectful libraries of
general-purpose data structures and algorithms.  In this paper, we
show how library specifications of the kind produced by these efforts
can also be effectively repurposed to guide complex component-based
synthesis tasks.

Our approach introduces a new specification-guided synthesis strategy
that interprets a library's specification (expressed in terms of
Hoare-style pre- and post-conditions) as type
specifications~\cite{htt}, using a bi-directional search strategy to
enable scalability and precision.  Specifically, we use strongest
postcondition forward-reasoning to build larger terms from existing
ones maintained by a synthesis search context, and weakest
precondition backward-reasoning from the postcondition (the synthesis
query) to enable goal-directed search.  Alone, each process has
important weaknesses - forward reasoning lacks knowledge about the
synthesis goal, while backward reasoning has an incomplete view of the
context from which terms must be synthesized.  We show how to mitigate
these weaknesses, and exploit their underlying synergies, by
integrating both within a unified synthesis framework.  For improved
efficiency and scalability, we additionally leverage a
\emph{conflict-driven} learning
strategy~\cite{cdcl-sat,dpll,cdcl-synthesis} \DEL{in the context of
  effectful program synthesis}, to build a knowledge base that records
\emph{discriminating propositions} associated with previously
encountered incorrect synthesized terms that can be used to more
intelligently guide the search process and safely prune the space of
possible candidates we need to consider.  The need for such careful
integration arises from the unbounded search space that must be
explored to satisfy a query - having effectful libraries means that
every call to a library method in a synthesized term can potentially
result in a new heap state that reflects the effectful behavior of the
method, leading to a potential explosion of possible candidate
programs that the synthesis procedure may have to consider.

%% Thus, it is
%% conceivable that a synthesized program may need to make many calls to
%% the same procedure since each such call potentially operates over (and
%% produces) a new state

%% Thus, the complexity of the problem
%% manifests both in the sophistication of the synthesis queries that can
%% be expressed and the specific order and form in which effects must be
%% generated in synthesized terms to satisfy the query.

%% We have applied our synthesis approach on an effectful core-language
%% $\lambda_{\eff}$ loosely patterned after~\cite{effect-monad}
%% that enables the synthesis of loop-free programs containing
%% conditionals, sequencing, local binding, and library function calls.
%% We have implemented the synthesis procedure in a tool named
%% \name\ that operates over OCaml libraries and clients, and our
%% evaluation
%% its effectiveness in several distinct domains, including {\it
%%   imperative data structure libraries}, {\it database} libraries for
%% different database instances, and {\it parser libraries} used to write
%% parsers in the manner of combinator-based parsing for different
%% different parser instances.  We evaluate the effectiveness and
%% expressiveness of \name by synthesizing {\it \#x } programs from these
%% domains using expressive synthesis queries that cannot be easily
%% synthesized using state-of-the-art component-based synthesis
%% techniques.  Synthesis times range from \emph{time in
%%   seconds/minutes}.

This paper makes the following contributions:
\begin{itemize}

\item We present a novel bi-directional deductive synthesis strategy
  for specification-guided component synthesis of effectful libraries.
  The synthesis strategy alternates between forward and backward
  enumerative search, seeking to compose terms consistent with library
  specifications.

\item We address scalability issues by additionally integrating a
  CDCL-style learning component that builds a knowledge base of failed
  candidate terms that can be used as search proceeds to avoid
  reconsideration of previously identified infeasible terms.
  
\item We present detailed experimental results on an implementation of
  these ideas called \name\ that enables component-based synthesis of
  OCaml libraries equipped with effectful specifications.  Our results
  demonstrate the utility of our approach over a range of different
  application domains with respect to both expressivity of the
  synthesis queries that can be handled, and scalability over the size
  of the search space that must be navigated.
\end{itemize}

The remainder of the paper is organized as follows.  In the next
section, we provide a detailed overview of our approach.
Section~\ref{sec:efflib-core} presents a declarative bi-directional
type-checking formulation of the synthesis procedure.
Section~\ref{sec:alg} formalizes the synthesis algorithm along with
details of the CDCL-learning approach used to improve enumerative
search.  Additional details about the implementation, along with
benchmark results, are presented in Section~\ref{sec:impl}.  Related
work is given in Section~\ref{sec:rel}, and conclusions are presented
in Section~\ref{sec:conc}.

\section{Overview}
\label{overview}

% \SJ{What assumptions are we making about the library; specifically,
%   wrt aliasing?}

\begin{wrapfigure}{l}{.58\textwidth}
\begin{center}
  \includegraphics[scale=.25]{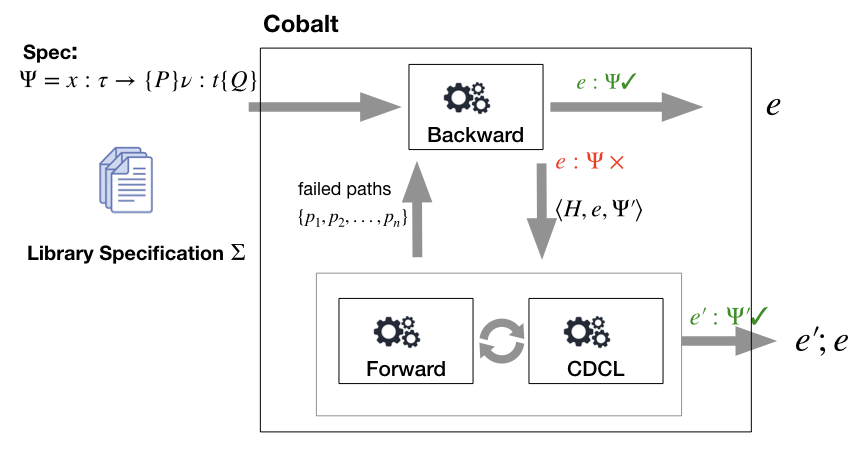}
\end{center}  
\caption{An overview of the \name\ synthesis process.}
\label{fig:overview}
\end{wrapfigure}
Figure~\ref{fig:overview} depicts \name's synthesis procedure and its
core components. \name takes as input a Hoare-triple style query
specification $\Psi$ and a set of library function specifications
$\Sigma$.  For $\Sigma$, we rely on available verified libraries like
~\cite{vocal,fstar} that come equipped with effectful specifications;
from a user's perspective, a Cobalt user thus only needs to provide a
declarative specification for the synthesis goal.

To motivate our approach, consider a string \textsf{Table} data
structure adopted from ~\cite{stateful-manifest-contract} 
and implemented using a mutable string list in an ML-like language. The
table maintains the invariant that its elements are pairwise
distinct. It provides a set of effectful library functions to add a
new string, to check membership, etc. on a table instance. These
components have associated specifications capturing their semantics as
shown in Figure~\ref{fig:lib}.

\begin{figure*}
\begin{subfigure}[b]{0.55\textwidth}
\begin{lstlisting}[basicstyle=\linespread{0.9}\small\sf,breaklines=true,language=ML]
type pair = Pair of float * int
type table = [string] ref

add_tbl : (tbl : table * s : string) -> 
{$\forall$ h, Tbl. sel (h, tbl) = Tbl $\wedge$ not (mem (Tbl, s)}
    v : unit 
$\forall$ h, v, h', Tbl, Tbl'. 
    sel (h', tbl) = Tbl' $\wedge$ sel (h, tbl) = Tbl $\wedge$
    mem (Tbl', s) $\wedge$ size (Tbl') == size (Tbl) + 1};

mem_tbl : (tbl : table * s : string) -> 
{true}    v : bool
{Tbl' = Tbl $\wedge$ ([v=true] <=> mem(Tbl', s)) $\wedge$
            ([v=false] <=> not (mem (Tbl', s)))};

size_tbl : (tbl : table) ->
{true} v : int {Tbl' = Tbl $\wedge$ v == size (Tbl)};

fresh_str : unit ->
{true} v : string {mem (Tbl', v) = false $\wedge$ Tbl' = Tbl};
 
avg_len_tbl : (tbl : table) ->
{size (Tbl) > 0} v : float  {Tbl' = Tbl $\wedge$ minmax (Tbl', v)};
 
clear_tbl : (tbl : table) -> {true} v : unit {size (Tbl') = 0};

(* remove less than *)
rlt_tbl : (tbl : table * s : string) ->
{true} v : unit  {size (Tbl') <= size (Tbl)};

(* remove greater than *)
rgt_tbl : (tbl : table * s : string) ->
{true} v : unit  {size (Tbl') <= size (Tbl)};
\end{lstlisting}
\caption{Effectful specifications for a \textsf{Table} library.}
\label{fig:lib}
\end{subfigure}
\begin{subfigure}[b]{0.35\textwidth}
\begin{lstlisting}[basicstyle=\small\sf,breaklines=true,language=ML]
(*A Safety Query-Spec*)
$\mathtt{goal1}$ :  (tbl : table * s : string) -> v:float

(*A Functional Query-Spec*)
$\mathtt{goal2}$ : (tbl : table * s : string) ->
{True}
   v : pair
{$\forall$ h, v, h, h', Tbl, Tbl'. 
  sel (h, tbl) = Tbl $\wedge$ sel (h', tbl) = Tbl' $\wedge$ 
  mem (Tbl', s) $\wedge$
  size (Tbl') = size (Tbl) + 1};

(*A Correct Solution*)
$\mathtt{goal2}$ (tbl : table * s : string) =  
  b1 <- mem_tbl (tbl, s);
  if (b1)
   then s1 <- fresh_str ();
         _ <- add_tbl(tbl, s1);
        x1 <- average_len_tbl (tbl);
        y1 <- size_tbl (tbl);  
        return Pair (x1, y1)
    else _ <- add_tbl (tbl, s);
        x1 <- average_len_tbl (tbl);
        y1 <- size_tbl (tbl);  
        return Pair (x1, y1)
\end{lstlisting}
\caption{Functional query-spec and a solution.}
\label{fig:solution}
\end{subfigure}
\caption{Effectful specifications for a \textsf{Table} library, Synthesis Queries and a Solution}
\end{figure*}

To capture the effectful behavior of these functions, we use
specifications that express pre- and post-conditions over abstract
heaps. For instance, the specification for the {\sf add\_tbl}
function (refer Figure~\ref{fig:lib} has a precondition that defines a stateful constraint on its
input string {\sf s}, requiring that it not be present in the table
referenced by {\sf tbl} in the input heap; specifications express
these constraints in terms of first-order predicate logic formulae
built from interpreted select/update operators~\cite{McCarthy1993} on
the heap (such as \textsf{sel}) and user-defined uninterpreted
function symbols like \textsf{mem} and \textsf{size} defined over
tables.  The postcondition captures the behavior of adding {\sf s} to
the table, assuming the precondition holds, by relating the state of
\begin{wrapfigure}{r}{.5\textwidth}
\vspace*{-.2in}  
\begin{center}
  \includegraphics[scale=.35]{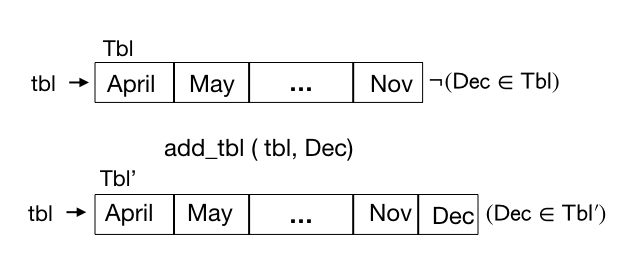}
\end{center}
\vspace*{-.2in}
\caption{A pictorial representation of the effects induced by a call to {\sf add\_tbl}.}
\label{fig:lib-pic}
\vspace*{-.2in}
\end{wrapfigure}
the table after the function completes (\textsf{Tbl'}) to its state on
entry (\textsf{Tbl}); specifically, it constrains the size of
\textsf{Tbl'} to be one more than \textsf{Tbl}, and asserts that
\textsf{s} is a member of \textsf{Tbl'}.  The \textsf{Tbl} and
\textsf{Tbl'} heap objects are accessed via input heap \textsf{h} and
output heap \textsf{h'}, respectively\footnote{We capitalize variables
  that correspond to ghost state in specifications; these are
intermediate computed heap values that do not appear as arguments or
  results of library functions.}.  The specifications given for other
library functions are similar but simplified to reduce clutter. For
instance, we drop quantifiers when obvious and assume {\sf Tbl} and
{\sf Tbl'} represent the table {\sf tbl} in pre- and post-heap {\sf h,
  h'}, respectively.

\DEL{Figure~\ref{fig:lib-pic} depicts the behavior of the
  {\sf add\_tbl (tbl, s)} library function over an input example table.
  The reference {\sf tbl} refers to a table (a list of strings) named
  {\sf Tbl} before the execution of the function.   An invocation {\sf
    add\_tbl (tbl, Dec) is executed provided the precondition that
    the string to be added (\textsf{Dec}) is not in {\sf
      Tbl} is true. If so, the functions adds the string to the
    table, yielding the post-state in which \textsf{tbl} now refers
    to a new table (labeled as {\sf Tbl'}). }  }
% \SJ{Could we also show how size changes in the figure?}

\paragraph{Synthesis Problems}
Given this library, there are two kinds of synthesis queries ({\it
  query-specs}) that can be made.  One is a type inhabitation query
similar to what is possible in other type-directed component-based
synthesis approaches~\cite{sypet, tygus, rbsyn}. We call these queries
{\it safety queries} since the synthesis goal is to generate a
type-safe term using library components.  Figure~\ref{fig:solution}
shows such a query (\texttt{goal1}) for a function that given a table
instance and as arguments returns a \textsf{float}.  A solution to
this query might be a function that applies \textsf{add\_tbl} to the
goal's arguments and returns the average length of the new table via
\textsf{avg\_len\_tbl}.

Our focus in this paper, however, is on solving richer queries that
exploit interesting effect-based functional correctness properties
desired of the client program. For instance, our goal might be to
synthesize a function that adds its input string to a table, returning
a pair of the average length and size of the new table. These kinds of
{\it effectful queries} can be specified using a {\it query-spec} such
as the one for \texttt{goal2} shown in Figure~\ref{fig:solution}.
% 
% \begin{figure}
% \begin{lstlisting}[basicstyle=\small\sf,breaklines=true,language=ML]
% (*A Safety Query-Spec*)
% $\mathtt{goal1}$ :  (tbl : table * s : string) -> v:float
% 
% (*A Functional Query-Spec*)
% $\mathtt{goal2}$ : (tbl : table * s : string) ->
% {True}
%    v : pair
% {$\forall$ h, v, h, h', Tbl, Tbl'. 
% sel (h, tbl) = Tbl $\wedge$ sel (h', tbl) = Tbl' $\wedge$ 
% mem (Tbl', s) $\wedge$ size (Tbl') = size (Tbl) + 1};
% 
% (*A Correct Solution*)
% $\mathtt{goal2}$ (tbl : table * s : string) =  
%   b1 <- mem (s);
%   if (b1) then  s1 <- fresh_str ();
%               _ <- add_tbl(tbl, s1);
%              x1 <- average_len_tbl (tbl);
%              y1 <- size_tbl (tbl);  
%              return Pair (x1, y1)
%   else  _ <- add_tbl s;
%       x1 <- average_len_tbl (tbl);
%       y1 <- size_tbl (tbl);  
%       return Pair (x1, y1)
% \end{lstlisting}
% \caption{Functional query-spec and a solution.}
% \label{fig:solution}
% \end{figure}

This goal specifies a function that takes a table instance
(\textsf{tbl}) and string parameter ({\sf s}) as arguments and whose
body satisfies the provided pre- and post-conditions. The
precondition imposes no constraints on the table or string.  The
postcondition is a relation between the table in the pre-heap
(\textsf{Tbl}) and the post-heap (\textsf{Tbl'}); it requires the
synthesized function to produce a \textsf{Tbl'} whose size is one
greater than \textsf{Tbl} and that contains the string \textsf{s}.
The result type of the function, however, additionally requires that
the function return a \emph{pair} of \textsf{float} and \textsf{int}
values.  Observe that there are no functions in the given interface
that explicitly returns a pair, although \textsf{size\_tbl} returns the
size of the table as an \textsf{int}, and \textsf{avg\_len\_tbl}
returns a value of type \textsf{float}, provided that its table argument
is not empty.

\subsection{Solution Overview}

To explain how \name\ synthesizes a suitable function (shown in
Figure~\ref{fig:solution}) given these various constraints, we first
explain the details of its bi-directional search strategy and 
CDCL-inspired search algorithm.

\subsubsection{Weakest Precondition Guided Search}

The synthesis procedure begins in a {\it backward} phase, starting
from the query's postcondition and return type.  \DEL{It maintains a
  {\it context}, a list of library functions and arguments provided by
  the user in the {\it query-spec}, as well as path conditions}.  The
procedure searches for a function {\sf f} under an initial context
that contains the arguments declared in the query (e.g. s) and
libraries, such that {\sf f} can be invoked in this context, leading
to the required postcondition. To make this decision, \name\ uses a
weakest precondition call rule and performs the following check,
assuming the library specification for {\sf f} is ($\overline{x_i :
  \tau_i}$) $\rightarrow$ \{$Pre_f$\} v : t \{$Post_f$\}:
\begin{lstlisting}[basicstyle=\small\tt,language=ML]
$\forall$ h. ([$y_i$/$x_i$]$Pre_f$) h => 
    ($\forall$ v:t. h'. ([$y_i$/$x_i$]$Post_f$) h v h' => Post h v h')
\end{lstlisting}
Given synthesized arguments $y_i$, it searches for a function $f$ to
which these arguments can be applied such that $f's$ precondition is
satisfied by the existing context, and $f$'s postcondition implies the
postcondition of the query.  \DEL{To illustrate, suppose we have a query goal}
\begin{lstlisting}[basicstyle=\small\tt,language=ML]
($y_i$ : int) -> { dom (h, i) = true} {v : 'a} { $y_i$ $\geq$ 5 $\wedge$ sel (h', i) $\leq$ 20 }    
\end{lstlisting}
\DEL{If the synthesis procedure context includes a condition ({\sf x}
$\geq$ 10), then given two functions with signatures,}
\begin{lstlisting}[basicstyle=\small\tt,language=ML]
$f$ : (x : int) $\rightarrow$ { true } v : int { x == 10 $\wedge$ sel (h', i) == 10 }
$g$ : (x : int) $\rightarrow$ { true } v : int { x == 4 }
\end{lstlisting}
\DEL{the synthesis procedure synthesizes the call, $f$(\textsf{x}), discarding the synthesis
candidate $g$(\textsf{x}).}
Note that this is a very strong requirement since the postcondition
might impose additional constraints not considered by $f$. For
instance we might have a situation where the above check does not
hold, but in which \textsf{Post} can be translated into a form {\sf R
  $\wedge$ Post'} where {\sf R} is a {\it frame}~\cite{separation}. In
such a case, we can choose a function candidate if it satisfies the
following weaker check:
\begin{lstlisting}[basicstyle=\small\tt,language=ML]
$\forall$ h. R $\wedge$ ([$y_i$/$x_i$]$Pre_f$) h => 
    ($\forall$ v:t. h'. ([$y_i$/$x_i$]$Post_f$) h v h' => Post' h v h')
\end{lstlisting}
This is an instance of a \emph{framing} problem and in
Section~\ref{sec:forward}, we discuss important optimizations that
allow us to soundly weaken this rule to allow partial satisfaction of
the query's goal.

If \name\ does not find any effectful function satisfying the check,
it searches for a pure function with the required return type and
generates a subprogram with holes called {\it hypotheses} and a
weakest precondition for this pure function call.  In our example,
there is no effectful library function that immediately satisfies the
goal.  Thus, \name\ chooses the pure {\sf Pair} constructor and
generates a term represented by the following derivation:
\begin{lstlisting}[caption={A backward (postcondition)-guided derivation; we omit
      introduction of the table instance \textsf{tbl} for perspicuity - \textsf{Tbl} and \textsf{Tbl'}
      represent \textsf{tbl}'s value in the pre- and post-heap, resp.},captionpos=b,label={lst:bwsearch},basicstyle=\small\tt,language=ML]
[(s: string)] $\vdash$ {True} (v : pair) 
{size (Tbl') = size (Tbl) + 1 $\wedge$ mem (Tbl' s)} 
    $\leadsto$
[(s: string)] $\vdash$ 
{ True }
 x1 <-  {(??) : float}; y1 <- {(??) : int} 
{size (Tbl') = size (Tbl) + 1 $\wedge$  mem (Tbl', s)} 
--------------------------
return Pair (x1, y1)}
\end{lstlisting}
The derivation is of the form,
\[
\Gamma \vdash \{\textsf{P}\} (v : \tau) \{\textsf{Q}\}) \leadsto (\Gamma\ \vdash \{\textsf{P}\} H; \textsf{WP} (t, \textsf{Q}) \mid t'\]
\noindent and yields a hypothesis $H$, a possibly \emph{holed} term,
along with a predicate (\textsf{WP}($t$, \textsf{Q})) constructed by
applying the weakest precondition semantics for the term chosen by the
synthesis algorithm; in the above example, $t'$ = \textsf{return Pair
  (x1, y1)}.  Since there are no available terms in the context
corresponding to the required constructor arguments (an \textsf{int}
and \textsf{float}), a new term is created with two holes of
appropriate argument types; these terms are bound to fresh variables
and used as arguments to the \textsf{Pair} constructor.  Thus, $H$
captures potential program shapes for the current synthesis choice.
Observe that neither \textsf{avg\_len\_tbl} nor \textsf{size\_tbl},
which can contribute to the appropriate return type, have
specifications that align with the currently synthesized term; for
example, \textsf{avg\_len\_tbl}, although returning a \textsf{float},
also requires that its input table size is greater than 0, a property
that is not ensured by the holed term's precondition (\textsf{True}).

Consequently, further progress on the backward derivation stalls.
Rather than aborting and searching for a new candidate, we instead
proceed to apply a forward search starting from the existing
precondition of the current term.  To make sure that we avail of
the information learnt from the backward derivation, we
pass $H$ (the sequence of holed terms constructing \textsf{x1} and \textsf{y1})
and the current weakest precondition - 
{\small\tt
  size (Tbl') = size (Tbl) +1 $\wedge$  mem (Tbl', s) -
} as the new postcondition for the forward search.

Although our backward search in this example simply does a pattern
matching over the data constructor {\sf Pair}, our synthesis procedure
can, in fact, generate interesting non-trivial programs of larger size
by searching for a function in the library iteratively at each step,
such that its specification allows for valid weakest-precondition
reasoning for the given post-condition.  For instance, consider a
simple synthetic example with two library functions, \textsf{m1} and
\textsf{m2}:\\
\begin{minipage}{0.5\textwidth}
\begin{lstlisting}[basicstyle=\linespread{0.9}\small\sf,breaklines=true,language=ML]
res : ref int;
flag : ref bool;
m1  : unit -> {res = 7}   v : int {res' = res + 3}      
\end{lstlisting}
\end{minipage}
\begin{minipage}{0.45\textwidth}
\begin{lstlisting}[basicstyle=\linespread{0.9}\small\sf,breaklines=true,language=ML]
m2 :  unit -> {res = 5 $\wedge$ flag = false} 
         v : int
      {res' = res + 2 $\wedge$ flag' = true}
\end{lstlisting}
\end{minipage}
\noindent
\newline These specifications highlight how the functions manipulate two mutable references:
an integer {\sf res}, and a boolean {\sf flag}.
Given this library and the following goal query:
\begin{lstlisting}[basicstyle=\small\tt,breaklines=true,language=ML]
   goal :  {res = 0 $\wedge$ flag = false} (v : unit) {res' = 10};
\end{lstlisting}    
backward synthesis starts by generating a partial term (a term
with holes) ($\lambda$ (). {\sf (??) : unit; m1 ()}) as the
specification for {\sf m1} satisfies the weakest-precondition check.
It then continues trying to fill the hole to find the next such
function with ({\sf WP( {\sf m1 ()}, \{res' = 10\}}) as an updated
postcondition.  Now, the specification for {\sf m2} satisfies this
check and thus a bigger program is created, consequently generating
the term ($\lambda$ ().{\sf \sf (??) : unit ; m2(); m1()}), before
passing the synthesis to the {\it forward-synthesis} component.

\begin{figure*}[!t]
%\begin{subfigure}[t]{0.57\textwidth}
\advance\leftskip-1cm
\centering 
    \includegraphics[width=1.0\textwidth]{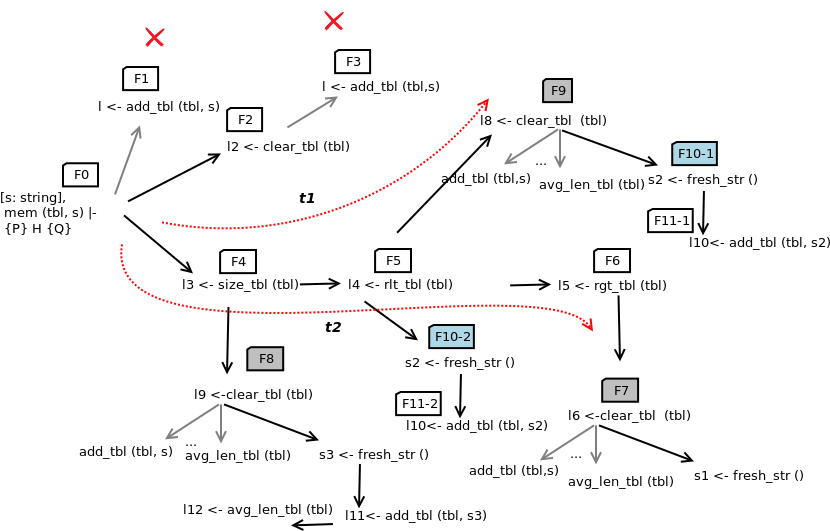}
    \caption{A partial forward enumeration. \textcolor{gray}{gray}
      edges show type- and/or specification-incorrect discarded
      transitions, {\bf black } edges show type-correct, potentially
      explorable transitions, and \textcolor{red}{red-dashed}
      edges give examples of type-correct paths that do not lead to required solution. Labeled nodes with same color show equivalent-modulo-stuckness nodes.}
    \label{fig:explore}
    %\end{subfigure}
\end{figure*}

 % 
% \begin{figure}[htbp]
% \resizebox{1.05\linewidth}{!}{\input{fig/overview_hl2.tex}}
% \caption{A transducer for parsing a string {\sf s} and returning the parsed string if successful. {\sf s} is a parameter, {\sf inp} is the input list being parsed.} 
% \label{fig:string-trans}
%  \end{figure}

Forward synthesis has two subcomponents (refer Figure~\ref{overview}) that collectively attempt to
synthesize a term $e'$, which when unified with the partial term $e$, synthesized by the backward synthesis, gives the required
solution. If it fails to find such a term, it again invokes the {\sf
 Backward} component with information about the failed program.
Forward synthesis uses strongest post-condition reasoning to
synthesize a term in a forward fashion. 
\subsubsection{Strongest Postcondition Guided Search}

Because the body of the function being synthesized may contain
top-level conditional branches or match-case statements, search begins
by first synthesizing the body of the function as a top-level
branching term.  Specifically, \name\ looks for functions with Boolean
return types, or arguments in the query-spec that have inductive type
(like lists, trees, etc.) and applies proof rules for {\it
  if-then-else} and {\it match} respectively to introduce top-level
conditional branching or matching synthesis sub-problems.  

For instance, given the specification in Listing~\ref{lst:bwsearch},
and the specification for Boolean-valued library function {\sf
  mem\_tbl} (refer Figure~\ref{fig:lib}), \name generates the
synthesis sub-queries shown in Figure~\ref{fig:forward-ex} by applying
an {\it if-then-else} synthesis rule and introducing true and false
postconditions from \textsf{mem\_tbl}'s specification to the
preconditions of the term's \emph{true} and \emph{false} branches
(shown in red).  We now explain how \name\ proceeds with the synthesis
query for the \emph{true} branch.  The enumeration procedure is
conceptually similar to the backward derivation, but uses the
strongest postcondition of the already-built term to guide the choice
of the next library component candidate.  Intuitively,
\name\ iteratively builds type-correct program terms of increasing
length in a depth-first manner until it finds a program that satisfies
the hypotheses (the holed terms), and the required specifications.
Not surprisingly, the domain of such terms is unbounded since we can
generate distinct unrolled looping terms like $t_1;{t_2}^{*};t_3$,
${t_1;{t_2};t_3}^{*}$, etc. of arbitrarily large depth.  To mitigate
this situation, we bound the maximum depth of search to a depth value
{\it k} on the length of program terms.  The value of {\it k} can be
chosen by the user and can be iteratively incremented.  For ease of
presentation, assume ({\it k} = 5) for our running example.
\begin{wrapfigure}{r}{.54\textwidth}
\begin{lstlisting}[basicstyle=\small\tt,breaklines=true,language=ML]
$\lambda$ (tbl : table, s : string).
 b1 <- mem_tbl (tbl, s);
 if (b1) then 
  {$\forall$ h. sel (h, tbl) = Tbl $\wedge$ $\textcolor{red}{(mem (Tbl, s)}$} 
     (x1 <- {(??) : float}; y1 <- {(??) : int} 
  {size (Tbl') = size (Tbl +1) $\wedge$ mem (Tbl' s)}
 else 
  {$\forall$ h. sel (h, tbl) = Tbl $\wedge$ $\textcolor{red}{not (mem (Tbl, s)}$} 
     (x1 <- {(??) : float}; y1 <- {(??) : int} 
  {size (Tbl') = size (Tbl +1) $\wedge$ mem (Tbl' s)}
\end{lstlisting}
\caption{A partial program with holes and top-level branching.}
\label{fig:forward-ex}
\end{wrapfigure}
%\noindent
Suppose during enumeration we have synthesized a term $t$ of size less
than {\it k}.  The procedure first tries to verify if $t$ is the
required solution by performing two checks that ascertain whether: a)
$t$ satisfies the hypothesis $H$ ($t$ $\prec$ $H$); and, b) 
\textsf{SP} (\textsf{P}, $t$) $\Rightarrow$
\textsf{Q'}, where \textsf{P} and \textsf{Q'} are pre- and
post-specifications for the query specification.  Informally,
$t\ \prec\ H$ if $t$ is a term that has the same shape as $H$
with every holed term replaced by a type-consistent concrete
one.

If $t$ satisfies these checks, the procedure returns $t$ as the
required solution. However, if either of the two checks fail, a search
commences to look for a component \textsf{f} that can be sequenced
with $t$. To guide this search, \name\ uses a strongest postcondition
call rule and performs the following check, assuming the library
specification for \textsf{f} is ($\overline{x_i : \tau_i}$)
$\rightarrow$ \{$Pre_f$\} v : t \{$Post_f$\}:\\
{\small\tt
\hspace*{.23in} $\forall$ h. (SP (P, t)) h => ([$y_i$/$x_i$]$Pre_f$) h}\\
\DEL{To illustrate, consider again the example scenario discussed in
  the previous section. Suppose we have a query; goal:}
\begin{lstlisting}[basicstyle=\small\tt,language=ML]
($y_i$ : int) $\rightarrow$ { dom (h, i) = true } v : 'a { $y_i$ $\geq$ 5 $\wedge$ sel (h',  i) $\leq$ 20 }
\end{lstlisting}
\DEL{and that the synthesis procedure context includes the
  condition, {\sf x} $\geq$ 10, possibly learnt from a 
conditional expression.  Given two functions with signatures:}
\begin{lstlisting}[basicstyle=\small\tt,language=ML]
$f$ : (x : int) $\rightarrow$ { x $\leq$ 20 } v : int { x == 10 $\wedge$ sel (h',  i) == 10 }
$g$ : (x : int) $\rightarrow$ { x $\leq$ 5 } v : int { x == 4 }
\end{lstlisting}  
\DEL{the synthesis procedure synthesizes the call $f$ (\textsf{x}),
    discarding the synthesis candidate $g$(\textsf{x}).}

Figure~\ref{fig:explore} presents a pictorial representation of this
enumeration process for some of the \emph{true} branch paths of the
synthesis query.  Each choice made by the search process is given a
label (\texttt{F$i$}). The edges include black edges representing
explored edges as well as gray edges showing discarded choices
encountered by the forward call-rule. For instance, edge (\texttt{F0
  -> F1}) is disallowed since the branch precondition
(\textsf{mem(Tbl, s)}) is inconsistent with the precondition for
({\sf add s}). Similarly, we have other discarded edges like
(\texttt{F2 -> F3}), etc.  The rule thus allows the search procedure
to discard multiple incorrect programs early on; for instance, given
the initial query above, \name prunes out incorrect paths such as
\textsf{\{\_ <- add\_tbl (s); ....\}} thereby significantly reducing
the search space of possible candidates.

\subsubsection{Conflict-driven learning based enumeration}
\label{sec:cdcl-overview}

While the forward (similarly backward) search process can prune out
many incorrect programs quickly, the number of candidate programs
(paths with black edges in the figure) is likely to be still very
large, making a na\"{\i}ve enumerative search over this space
infeasible.  A primary reason is the likelihood of repeated
exploration of previously seen paths allowed by forward reasoning that
do not lead to the goal postcondition. For instance, consider the path
(\texttt{F0 -> F4 -> F5 -> F6 -> F7}).  Here, the corresponding
program term is not a solution since the postcondition of the
synthesized term:
%\newpage
\begin{lstlisting}[basicstyle=\small\tt,breaklines=true,language=ML]
 $\exists$ Tbl1, Tbl2, Tbl3. $\forall$ Tbl, Tbl'. 
    Tbl1 = Tbl $\wedge$ size (Tbl2) <= size (Tbl1) $\wedge$ 
    size (Tbl3) <= size (Tbl2) $\wedge$	
    size (Tbl') = 0  
\end{lstlisting}
does not imply the postcondition of the true branch of the current
synthesis candidate, which requires that\\
{\small\tt
\hspace*{.23in}size (Tbl') = size (Tbl +1) $\wedge$ mem (Tbl'
s)}\\ Further, since the depth of the path when reaching \textsf{F7}
(here, 4) is less than the depth-bound (k=5), the search process seeks
a component that can satisfy the required postcondition.
Unfortunately, no such choice is possible. Thus, the algorithm
backtracks and makes a different choice at one of the earlier
nodes. We call a node like \texttt{F7} that can no longer make
progress for a given {\it k} as a {\it k-bound-stuck-node}, inspired
by the notion of a {\it conflict-node} in conflict-driven learning
approaches~\cite{cdcl-sat,ZMC+01}.

Suppose, the algorithm backtracks to node \texttt{F5} (over multiple
steps); unfortunately, it again faces a choice similar to the choice
at the stuck-node \texttt{F7} to select/discard the function {\sf
  clear\_tbl}.  Since the term corresponding to path (\texttt{F0 -> F4 ->
  F5 -> F9}) (call it $t_1$) is syntactically distinct from the term
corresponding to (\texttt{F0 -> F4 -> F5 -> F6 -> F7}) (call it
$t_2$), the algorithm cannot trivially prune out this choice.  But, by
choosing \texttt{F9}, it must again performs checks similar to those
performed under the $k$-bound-stuck-node \texttt{F7}. For instance,
the algorithm would again visit discarded terms (gray edges) like
({\sf add\_tbl(tbl, s), avg\_len\_tbl (tbl), etc.}). Finally, it will
eventually find itself at another $k$-bound-stuck node
(\texttt{F10-1}) causing it to backtrack to \texttt{F5}.  Once again,
it faces a choice (\texttt{F5 -> F10-2}), which is similar to the new
stuck node found at (\texttt{F10-1}).  Figure~\ref{fig:explore}
depicts how the algorithm repeatedly traverses many paths similar to
an already visited stuck path, and must therefore eventually
backtrack.

%% Intuitively, the main problem we
%% face (as do other enumerative search-based synthesis approaches) is
%% that simple enumeration fails to learn from past unsuccessful
%% exploration, and is therefore unable to avoid repeatedly making the
%% same ineffective choices.

Even though concluding the similarity between $t_1$ and $t_2$ is not
possible syntactically, we observe that this similarity arises because
the algorithm is making the same function choice (i.e. \textsf{clear\_tbl}) and the strongest postconditions \textsf{SP (P, $t_1$)} and
\textsf{SP (P, $t_2$)} that guide the search process under \textsf{F9}
and \textsf{F7} are related by \textsf{(SP (P, $t_1$) $\Rightarrow$ SP
  (P, $t_2$))}.  In an unbounded depth-first search (k=$\infty$), for
any path explored under \textsf{F9}, there exists a path explored
under \textsf{F7}, and hence by knowing that \textsf{F7} does not lead
to a solution (i.e. is a $k$-bound-stuck-node), we can conclude that
\textsf{F9} will also not lead to a solution.  We can, therefore,
discard the exploration of the tree rooted at \textsf{F9} by learning
the postconditions at \textsf{F7} and \textsf{F9}. We call nodes like
(\textsf{F7} and \textsf{F9}) as {\it equivalent-modulo-stuckness},
and highlight their similarity by depicting them with the same color
in Figure~\ref{fig:explore}.

However, since our exploration is bounded by terms of size {\it k},
there may be paths that were prematurely truncated under \textsf{F7}
but can make progress under \textsf{F9}. For example, consider a path
(\textsf{F10-1 -> F11-1}) under \textsf{F9}.  Eagerly discarding
\textsf{F9} would lead us to miss such paths and may result in failure
to satisfy a feasible synthesis query under a given \emph{k} bound.
Notice, however, that for each such path under \textsf{F9}, there is a
smaller path e.g. (\textsf{F10-2 -> F11-2}) that is also reachable and
can lead to a solution if the longer path under \textsf{F9} can.  Thus
\textsf{F7} and \textsf{F9} can be assumed to be logically
equivalent-modulo-stuckness; we can thus safely discard the
exploration of the tree rooted at \textsf{F9}, given that there is an
equivalent path at \textsf{F10-2}.

Based on the above observations, we equip our search procedure with a
conflict-driven learning component called {\it CDCL-search} that
learns {\it discriminating propositions} ($D^{k}$ (\textsf{Fi}))
associated with each visited {\it k-bound-stuck-node} {\sf Fi} and
uses them to discard future exploration of nodes that are logically
equivalent to an earlier k-bound-stuck-node, modulo the stuckness
property.

% \SJ{This section seems unnecessarily detailed.  We should be able to
%   state what CDCL is doing on this example without being so low-level.}
We explain the working of the algorithm using our running example.
Upon encountering a {\it k-bound-stuck-node} (e.g. node F7), our CDCL
search procedure learns two propositions.  First, it learns a
proposition $S_p$, which we call the {\it stuck-path proposition} that
captures the post-state for the k-bound-stuck-node (e.g. {\sf F7}).
Second, it creates a disjunctive formula $T_p$ called {\it truncated
  proposition} containing a disjunct for each call truncated
prematurely for the k-bound-stuck-node (e.g. {\sf F7} -> fresh\_str).
The idea is to learn information about paths that were taken but were
prematurely left unexplored due to the bound $k$.  The
\emph{discriminating proposition} for a k-bound-stuck node $D^{k}$
(k-bound-stuck-node) is given by a tuple ($S_p$, $T_p$).
% \SJ{This is
%   confusing - a tuple is not a set, and neither the rest of the discussion
%   nor the formal semantics uses the tuple notation. By tuple, do you
%   mean disjunct?}

% 
% First, it learns the
% strongest postcondition for the path leading to stuck nodes, i.e.,
% \textsf{SP (P, $t_2$)}, as given above. Let us call this a {\it
%   stuck-path proposition} ($S_p$).  Next, for each allowed choice {\sf
%   fj (...)} (black edges under F7), the algorithm learns the
% precondition for {\sf fj (...)}.  For instance, we learn
% preconditions for {\sf fresh\_str ()}.  The idea is to learn
% information about paths that were taken but were prematurely left
% unexplored due to the bound $k$.  This gives us the following
% \emph{truncated proposition} ($T_p$): $\forall$ \texttt{Tbl.  true}.
% The \emph{discriminating proposition set} $D^{k}$ (\textsf{F7}) is thus
% given by a tuple ($S_p$, $T_p$).

%% Additionally, it also learns two
%% disjunctions - (a) for each choice {\sf fi (...)} that is discarded
%% (gray edges under F7) during forward search, the algorithm learns the
%% precondition for {\sf fi (...)} as a disjunct. For instance, we learn
%% preconditions for {\sf add\_tbl (tbl,s), avg\_len (tbl), etc.}.  This
%% gives us the following \emph{failure-predicate} disjunction, let us
%% call it the {\it failure-predicate} ($F_p$):
%% \begin{lstlisting}[basicstyle=\small\tt,breaklines=true,language=ML]
%%   $\forall$ Tbl.  mem (Tbl, s) = false $\vee$ size (Tbl) > 0 
%% \end{lstlisting}    

\begin{wrapfigure}{l}{.45\textwidth}
  %\SJ{Too complicated}
\begin{lstlisting}[basicstyle=\small\tt,breaklines=true,language=ML]
$\Bigg\{$not ( $\forall$ Tbl'. size (Tbl') = 0  =>  size (Tbl') = 0) $\Bigg\}$ $\vee$ 
$\Bigg\{$($\forall$ Tbl, Tbl'. ... $\wedge$ size (Tbl') = 0  => true $\wedge$ 
	  (not ($\forall$ Tbl, Tbl'. size (Tbl') <= size (Tbl) => true)) $\Bigg\}$
\end{lstlisting}
\caption{Checks derived using the discriminating proposition $D^{k}$ (\textsf{F7}) in CDCL-search.}
\label{fig:checks}
\end{wrapfigure}

% \begin{figure}[hbt]
%   %\SJ{Too complicated}
% \begin{lstlisting}[basicstyle=\small\tt,breaklines=true,language=ML]
% $\Bigg\{$not ( $\forall$ Tbl'. size (Tbl') = 0  =>  size (Tbl') = 0) $\Bigg\}$ $\vee$ 
% $\Bigg\{$($\forall$ Tbl, Tbl'. ... $\wedge$ size (Tbl') = 0  => true $\wedge$ 
% 	  (not ($\forall$ Tbl, Tbl'. size (Tbl') <= size (Tbl) => true)) $\Bigg\}$
% \end{lstlisting}
% \caption{Using the discriminating proposition $D^{k}$ (\textsf{F7}) in CDCL-search.}
% \label{fig:checks}
% \end{figure}

$D^{k}$ (\textsf{F7}) can help us to discard logically equivalent-modulo
stuck nodes: the algorithm backtracks with this learned information to
the earlier decision node \textsf{F5}; while making the decision at
edge (\textsf{F5 -> F9}) with the $D^{k}$ (\textsf{F7}) information in hand,
the algorithm checks if the decision node \textsf{F9} is logically
equivalent-modulo-stuckness with the earlier encounter of {\sf clear\_tbl(tbl)}.
The algorithm performs the following checks, where $t_3$ is the term
corresponding to path (\textsf{F0 -> F4 -> F5}), and $\den{\textsf{F7}}$ =
\textsf{clear()}, the function invoked at node \textsf{F7} in Figure~\ref{fig:explore}:\\
%\begin{lstlisting}[basicstyle=\small\tt,breaklines=true,language=ML]
\hspace*{.23in}{\small\tt
  (not ( $D^{k}$($\den{\textsf{F7}}$).$S_p$ => SP (P, $t_1$)))  $\vee$ \\
\hspace*{.23in}  (SP (P, $t_1$) => $D^{k}$($\den{\textsf{F7}}$).$T_p$ $\wedge$ \\
\hspace*{.23in}      not (SP (P, $t_3$) => $D^{k}$($\den{\textsf{F7}}$).$T_p$))\\
}
%\end{lstlisting}    
Intuitively, these two disjuncts check the two observations discussed
earlier.  The first disjunct captures the fact that any path that can
be explored by making this choice was already visited and seen to be
leading to a stuck-node under the earlier exploration of {\sf clear\_tbl(tbl)}
at \textsf{F7}.  The second disjunct verifies that for any node that
was truncated prematurely under the stuck-node and that can make
progress under the current choice, there is an equivalent path in a
tree rooted outside \textsf{F9}.  For our running example, this
translates and simplifies to checks shown in Figure~\ref{fig:checks}.
Since both these disjuncts are false, the CDCL-search algorithm
decides that for the current value of {\it k}, the two nodes are
logically equivalent-modulo-stuckness and  it can thus safely discard
the exploration of \textsf{F9}.
% 
% \begin{figure}[htb]
% \begin{center}
%   \includegraphics[scale=.25]{fig/cobalt-arch.pdf}
% \end{center}  
% \caption{An overview of the \name\ synthesis process.}
% \label{fig:overview}
% \end{figure}

The forward-algorithm continues the exploration with learning until it
finds a solution for the given value $k$.  If it fails to find a
program, it returns the failed paths of lengths upto {\it k} to the
backward search again, in a \emph{handshaking} step.  At this point,
the backward algorithm may need to backtrack and make different
choices.  By supplying failing information about paths, the backward
search can avoid choosing equivalent terms modulo these failures.  We
depict the various components of the synthesis procedure in
Figure~\ref{fig:overview}.  Applying these mechanisms (backward+forward+cdcl) to the original
goal (\textsf{goal2}), \name\ synthesizes the solution shown in
Figure~\ref{fig:solution} in approximately 7 seconds. 
%\SJ{Yes, I think this
 %s would be helpful.}  The forward-alone synthesis
(i.e. forward+cdcl) finds a solution in 10 seconds; a forward-no-cdcl
synthesis strategy explores many more paths (compared to forward+cdcl)
and takes 28 seconds, while a backward-alone synthesis strategy fails
to find a solution within a 10 minute time-bound.

\section{\name\ Synthesis}
\label{sec:efflib-core}

\newcommand{\ottdrule}[4][]{{\displaystyle\small\frac{\begin{array}{c}#2\end{array}}{#3}\quad\ottdrulename{#4}}}
\newcommand{\ottusedrule}[1]{\[#1\]}
\newcommand{\ottpremise}[1]{#1 \\}
\newenvironment{ottdefnblock}[3][]{ \framebox{\mbox{#2}} \quad #3 \\[0pt]}{}
\newenvironment{ottfundefnblock}[3][]{ \framebox{\mbox{#2}} \quad #3 \\[0pt]\begin{displaymath}\begin{array}{c}}{\end{array}\end{displaymath}}
\newcommand{\ottfunclause}[2]{ #1 \equiv #2 \\}
\newcommand{\ottnt}[1]{\mathit{#1}}
\newcommand{\ottmv}[1]{\mathit{#1}}
\newcommand{\ottkw}[1]{\mathbf{#1}}
\newcommand{\ottsym}[1]{#1}
\newcommand{\ottcom}[1]{\text{#1}}
\newcommand{\ottdrulename}[1]{\textsc{#1}}
\newcommand{\ottcomplu}[5]{\overline{#1}^{\,#2\in #3 #4 #5}}
\newcommand{\ottcompu}[3]{\overline{#1}^{\,#2<#3}}
\newcommand{\ottcomp}[2]{\overline{#1}^{\,#2}}
\newcommand{\ottgrammartabular}[1]{\begin{supertabular}{llcllllll}#1\end{supertabular}}
\newcommand{\ottmetavartabular}[1]{\begin{supertabular}{ll}#1\end{supertabular}}
\newcommand{\ottrulehead}[3]{$#1$ & & $#2$ & & & \multicolumn{2}{l}{#3}}
\newcommand{\ottprodline}[6]{& & $#1$ & $#2$ & $#3 #4$ & $#5$ & $#6$}
\newcommand{\ottfirstprodline}[6]{\ottprodline{#1}{#2}{#3}{#4}{#5}{#6}}
\newcommand{\ottlongprodline}[2]{& & $#1$ & \multicolumn{4}{l}{$#2$}}
\newcommand{\ottfirstlongprodline}[2]{\ottlongprodline{#1}{#2}}
\newcommand{\ottbindspecprodline}[6]{\ottprodline{#1}{#2}{#3}{#4}{#5}{#6}}
\newcommand{\ottprodnewline}{\\}
\newcommand{\ottinterrule}{\\[5.0mm]}
\newcommand{\ottafterlastrule}{\\}
\newcommand{\ottmetavars}{
\ottmetavartabular{
 $ \ottmv{variables} ,\, \ottmv{x} ,\, \ottmv{y} ,\, \ottmv{z} ,\, \ottmv{h}  ,\, \nu  $ & \ottcom{variables} \\
 $ \ottmv{constructors} ,\, \mathsf{D_i}  $ & \ottcom{constructors} \\

 $ \ottmv{function names} ,\, \ottmv{f} ,\, \ottmv{g} $ & \ottcom{function 
names} \\
 $ \ottmv{constants} ,\, \ottmv{c} ,\, \ottmv{true} ,\, \ottmv{false}$ & 
\ottcom{constants} \\
 $ \ottmv{qualifiers} ,\, \ottmv{P} ,\, \ottmv{Q} ,\, \ottmv{R} $ & 
\ottcom{qualifiers} \\
 $ \ottmv{index} ,\, \ottmv{i} ,\, \ottmv{j} ,\, \ottmv{n} ,\, \ottmv{m} $ & \ottcom{indices} \\
}}

\newcommand{\otte}{
\ottrulehead{\ottnt{e}}{::=}{\ottcom{monExp}}\ottprodnewline
\ottfirstprodline{|}{\ottmv{x}}{}{}{}{\ottcom{variable}}\ottprodnewline
\ottprodline{|}{\lambda  \ottmv{x}  \ottsym{:}  \tau  \ottsym{.}  \ottnt{e}}{}{}{}{\ottcom{abstraction}}\ottprodnewline
\ottprodline{|}{\ottnt{e} \, \ottnt{e'}}{}{}{}{\ottcom{application}}\ottprodnewline
\ottprodline{|}{\ottkw{skip}}{}{}{}{\ottcom{skip}}\ottprodnewline
\ottprodline{|}{\ottsym{!}  \ottmv{x}}{}{}{}{\ottcom{deref}}\ottprodnewline
\ottprodline{|}{\ottmv{x}  \ottsym{:=}  \ottnt{v}}{}{}{}{\ottcom{assign}}\ottprodnewline
\ottprodline{|}{\ottkw{ref} \, \ottnt{v}}{}{}{}{\ottcom{alloc}}\ottprodnewline
\ottprodline{|}{\ottkw{match} \, \ottnt{e} \, \ottkw{with} \, \ottkw{Di} \, \ottsym{(} \, \ottcomp{\ottmv{x_{\ottmv{i}\,\ottmv{j}}}}{\ottmv{j}} \, \ottsym{)}  \rightarrow  \ottnt{e_{\ottmv{i}}}}{}{}{}{\ottcom{match}}\ottprodnewline
\ottprodline{|}{\ottkw{if} \, \ottnt{e} \, \ottkw{then} \, \ottnt{e_{{\mathrm{0}}}} \, \ottkw{else} \, \ottnt{e_{{\mathrm{1}}}}}{}{}{}{\ottcom{if}}\ottprodnewline
\ottprodline{|}{\ottkw{return} \, \ottnt{e}}{}{}{}{\ottcom{ret}}\ottprodnewline
\ottprodline{|}{\ottmv{x}  \leftarrow  \ottsym{(}  \ottnt{e_{{\mathrm{1}}}}  \ottsym{)}  \ottsym{;}  \ottsym{(}  \ottnt{e_{{\mathrm{2}}}}  \ottsym{)}}{}{}{}{\ottcom{mon-bind}}\ottprodnewline
\ottprodline{|}{\Box}{}{}{}{\ottcom{typed-hole}}}

\newcommand{\otthole}{
\ottrulehead{\Box}{::=}{\ottcom{holedExp}}\ottprodnewline
\ottfirstprodline{|}{\ottsym{(}  \ottsym{\mbox{?}\mbox{?}}  \ottsym{)}  \ottsym{:}  \tau}{}{}{}{\ottcom{typed-hole}}}

\newcommand{\ottv}{
\ottrulehead{\ottnt{v}}{::=}{\ottcom{value}}\ottprodnewline
\ottfirstprodline{|}{\ottmv{x}}{}{}{}{\ottcom{var}}\ottprodnewline
\ottprodline{|}{\ottmv{c}}{}{}{}{\ottcom{constant}}\ottprodnewline
\ottprodline{|}{\ottkw{Di} \, \ottsym{(} \, \ottcomp{\ottmv{x_{\ottmv{i}\,\ottmv{j}}}}{\ottmv{j}} \, \ottsym{)}}{}{}{}{\ottcom{cons-app}}}

\newcommand{\ottt}{
\ottrulehead{\ottnt{t}}{::=}{\ottcom{base-types}}\ottprodnewline
\ottfirstprodline{|}{\ottkw{heap}}{}{}{}{\ottcom{heaptype}}\ottprodnewline
\ottprodline{|}{\alpha}{}{}{}{\ottcom{typevar}}\ottprodnewline
\ottprodline{|}{\ottkw{int}}{}{}{}{\ottcom{int-type}}\ottprodnewline
\ottprodline{|}{\ottkw{bool}}{}{}{}{\ottcom{bool-type}}\ottprodnewline
\ottprodline{|}{\ottkw{TN} \, \ottsym{(}  \ottnt{t}  \ottsym{)}}{}{}{}{\ottcom{user defined type}}\ottprodnewline
\ottprodline{|}{\ottkw{ref} \, \ottnt{t}}{}{}{}{\ottcom{reference}}}

\newcommand{\ottTau}{
\ottrulehead{\tau}{::=}{\ottcom{type}}\ottprodnewline
\ottfirstprodline{|}{\ottnt{t}}{}{}{}{\ottcom{base}}\ottprodnewline
\ottprodline{|}{\ottsym{\{}  \nu  \ottsym{:}  \ottnt{t}  \mid  \phi  \ottsym{\}}}{}{}{}{\ottcom{scalar}}\ottprodnewline
\ottprodline{|}{\ottsym{(}  \ottmv{x}  \ottsym{:}  \tau_{{\mathrm{1}}}  \ottsym{)}  \rightarrow  \tau_{{\mathrm{2}}}}{}{}{}{\ottcom{dependent arrow}}\ottprodnewline
\ottprodline{|}{\ottsym{\{}  \phi_{{\mathrm{1}}}  \ottsym{\}}  \nu  \ottsym{:}  \ottnt{t}  \ottsym{\{}  \phi_{{\mathrm{2}}}  \ottsym{\}}}{}{}{}{\ottcom{computation type}}}

\newcommand{\ottPhi}{
\ottrulehead{\phi}{::=}{\ottcom{proposition}}\ottprodnewline
\ottfirstprodline{|}{\ottkw{true}}{}{}{}{\ottcom{true}}\ottprodnewline
\ottprodline{|}{\ottkw{false}}{}{}{}{\ottcom{true}}\ottprodnewline
\ottprodline{|}{\neg \, \ottsym{(}  \phi  \ottsym{)}}{}{}{}{\ottcom{not}}\ottprodnewline
\ottprodline{|}{\ottmv{Q} \, \ottcomp{\ottmv{x_{\ottmv{i}}}}{\ottmv{i}}}{}{}{}{\ottcom{param\_qualifier}}\ottprodnewline
\ottprodline{|}{\phi_{{\mathrm{1}}}  \wedge  \ottsym{(}  \phi_{{\mathrm{2}}}  \ottsym{)}}{}{}{}{\ottcom{disjunction}}\ottprodnewline
\ottprodline{|}{\phi_{{\mathrm{1}}}  \vee  \ottsym{(}  \phi_{{\mathrm{2}}}  \ottsym{)}}{}{}{}{\ottcom{conjunction}}\ottprodnewline
\ottprodline{|}{\phi_{{\mathrm{1}}}  \Rightarrow  \ottsym{(}  \phi_{{\mathrm{2}}}  \ottsym{)}}{}{}{}{\ottcom{implication}}\ottprodnewline
\ottprodline{|}{\forall \, \ottmv{x}  \ottsym{.}  \ottsym{(}  \phi  \ottsym{)}}{}{}{}{\ottcom{forall prop}}\ottprodnewline
\ottprodline{|}{\exists \, \ottmv{x}  \ottsym{.}  \ottsym{(}  \phi  \ottsym{)}}{}{}{}{\ottcom{exists prop}}\ottprodnewline
\ottprodline{|}{\ottsym{\{}  \ottmv{x_{{\mathrm{1}}}}  \ottsym{/}  \ottmv{x}  \ottsym{\}}  \phi}{}{}{}{\ottcom{subst}}\ottprodnewline
\ottprodline{|}{\ottsym{(}  \phi  \ottsym{)}  \ottmv{x}}{}{}{}{\ottcom{propred}}\ottprodnewline
\ottprodline{|}{\ottkw{SP} \, \ottsym{(}  \phi  \ottsym{,}  \ottnt{e}  \ottsym{)}}{}{}{}{\ottcom{sp}}\ottprodnewline
\ottprodline{|}{\ottkw{WP} \, \ottsym{(}  \ottnt{e}  \ottsym{,}  \phi  \ottsym{)}}{}{}{}{\ottcom{wp}}}

\newcommand{\ottG}{
\ottrulehead{\Gamma}{::=}{\ottcom{type environment}}\ottprodnewline
\ottfirstprodline{|}{\varnothing}{}{}{}{}\ottprodnewline
\ottprodline{|}{\Gamma  \ottsym{,}  \phi}{}{}{}{}\ottprodnewline
\ottprodline{|}{\Gamma  \ottsym{,} \, \ottcomp{\ottmv{x_{\ottmv{i}}}  \ottsym{:}  \tau_{\ottmv{i}}}{\ottmv{i}}}{}{}{}{}}

\newcommand{\ottS}{
\ottrulehead{\Sigma}{::=}{\ottcom{library environment}}\ottprodnewline
\ottfirstprodline{|}{\varnothing}{}{}{}{}\ottprodnewline
\ottprodline{|}{\Sigma  \ottsym{,}  \ottmv{f}  \ottsym{:}  \ottsym{(}  \ottmv{x}  \ottsym{:}  \tau  \ottsym{)}  \rightarrow  \tau_{{\mathrm{1}}}}{}{}{}{\ottcom{function libraries}}\ottprodnewline
\ottprodline{|}{\Sigma  \ottsym{,} \, \ottkw{Di} \, \ottsym{:} \, \ottcomp{\ottmv{x_{\ottmv{i}}}  \ottsym{:}  \tau_{\ottmv{i}}}{\ottmv{i}} \, \rightarrow  \tau}{}{}{}{\ottcom{constuctor defs}}}

\newcommand{\ottgrammar}{\ottgrammartabular{
\otte\ottinterrule
\otthole\ottinterrule
\ottv\ottinterrule
\ottt\ottinterrule
\ottTau\ottinterrule
\ottPhi\ottinterrule
\ottG\ottinterrule
\ottS\ottafterlastrule
}}

% defnss
% defns Forward
%% defn FW
% \newcommand{\ottdruleFWXXSXXvar}[1]{\ottdrule[#1]{%
% \ottpremise{\ottmv{x}  \ottsym{:}  \tau_{{\mathrm{1}}} \, \in \, \Gamma}%
% \ottpremise{\ottsym{G;}  \Sigma  \vdash  \tau_{{\mathrm{1}}}  <:  \tau}%
% }{
% \Gamma  \ottsym{;}  \Sigma  \vdash  \tau  \twoheadrightarrow  \ottmv{x}}{%
% {\tiny \ottdrulename{FW\_var}}{}%
% }}

\newcommand{\ottdruleFWXXSXXvar}[1]{\ottdrule[#1]{%
\ottpremise{\ottmv{x}  \ottsym{:}  \tau \, \in \, \Gamma}%
}{
\Gamma  \ottsym{;}  \Sigma  \vdash  \tau  \twoheadrightarrow  \ottmv{x}}{%
{\tiny \ottdrulename{FW\_var}}{}%
}}

\newcommand{\ottdruleFWXXSXXmatch}[1]{\ottdrule[#1]{%
\ottpremise{\Gamma  \ottsym{;}  \Sigma  \vdash  \ottsym{\{}  \nu  \ottsym{:}  
\ottkw{TN} \,  \mid  \phi  \ottsym{\}}  \twoheadrightarrow  \ottnt{e}}%

\ottpremise{\ottkw{Di} \, \ottcomp{(\ottmv{x_{\ottmv{j}}} 
 \ottsym{:}  \tau_{\ottmv{j}})} \, \rightarrow  
 \ottsym{\{} 
 \nu  \ottsym{:}  \ottkw{TN}  \mid  
\phi_{\ottmv{i}}  \ottsym{\}} \, \in \, \Sigma}%

\ottpremise{\Gamma_{\ottmv{i}} \, \equiv \, \Gamma  \ottsym{,} \, 
\ottcomp{\ottmv{x_{\ottmv{j}}}  \ottsym{:}  
\tau_{\ottmv{j}}}{}  \ottsym{,}  \ottsym{\{}  \ottmv{x'}  
\ottsym{/}  \nu  \ottsym{\}}  \phi_{\ottmv{i}}}%

\ottpremise{
\Gamma  \ottsym{,} \, 
\ottsym{\{}  \ottmv{x'}  
\ottsym{/}  \nu  \ottsym{\}}  \phi \ottsym{,} \,
\Gamma_{\ottmv{i}}  \ottsym{;}  \Sigma  \vdash  \ottsym{\{}  
\ottmv{P}  \,  \ottsym{\}}  \nu  \ottsym{:}  \ottnt{t}  \ottsym{\{}  \ottmv{Q}  
\,  \ottsym{\}}  \twoheadrightarrow  \ottnt{e_{\ottmv{i}}}}%
}{
\Gamma  \ottsym{;}  \Sigma  \vdash  \ottsym{\{}  \ottmv{P}  \,  \ottsym{\}}  \nu  \ottsym{:}  \ottnt{t}  \ottsym{\{}  \ottmv{Q}  \,  \ottsym{\}}  \twoheadrightarrow \newline   
\ottkw{match} \, \ottnt{e} \, \ottkw{with} \, \ottkw{Di} \, \ottsym{(} \, 
\ottcomp{\ottmv{x_{\ottmv{j}}}}{} \, \ottsym{)}  \rightarrow  
\ottnt{e_{\ottmv{i}}}}{%
{\tiny \ottdrulename{FW\_match}}{}%
}}

\newcommand{\ottdruleFWXXSXXif}[1]{\ottdrule[#1]{%
\ottpremise{\Gamma  \ottsym{;}  \Sigma  \vdash  \ottsym{\{}  \nu  \ottsym{:}  
\ottkw{bool}  \mid  \phi_{{\mathrm{t}}}  \wedge  \phi_{{\mathrm{f}}}   
 \ottsym{\}}  \twoheadrightarrow  \ottnt{e}}%
\ottpremise{\Gamma  \ottsym{,}  \ottsym{\{}  \mathsf{true}  \ottsym{/}  \nu  
\ottsym{\}}  \phi_{{\mathrm{t}}}  \ottsym{;}  \Sigma  \vdash  \ottsym{\{}  
\ottmv{P}  \,  \ottsym{\}}  \nu  \ottsym{:}  \ottnt{t}  \ottsym{\{}  \ottmv{Q}  
\,  \ottsym{\}}  \twoheadrightarrow  \ottnt{e_{{\mathrm{t}}}}}%
\ottpremise{\Gamma  \ottsym{,}  \ottsym{\{}  \mathsf{false}  \ottsym{/}  \nu  
\ottsym{\}}  \phi_{{\mathrm{f}}}  \ottsym{;}  \Sigma  \vdash  \ottsym{\{}  
\ottmv{P}  \,  \ottsym{\}}  \nu  \ottsym{:}  \ottnt{t}  \ottsym{\{}  \ottmv{Q}  
\,  \ottsym{\}}  \twoheadrightarrow  \ottnt{e_{{\mathrm{f}}}}}%
}{
\Gamma  \ottsym{;}  \Sigma  \vdash  \ottsym{\{}  \ottmv{P}  \,  \ottsym{\}}  \nu  \ottsym{:}  \ottnt{t}  \ottsym{\{}  \ottmv{Q}  \,  \ottsym{\}}  \twoheadrightarrow  \ottkw{if} \, \ottnt{e} \, \ottkw{then} \, \ottnt{e_{{\mathrm{t}}}} \, \ottkw{else} \, \ottnt{e_{{\mathrm{f}}}}}{%
{\tiny \ottdrulename{FW\_if}}{}%
}}

\newcommand{\ottdruleFWXXSXXcall}[1]{\ottdrule[#1]{%
\ottpremise{\ottmv{f}  \ottsym{:} \, \ottcomp{\ottmv{x_{\ottmv{i}}}  \ottsym{:}  
\tau_{\ottmv{i}}}{} \, \rightarrow  \ottsym{\{}  
\ottmv{P_{{\mathrm{1}}}}  \,  \ottsym{\}}  \nu  \ottsym{:}  \ottnt{t'}  
\ottsym{\{}  \ottmv{Q_{{\mathrm{1}}}}  \,  \ottsym{\}} \, \in \, \Sigma}%

\ottpremise{\Gamma  \ottsym{;}  \Sigma  \vdash  \tau_{\ottmv{i}}  
\twoheadrightarrow  \ottmv{y_{\ottmv{i}}}}%

\ottpremise{\Gamma  \ottsym{;}  \Sigma  \vdash  \ottmv{P}   \Rightarrow  
\ottmv{P_{{\mathrm{1}}}}}%

\ottpremise{  \ottmv{Q'} \, \equiv \, \ottkw{SP} \, \ottsym{(}  
\ottmv{P}  \,  \ottsym{,}  \ottmv{f}  \ottsym{(} \, 
\ottcomp{\ottmv{y_{\ottmv{i}}}}{} \, \ottsym{)}  \ottsym{)}   
 \!\! = \!\!  \ottmv{P}  \wedge  \ottmv{Q_{{\mathrm{1}}}}}%
\ottpremise{\Gamma  \ottsym{,}  \ottcomp{\ottmv{y_{\ottmv{i}}}  \ottsym{:}  
\tau_{\ottmv{i}}}{}
 \ottsym{;}  \Sigma  \vdash  \ottsym{\{}  \ottmv{Q'}  \,  \ottsym{\}}  \nu  
\ottsym{:}  \ottnt{t}  \ottsym{\{}  \ottmv{Q}  \,  \ottsym{\}}  
\twoheadrightarrow  \ottnt{e}}%
}{
\Gamma  \ottsym{;}  \Sigma  \vdash  \ottsym{\{}  \ottmv{P}  \,  \ottsym{\}}  \nu 
 \ottsym{:}  \ottnt{t}  \ottsym{\{}  \ottmv{Q}  \,  \ottsym{\}}  
\twoheadrightarrow  \ottmv{z}  \leftarrow  \ottsym{(}  \ottmv{f}  \ottsym{(} \, 
\ottcomp{\ottmv{y_{\ottmv{i}}}}{} \, \ottsym{)}  \ottsym{)}  \ottsym{;} 
 \ottsym{(}  \ottnt{e}  \ottsym{)}}{%
{\tiny \ottdrulename{FW\_call}}{}%
}}

\newcommand{\ottdruleFWXXSXXframe}[1]{\ottdrule[#1]{%
\ottpremise{\Gamma  \ottsym{,}  \ottmv{R}  \,  \ottsym{;}  \Sigma  \vdash  
\ottsym{\{}  \ottmv{P}  \,  \ottsym{\}}  \nu  \ottsym{:}  \ottnt{t}  \ottsym{\{} 
 \ottmv{Q}  \,  \ottsym{\}}  \twoheadrightarrow  \ottnt{e}}%
\ottpremise{\ottsym{(}  \ottsym{(}  \ottkw{Vars} \, \ottsym{(}  \ottmv{R}  \,  
\ottsym{)}  \ottsym{)} \, \cap \, \ottsym{(}  \ottkw{EVars} \, \ottsym{(}  
\ottmv{P}  \,  \ottsym{,}  \ottmv{Q}  \,  \ottsym{)}  \ottsym{)}  \ottsym{)}  
\!\! = \!\! \, \varnothing  }%
\ottpremise{\ottsym{(}  \ottsym{(}  \ottkw{Qual} \, \ottsym{(}  \ottmv{R}  \,  
\ottsym{)}  \ottsym{)} \, \cap \, \\ %NEWLINE%
\ottsym{(}  \ottsym{(}  \ottkw{Qual} \, 
\ottsym{(}  \ottmv{P}  \,  \ottsym{)}  \ottsym{)} \, \cup \, \ottsym{(}  
\ottkw{Qual} \, \ottsym{(}  \ottmv{Q}  \,  \ottsym{)}  \ottsym{)}  \ottsym{)}  
\ottsym{)}  \!\! = \!\!  \ottsym{(}  \varnothing  \ottsym{)}}%
}{
\Gamma  \ottsym{;}  \Sigma  \vdash  \ottsym{\{}  \ottmv{P}  \,  \wedge  
\ottsym{(}  \ottmv{R}  \,  \ottsym{)}  \ottsym{\}}  \nu  \ottsym{:}  \ottnt{t}  
\ottsym{\{}  \ottmv{Q}  \,  \wedge  \ottsym{(}  \ottmv{R}  \,  \ottsym{)}  
\ottsym{\}}  \twoheadrightarrow  \ottnt{e}}{%
{\tiny \ottdrulename{FW\_frame}}{}%
}}

\newcommand{\ottdruleFWXXSXXsub}[1]{\ottdrule[#1]{%
\ottpremise{\Gamma  \ottsym{;}  \Sigma  \vdash  \ottsym{\{}  \ottmv{P}  \,  \ottsym{\}}  \nu  \ottsym{:}  \ottnt{t}  \ottsym{\{}  \ottmv{Q_{{\mathrm{1}}}}  \,  \ottsym{\}}  \twoheadrightarrow  \ottnt{e}}%
\ottpremise{\Gamma  \ottsym{;}  \Sigma  \vdash  \ottkw{SP} \, \ottsym{(}  \ottmv{P}  \,  \ottsym{,}  \ottnt{e}  \ottsym{)}  \Rightarrow  \ottsym{(}  \ottmv{Q}  \,  \ottsym{)}}%
}{
\Gamma  \ottsym{;}  \Sigma  \vdash  \ottsym{\{}  \ottmv{P}  \,  \ottsym{\}}  \nu  \ottsym{:}  \ottnt{t}  \ottsym{\{}  \ottmv{Q}  \,  \ottsym{\}}  \twoheadrightarrow  \ottnt{e}}{%
{\tiny \ottdrulename{FW\_sub}}{}%
}}

\newcommand{\ottdefnFW}[1]{\begin{ottdefnblock}[#1]{$\Gamma  \ottsym{;}  \Sigma  
\vdash  \tau  \twoheadrightarrow  \ottnt{e}$}{\ottcom{Forward Synthesis}}
\ottusedrule{\ottdruleFWXXSXXmatch{}} \\
\ottusedrule{\ottdruleFWXXSXXvar{}} \\
\ottusedrule{\ottdruleFWXXSXXif{}} \\
\ottusedrule{\ottdruleFWXXSXXcall{}} \\
\ottusedrule{\ottdruleFWXXSXXframe{}} \\
\ottusedrule{\ottdruleFWXXSXXsub{}} 
\end{ottdefnblock}}

\newcommand{\ottdefnsForward}{
\ottdefnFW{}}

\newcommand{\ottdruleBWXXSXXbackXXhole}[1]{\ottdrule[#1]{%
\ottpremise{\ottmv{y} \, \notin \, \ottkw{Dom} \, \ottsym{(}  \Gamma  
\ottsym{)}}%
}{
\Gamma  \ottsym{;}  \Sigma  \vdash  
 \tau  \twoheadleftarrow  \ottmv{y}  \leftarrow 
 \ottsym{(}  \ottsym{(}  \ottsym{\mbox{?}\mbox{?}}  \ottsym{)}  \ottsym{:}  
\tau  \ottsym{)}  \ottsym{;}  \ottsym{(}  \ottkw{skip}  
\ottsym{)}}{%
{\tiny \ottdrulename{BW\_hole}}{}%
}}

\newcommand{\ottdruleBWXXSXXcall}[1]{\ottdrule[#1]{%
\ottpremise{\ottmv{f}  \ottsym{:} \, \ottcomp{\ottmv{x_{\ottmv{i}}}  \ottsym{:} 
\tau_{\ottmv{i}}}{} \, \rightarrow  \ottsym{\{}  
\ottmv{P_{{\mathrm{1}}}}  \,  \ottsym{\}}  \nu  \ottsym{:}  \ottnt{t'}  
\ottsym{\{}  \ottmv{Q_{{\mathrm{1}}}}  \,  \ottsym{\}} \, \in \, \Sigma}%

\ottpremise{\Gamma  \ottsym{;}  \Sigma  \vdash  \tau_i \twoheadleftarrow  
\ottmv{y_{\ottmv{i}}}}%
\ottpremise{\Gamma  \ottsym{;}  \Sigma  \vdash  
 \ottmv{P_{{\mathrm{1}}}}  \Rightarrow (\ottmv{Q_{{\mathrm{1}}}}  
\Rightarrow \ottmv{Q})}%

\ottpremise{\ottmv{P'} \, \equiv \, \ottkw{WP} \, \ottsym{(}  
\ottmv{f}  \ottsym{(} \, \ottcomp{\ottmv{y_{\ottmv{i}}}}{} \, 
\ottsym{)}  \ottsym{,}  \ottmv{Q}  \,  \ottsym{)}    \!\! = \!\!  
\ottmv{P_{{\mathrm{1}}}} \wedge \ottsym{(} \,  \ottmv{Q_{{\mathrm{1}}}}  \,  
\Rightarrow  
\ottmv{Q}  \,  \ottsym{)}}%
\ottpremise{\Gamma  \ottsym{;}  \Sigma  \vdash  \ottsym{\{}  \ottmv{P}  \,  
\ottsym{\}}  \ottsym{(}  \ottsym{\mbox{?}\mbox{?}}  \ottsym{)}  \ottsym{:}  
\tau_{\ottmv{i}}  \ottsym{\{}  \ottmv{P'}  \,  \ottsym{\}}  \twoheadleftarrow  
\ottnt{e_{\ottmv{i}}}}%
\ottpremise{\ottcomp{\ottmv{y_{\ottmv{i}}}  \leftarrow  \ottsym{(}  \ottnt{e_{\ottmv{i}}}  
\ottsym{)}}{} 
 \ottsym{;}  \ottsym{(}  \ottmv{f}  \ottsym{(} \, 
\ottcomp{\ottmv{y_{\ottmv{i}}}}{} \, \ottsym{)}  \ottsym{)} \notin F}%

}{
\Gamma  \ottsym{;}  \Sigma  \vdash  \ottsym{\{}  \ottmv{P}  \,  \ottsym{\}}  
\ottsym{(}  \ottsym{\mbox{?}\mbox{?}}  \ottsym{)}  \ottsym{:}  \ottnt{t'}  
\ottsym{\{}  \ottmv{Q}  \,  \ottsym{\}}  \twoheadleftarrow  
 \ottcomp{\ottmv{y_{\ottmv{i}}}  \leftarrow  \ottsym{(}  \ottnt{e_{\ottmv{i}}}  
\ottsym{)}}{} 
 \ottsym{;}  \ottsym{(}  \ottmv{f}  \ottsym{(} \, 
\ottcomp{\ottmv{y_{\ottmv{i}}}}{} \, \ottsym{)}  \ottsym{)}}{%
{\tiny \ottdrulename{BW\_call}}{}%
}}

\newcommand{\ottdruleBWXXSXXconseqXXframe}[1]{\ottdrule[#1]{%
\ottpremise{\ottmv{P}  \,  \vdash  \ottmv{P_{{\mathrm{1}}}}  \,  \wedge  
\ottsym{(}  \ottmv{R}  \,  \ottsym{)}}%
\ottpremise{\ottmv{Q_{{\mathrm{1}}}}  \,  \wedge  \ottsym{(}  \ottmv{R}  \,  
\ottsym{)}  \vdash  \ottmv{Q}  \,}%
\ottpremise{\ottsym{(}  \ottsym{(}  \ottkw{Vars} \, \ottsym{(}  \ottmv{R}  \,  
\ottsym{)}  \ottsym{)} \, \cap \, \ottsym{(}  \ottkw{EVars} \, \ottsym{(}  
\ottmv{P_{{\mathrm{1}}}}  \,  \ottsym{,}  \ottmv{Q_{{\mathrm{1}}}}  \,  
\ottsym{)}  \ottsym{)}  \ottsym{)}  \!\! = \!\!  \ottsym{(}  \varnothing  
\ottsym{)}}%
\ottpremise{\ottsym{(}  \ottsym{(}  \ottkw{Qual} \, \ottsym{(}  \ottmv{R}  \,  
\ottsym{)}  \ottsym{)} \, \cap \, \ottsym{(}  \ottsym{(}  \ottkw{Qual} \, 
\ottsym{(}  \ottmv{P_{{\mathrm{1}}}}  \,  \ottsym{)}  \ottsym{)} \, \cup \, 
\ottsym{(}  \ottkw{Qual} \, \ottsym{(}  \ottmv{Q_{{\mathrm{1}}}}  \,  \ottsym{)} 
 \ottsym{)}  \ottsym{)}  \ottsym{)}  \!\! = \!\!  \ottsym{(}  \varnothing  
\ottsym{)}}%
\ottpremise{\Gamma  \ottsym{;}  \Sigma  \vdash  \ottsym{\{}  
\ottmv{P_{{\mathrm{1}}}}  \,  \ottsym{\}}  \ottsym{(}  \ottsym{\mbox{?}\mbox{?}} 
 \ottsym{)}  \ottsym{:}  \ottnt{t}  \ottsym{\{}  \ottmv{Q_{{\mathrm{1}}}}  \,  
\ottsym{\}}  \twoheadleftarrow  \ottnt{e}}%
}{
\Gamma  \ottsym{;}  \Sigma  \vdash  \ottsym{\{}  \ottmv{P}  \,  \ottsym{\}}  
\ottsym{(}  \ottsym{\mbox{?}\mbox{?}}  \ottsym{)}  \ottsym{:}  \ottnt{t}  
\ottsym{\{}  \ottmv{Q}  \,  \ottsym{\}}  \twoheadleftarrow  \ottnt{e}}{%
{\tiny \ottdrulename{BW\_frame}}{}%
}}

\newcommand{\ottdruleBWXXSXXbackXXfor}[1]{\ottdrule[#1]{%
\ottpremise{\Gamma  \ottsym{;}  \Sigma  \vdash  \ottsym{\{}  \ottmv{P}  \,  \ottsym{\}}  \nu  \ottsym{:}  
\ottnt{t}  \ottsym{\{}  \ottmv{Q}  \,  \ottsym{\}}  \twoheadrightarrow  \ottnt{e}}%
}{
\Gamma  \ottsym{;}  \Sigma  \vdash  \ottsym{\{}  \ottmv{P}  \,  \ottsym{\}}  \ottsym{(}  \ottsym{\mbox{?}\mbox{?}}  \ottsym{)}  \ottsym{:}  \ottnt{t}  \ottsym{\{}  \ottmv{Q}  \,  \ottsym{\}}  \twoheadleftarrow  \ottnt{e}}{%
{\tiny \ottdrulename{BW\_fw}}{}%
}}

\newcommand{\ottdruleBWXXSXXsub}[1]{\ottdrule[#1]{%
\ottpremise{\Gamma  \ottsym{;}  \Sigma  \vdash  \ottsym{\{}  \ottmv{P_{{\mathrm{1}}}}  \,  \ottsym{\}}  \ottsym{(}  \ottsym{\mbox{?}\mbox{?}}  \ottsym{)}  \ottsym{:}  \ottnt{t}  \ottsym{\{}  \ottmv{Q}  \,  \ottsym{\}}  \twoheadleftarrow  \ottnt{e}}%
\ottpremise{\ottsym{(}  \ottkw{Holes} \, \ottsym{(}  \ottnt{e}  \ottsym{)}  \ottsym{)}  \!\! = \!\!  \ottsym{(}  \varnothing  \ottsym{)}}%
\ottpremise{\Gamma  \ottsym{;}  \Sigma  \vdash  \ottmv{P}  \,  \Rightarrow  \ottsym{(}  \ottkw{WP} \, \ottsym{(}  \ottnt{e}  \ottsym{,}  \ottmv{Q}  \,  \ottsym{)}  \ottsym{)}}%
}{
\Gamma  \ottsym{;}  \Sigma  \vdash  \ottsym{\{}  \ottmv{P}  \,  \ottsym{\}}  \ottsym{(}  \ottsym{\mbox{?}\mbox{?}}  \ottsym{)}  \ottsym{:}  \ottnt{t}  \ottsym{\{}  \ottmv{Q}  \,  \ottsym{\}}  \twoheadleftarrow  \ottnt{e}}{%
{\tiny \ottdrulename{BW\_sub}}{}%
}}

\newcommand{\ottdefnBW}[1]{\begin{ottdefnblock}[#1]{%
      $\begin{array}{c}\Gamma  \ottsym{;}  \Sigma  \vdash  \{ \phi_{{\mathrm{1}}} \}  \diamondsuit  \{ \phi_{{\mathrm{2}}}  \} \twoheadleftarrow \ottnt{e}\\
      \Gamma  \ottsym{;}  \Sigma  \vdash  \tau  \twoheadleftarrow  \ottnt{e}
        \end{array}$
   }{\ottcom{Backward Synthesis}}
\ottusedrule{\ottdruleBWXXSXXbackXXfor{}}\\
\ottusedrule{\ottdruleBWXXSXXbackXXhole{}}\\
\ottusedrule{\ottdruleBWXXSXXcall{}}\\
\ottusedrule{\ottdruleBWXXSXXsub{}}\\
\ottusedrule{\ottdruleBWXXSXXconseqXXframe{}}\\
\end{ottdefnblock}}

\newcommand{\ottdefnsBackward}{
\ottdefnBW{}}

\newcommand{\ottdefnss}{
\ottdefnsForward
\ottdefnsBackward
}

\newcommand{\ottall}{\ottmetavars\\[0pt]
\ottgrammar\\[5.0mm]
\ottdefnss}

We now present a set of {\it bi-directional} search rules and the
CDCL-search algorithm presented in the last section that formalizes our
specification-guided synthesis strategy.

\paragraph{Synthesis Language.}
Our synthesis procedure operates over a core-calculus
$\lambda_{\eff}$~\cite{effect-monad}, an extension of
the call-by-value simply-typed $\lambda$-calculus tailored to support
specification-guided component-based synthesis.  The language
differentiates between pure and impure expressions, the latter being
those whose computation can induce effects.  Values are constants of
base type, type constructor applications, (closed) lambda expressions,
and locations.  Pure expressions are values and variables.  Impure
expressions include calls to effectful library functions, expressions
that create references, pattern-matching and conditional expressions
whose bodies may introduce effects, a monadic return expression, and
and a monadic sequencing expression (\emph{x} $\leftarrow$ \eip$_1$ ;
\eip$_2$) that evaluates $\eip_1$ and binds its result to a variable
    {\sf x} in $\eip_2$.

%% {\it Secondly}, even though the language supports effectful 
%% computations, the function arguments and local variables are {\it immutable}, 
%% and mutations occur only on global heap variables. {\it Thirdly}, 
%% the language 
%% does not provide explicit terms for reference creation, update or 
%% dereferencing as available in a ML-core and effects are 
%% materialized only in the library of functions. An environment $\Sigma$ 
%% maps the function names to their effectful specifications.

%% We observed that these 
%% restrictions are not very taxing and have limited effect on component-based 
%% synthesis over effectful 
%% libraries, this is due to the fact that the libraries usually provide 
%% effectful operations over a set of 
%% global mutable references and the terms required to be synthesized are 
%% intelligent compositions of library functions obviating the need for explicit 
%% mutable reference manipulations.

%% \paragraph{Non expressible terms}
%% \AM{May be, it is better to move this  to limitations section.}
%% However, this limits the expressiveness of our language, for instance, we 
%% cannot synthesize a function f having the following basic type f : (ref int) -> 
%% (ref int) or a higher order function like f : (int -> bool) -> int.

\begin{figure}[t]
\small
\centering % used for centering table
%\ottmetavars\\[0pt]
\small \begin{tabular}{l l l l} % centered columns (4 columns)
\emph{c} $\in$ \emph{Constants}\\
\textsf{x} $\in$ \emph{Variables}\\
$\ell$ $\in$ \emph{Locations}\\
{\sf v} $\in$ \emph{Value} & ::= & \emph{c} $\mid$ $\ell$ $\mid$ $\lambda$ (\textsf{x}:$\tau$). \eip $\mid$ $\mathsf{D_i}$ $\overline{x_{j}}$ \\
$\ep$ $\in$ \emph{Pure Exp} & ::= & \textsf{x} $\mid$ \textsf{v} \\
$\eip$ $\in$ \emph{Impure Exp} & ::= &  \textsf{f} ($\overline{\ep}$) $\mid$ \textsf{ref v} $\mid$  {\bf match} \ep {\bf with} $\mathsf{D_i}$ $\overline{\mathsf{x}_{j}}$ $\rightarrow$ \eip \\
&& 					$\mid$ {\bf if} \ep\ {\bf then} \eip\ {\bf else} \eip  $\mid$ {\bf return} \ep $\mid$ \textsf{x} $\leftarrow$ \eip$_1$; \eip$_2$ $\mid$ $\diamondsuit$ \\
\textsf{f} $\in$ \emph{Library Function}	\\ % & ::= & $\lambda$ (x:$\tau$). e   \\	
{$\diamondsuit$} $\in$ \emph{hole}	& ::= &  (??) : $\tau$ \\	
{\sf TN} $\in$ \emph{TypeNames} & ::= &  {\sf list}, {\sf tree}, {\sf pair}, 
\ldots\\
{\sf t} $\in$ \emph{Base-Type} & ::= & {\sf int} $\mid$ {\sf bool} $\mid$ \ldots $\mid$ {\sf heap} $\mid$ {\sf TN } $\mid$ \textsf{t ref} \\
$\tau$ $\in$ \emph{Type} & ::= & \{$\nu$ : {\sf t} | $\phi$ \} $\mid$ (\textsf{x} : $\tau$) $\rightarrow$ $\tau$  $\mid$ \{ $\phi_1$ \} $\nu$ : {\sf t} \{ $\phi_2$ \}  \\
% 
% $\rho$ $in$ {\sf Regions} & & $\gamma$ $\in 2^{\rho}$ \\
% $\mathbf{\sigma}$ $\in$ \emph{Effect}  & ::= &  $\langle$ $\gamma_r, \gamma_w$ $\rangle$ \\ 
$\phi$,$P$,$Q$ $\in$ \emph{Propositions} & ::= & \textsf{true} $\mid$ \textsf{false} $\mid$ 
$Q(\overline{x_i})$ \\
	  && $\mid$ $\neg$ $\phi$ $\mid$ $\phi$ $\wedge$ $\phi$  $\mid$ $\phi \lor \phi$  $\mid$ $\phi$ $\Rightarrow$ $\phi$ $\mid$ $\forall$ (x : t). $\phi$ $\mid$ $\exists$ (x : t). $\phi$ 
\\
	%  && $\mid$ \{ $x_1$ / x \} $\phi$ $\mid$ {\bf SP} ($\phi$, e) $\mid$ 
%{\bf WP} (e, $\phi$)\\
	  
$\Gamma$ $\in$ \emph{Type Context} & ::= & $\varnothing$ $\mid$ $\Gamma$, x : $\tau$ $\mid$ $\Gamma$, $\phi$ \\
$\Sigma$ $\in$ \emph{Library} & ::= & $\varnothing$ $\mid$ $\Sigma$, $f$ : 
($\overline{x_i : \tau_{i}}$) $\rightarrow$ $\tau$ \\
        &&      $\mid$ $\Sigma$, $\mathsf{D_i}$ $\overline{x_j : 
\tau_{j}}$ $\rightarrow$ $\tau$\\
\hline
\end{tabular}%
\caption{$\lambda_{eff}$ Expressions and Types} % title of Table
\label{fig:syntax} % is used to refer this table in the text
\end{figure}

As we have seen in our earlier examples, the language also allows
typed holed-expressions of a given type $\tau$ that takes the form
((??) : $\tau$).  Such a term represent an unknown expression in a
program that must be constrained by the type $\tau$; our synthesis
procedure transforms such expressions by replacing these holes with
concrete terms.

\paragraph{Types and Environments.}

The type language includes support for {\it base types} such as types
for integers, Booleans, strings, etc., as well as a special {\sf heap}
type to denote the type of abstract heap variables like {\sf h, h'}
found in specifications.  There are additionally user-defined data
types \textsf{TN}, and type constructors used to type references that
hold values of some base type.  More interestingly, base types can be
refined with \emph{propositions}, and effectful computations have
types defined in terms of Hoare-style pre and postconditions of the
form {\sf \{$\phi_1$\} $\nu$ : t \{$\phi_1'$\}} that represents an
effectful computation, which when executed in a pre-state satisfying
proposition \{$\phi_1$\}, upon termination, returns a value $\nu$ of
base type {\sf t} along with a post-state satisfying \{$\phi_1'$\}.

Propositions ($\phi$) are first-order predicate logic formulae over
base-typed variables.  Propositions also include a set of {\it
  Qualifiers} which are user-defined uninterpreted functions symbols
such as {\sf mem, size} etc. used in our example; qualifiers also
include two special interpreted function symbols (\textsf{sel} and
\textsf{update}) used to model access and modification to the global
heap\footnote{Details about the language's type system can be found in
  the supplemental material.}  The type language also includes
dependent-function types since arguments and return values of library
functions can be associated with types that are refined by
propositions.

Propositions in pre- and post-conditions capture {\it non-spatial}
properties of the pre- and post-abstract heaps respectively.  These
properties capture actions involving accesses and modifications to
heap objects associated with a heap variable ({\sf sel} and {\sf
  update}), or describe shallow structural properties of heap objects,
e.g., {\sf length, head, etc.} for a list.  Our current implementation
currently does \emph{not} allow expression of spatial properties that
describe disjointedness of heap fragments.  Consequently, we assume that
each heap object is always referenced by a unique path (variable {\sf
  x} or {\sf x.f.y}) and that there is no sharing of heap objects.  We
have found that these assumptions are not particularly onerous in the
context of the OCaml libraries we have examined.

%%  This may feel like very strong restriction, however, note
%% that in practice this still allows us to synthesize programs from
%% numerous interesting OCaml libraries and applications from across
%% domains. One key advantage of this restriction is that we do not
%% require rich separation logic formulas ~\cite{separation} to
%% explicitly state disjointness of the heap.  On the downside, it
%% restricts us to synthesize programs which require heap sharing, for
%% instance although we can synthesize programs using Singly-linked lists
%% we cannot synthesize several program which needs to works over a
%% circular linked list, like inserting into such a
%% linked-list. Similarly we cannot synthesize programs over graphs which
%% has explicit sharing on nodes.}

There are two environments maintained by \name, of particular interest
to our synthesis procedure: (1) environment $\Gamma$ records the types
of variables along with a set of propositions relevant to a specific
context, and (2) and, environment $\Sigma$ maps library functions and
datatype constructors to their signatures.

\subsection{A \name\ Synthesis Problem}
A \name\ synthesis problem can be described as follows: Given a
library $\Sigma$ of functions and data constructors, annotated with a
suitable types, a type environment $\Gamma$, and a goal specification
($\Psi$), which is a dependent-function type of the form
\[ (\mathsf{x} : \tau) \rightarrow (\{P\} \mathsf{v} : \mathsf{t} \{Q\}) \]
\DEL{where {\sf v} is a free variable denoting the return value of the program and pre and post-conditions $P, Q$ may contain argument variable {\sf x}, heap locations and the return variable {\sf v}},
the synthesis problem seeks to synthesize an expression {$e \in e_p \cup e_{ip}$} in $\lambda_{\mathit{eff}}$
such that
\[\Gamma; \Sigma \vdash e :(\mathsf{x} : \tau) \rightarrow \{P\} \mathsf{v} : \mathsf{t} \{Q\} \]

\subsection{Bi-directional Deductive Component-Based Synthesis}
Given a \name\ synthesis problem, the synthesis procedure is a {\it
  bi-directional} deductive proof-search~\cite{fiat, myth, synquid} over
library functions and the given query specification. We next explain
each of these modes of the synthesis procedure.
%Figure~\ref{fig:overview} shows an overview of the process. 
% It begins with a {\it backward-synthesis} which uses the  problem 
% specification's postcondition and applies a weakest-precondition 
% ~\cite{wp-reasoning} based backward-reasoning over procedure-calls to search for 
% a proof and consequently a $\lambda_{eff}$ expression. If it fails to find such 
% a term, it passes the baton to the {\it forward-synthesis} in 
% a manner similar to bidirectional-typing~\cite{bidirectional-typing},
% however, unlike standard bidiretional typing, it also passes the partial 
% program synthesized ({\sf t}), a possible holed 
% expression {\sf H} and a weakest-precondition generated for {\sf t} against 
% the original postcondition. 
% The {\it forward-synthesis} creates a new synthesis specification ($\Psi$') 
% using the information passed by the backward procedure and applies a 
% strongest-postcondition~\cite{sp-reasoning} based forward reasoning over 
% procedure calls to search for a proof. In 
% case the forward procedure again 
% fails to find a solution it passes back the baton to the backward procedure 
% along with a {\it knowledge} of failed programs. The backward search then  
% backtracks 
% from its most recent choice and uses the information from the forward search to 
% make a different choice. 
% In case the {\it backward-search} has exhausted all possible choices, the 
% process terminates without a result.

\subsubsection{Forward Synthesis}
\label{sec:forward}

\begin{figure}[t!]
\begin{subfigure}[t]{0.45\textwidth}
\small
\ottdefnsForward{}
\caption{Forward Type Synthesis Rules}
\label{fig:forward}
\end{subfigure}
\begin{subfigure}[t]{0.50\textwidth}
\small
\ottdefnsBackward{}
\caption{Backward Type Synthesis Rules}
\label{fig:backward}
\end{subfigure}
\caption{Forward and Backward Type Synthesis Rules}
\end{figure}

% \begin{figure}[htbp]
% \small
% \ottdefnsForward{}
% \caption{Forward Type Synthesis Rules}
% \label{fig:forward}
% \end{figure}

Figure~\ref{fig:forward} shows our forward synthesis system using  synthesis rules of the following form:
\[
 \Gamma  \ottsym{;}  \Sigma  \vdash  \tau  \twoheadrightarrow  \ottnt{e}
\]
\noindent Each such rule defines a declarative judgment explaining the
generation (along with a proof-derivation) of a conclusion term {\sf
  e} in an environment of types ($\Gamma$) and libraries ($\Sigma$)
against a given type $\tau$, using the derivation of other {\it
  well-typed} subterms in the rule's premise. Generating a variable
({\small{\textsc{FW\_Var}}}) simply requires choosing the variable of
the required type from the environment.  To generate a {\it match}
expression ({\small{\textsc{FW\_Match}}}) the procedure first
recursively generates a term {\sf e} using the
({\small{\textsc{FW\_Var}}}) rule for a datatype \textsf{TN} from the
environment. Second, it creates an extended environment $\Gamma_i$ for
each case branch {\sf i} using constructors ($\mathsf{D_i}$
($\overline{\mathsf{x}_j : \tau_j}$) $\rightarrow$ \{ $\nu$ : TN |
$\phi_i$ \} for the \textsf{TN}, while replacing the bound variable
$\nu$ in each $\phi_i$ with an existential match variable
$x'$. Finally, it recursively generates expressions ($\mathsf{e}_i$)
for the original synthesis problem specification in each of these
extended environments. Thus, the rule allows us to break the original
synthesis problem into $i$ subproblems that can be solved in
stronger environments, thereby pushing type information from a
constructor's specification (${\mathsf{D_i}}$ $\in$ $\Sigma$) to the
synthesis query.

The generation of a conditional {\it if-then-else} expression
({\small{\textsc{FW\_IF}}}) is similar to \textsf{match} with a few
important differences. It first requires the generation of the Boolean
test expression {\sf e}. Since our focus is component-based synthesis,
the synthesis procedure only has access to the library specifications
($\Sigma$) and the goal specification $\Psi$ at its disposal. Thus,
only way to generate a Boolean-typed expression is via function
calls. Consequently, the procedure searches for a library function
call with Boolean return type using the {\small{\textsc{FW\_Call}}}
rule described below and postconditions ($\phi_t$ and $\phi_f$) for
true and false return values, resp.
%% For instance, consider the specification for the {\sf mem\_tbl}
%% function (refer Figure~\ref{fig:lib}) with $\phi_1$ as ({\sf [$\nu$ =
%%     true] <=> mem (Tbl', s)}) and $\phi_0$ as ({\sf [$\nu$ = false]
%%   <=> not (mem (Tbl', s))}).  Note that we do not require the type to
%% be of this form, rather, the predicate for false branch is simply the
%% negation of the true branch predicate.
It then creates extended environments to recursively synthesize the
true and false branch by substituting {\sf true} and {\sf false} for
the bounded variable $\nu$, and synthesizes terms for the true and
false branches in their extended environments.

The {\small{\textsc{FW\_Call}}} rule defines the strongest-post
condition forward search procedure.  The rule depicts a scenario when
a single function-call does not suffice to generate a term for the
required goal specification $\Psi$. It breaks the original synthesis
into two sub-synthesis problems: First, it searches for a
function $f$ in the library with a type, such that a) the synthesis
procedure can synthesize expressions $\overline{y_i}$ as its arguments (see
second premise); b) with appropriate mapping for the heap and
arguments variables\footnote{We drop variable substitutions in
propositions to reduce clutter in rules}, it can satisfy the forward
rule for Hoare-style reasoning, i.e. the goal's precondition {\sf P}
implies the required precondition {$\mathsf{P_1}$} for $f$ (see third
premise).  Successfully checking these two conditions implies that a
function call expression ($f$ ($\overline{y_i}$)) in the current
environment is well-typed.  Second, it calculates the strongest-post
condition ({\bf SP} (P, ($f$ ($\overline{y_i}$)))) for this term, and
recursively synthesizes an expression {\sf e} with this as the new
precondition.  The expression synthesized for the original
specification is a monadic sequencing of the function call ($f$
($\overline{y_i}$)) and {\sf e}.

\paragraph{Framing.}
The rule ({\small\textsc{FW\_Frame}}) is concerned with expression
synthesis taking into account \emph{frames}, heap/store fragments that
do not change during the evaluation of a program
expression~\cite{separation}.

%% The problem arises in our setting as
%% each library function has a small footprint specification and the
%% synthesis rules perform local reasoning over them.
%% Traditionally, a frame rule is defined using a separating conjunct
%% (\{P $\star$ R \} v : t \{ Q $\star$ R\}) that intuitively ensures
%% that a computation having effects prescribed by a local specification
%% (\{P\} v : t \{ Q\}) will not effect predicates on the heap that are
%% disjoint from the heap fragment over which $P$ and $Q$ operate.
%% The rule is a simplified version of the frame rule 

%% made possible by the restrictions over mutable variables in
%% $\lambda_{eff}$.  Recall that $\lambda_{eff}$ disallows implicit
%% aliasing by requiring that formal arguments to functions, as well as
%% local variables are immutable and only the shared heap variables
%% (e.g. the tbl) are mutable.  This allows us to fully explicate the
%% references mutated by each function in its specification's pre and
%% postcondition. For instance, consider a function like {\sf
%%   fresh\_str} (refer Figure~\ref{fig:lib}), since it does not use the
%% qualifier {\sf size} in its specification, thus we can soundly assume
%% that the size of the table in the pre-heap does not change in the
%% post-heap.  Under this assumption we can effectively translate the
%% traditional disjoint heap requirement (P $\star$ Q) to a simpler
%% requirement over disjointness of the mutable variables and qualifiers.

The auxiliary function {\bf Vars}(\emph{R}) gives the set of
  reference used in $R$. The auxiliary function {\bf EVars} takes a
  list of propositions and returns the set of existential references
  found in the environment used in these propositions; these
  existentials are introduced when computing the strongest
  postcondition in the {\small{\textsc{FW\_Call}}} rule. The function
  {\bf Qual}(\emph{R}) gives the set of qualifiers (like
    \textsf{size}, \textsf{mem}, etc.) used in $R$.  The premise in
    {\small{\textsc{FW\_Frame}}} checks that the references found in the
    frame $R$ are disjoint from the existential references in $P$ and
    $Q$ and that $R$'s qualifier set is also disjoint from $P$ and
    $Q$.

The subtype rule ({\small{\textsc{FW\_Sub}}}) defines the condition
for the successful termination of the forward proof search process
using the standard strongest postcondition-based verification
condition check.

% 
% \begin{figure*}[!h]
% \centering
% \begin{tabular}[c]{cc}
% \begin{subfigure}[t]{0.50\linewidth}
% \centering
% \ottdefnFW
% \caption{Forward Type Synthesis Rules}
% \label{fig:forward}
% \end{subfigure}&
% \begin{subfigure}[t]{0.50\linewidth}
% \centering
% \ottdefnBW
% \caption{Backward Type Synthesis Rules}
% \label{fig:backward}
% \end{subfigure}
% \end{tabular}    
% \caption{Compensation measures}
% \label{fig:ceo}
% \end{figure*}

\subsubsection{Backward Synthesis}
% 
% \begin{figure}[t]
% \small
% \ottdefnsBackward{}
% \caption{Backward Type Synthesis Rules}
% \label{fig:backward}
% \end{figure}

Figure~\ref{fig:backward} presents the backward synthesis inference 
rules whose judgments are either of the form:
\[
 \Gamma  \ottsym{;}  \Sigma  \vdash  \tau  \twoheadleftarrow  \ottnt{e}
\]
for introducing a holed subterm into, or
\[
 \Gamma  \ottsym{;}  \Sigma  \vdash  \{\phi_1\}\, \diamondsuit\, \{\phi_2\}  \twoheadleftarrow  \ottnt{e}
\]
for eliminating a holed subterm from, the term being synthesized.

%% Each such rule defines a judgment which uses the environments, the given pre- 
%% and postconditions and a typed-hole expression to generate an expression using 
%% the weakest-precondition backward reasoning.

Backward synthesis can invoke forward-synthesis non-deterministically
at any time (see rule {\small\textsc{BW\_FW}}). In practice, we invoke
the forward rule when the backward synthesis cannot make any progress,
i.e. when no backward rule applies.

The backward hole rule ({\small\textsc{BW\_hole}}) generates a holed
expression bound to a fresh variable $y$ for an arbitrary synthesis
query. This is the introduction rule for a holed expression that
allows the procedure to create hypotheses when backward synthesis
cannot find a required term in the context.

The main rule for backward enumeration is
{\small\textsc{BW\_Call}}. The rule requires searching for a function
$f$ in the library with a return type matching the hole. Note the
difference from the {\small\textsc{FW\_Call}} rule, where we looked
for any allowed function call; here, we use goal directed search
instead.  Once such a function is found, the rule generates arguments
$y_i$ for $f$ by either introducing holed terms for each argument,
effectively yielding new synthesis sub-queries, or finding suitable
variables (using {\small\textsc{BW\_Var}}, similar to
{\small\textsc{FW\_Var}}, not shown here) in the environment) of the
required type. This is an instance where the effect of having an incomplete view
of the context becomes apparent during the backward search.  The rule
ensures that the function call can be soundly made using the weakest
precondition check.  This check verifies that, assuming the
precondition for the function ($P_1$) in the given environment
($\Gamma$) holds, that the postcondition for the function ($Q_1$)
implies the goal postcondition ($Q$) with appropriate substitution for
heap variables and arguments\footnote{We elide variable substitutions
  in the rules for perspicuity.}  If this check succeeds, 
it further checks that the resulting term is not already seen as a failed program using the 
set of learned failed programs {\sf F}.  If successful,
the weakest
precondition predicate ({\bf WP}(($f$ ($\overline{y_i}$)), $Q$)) for the
function call using $Q$ and the function's argument and specifications
is used.  Finally, it creates new subproblems using this weakest
precondition as the postcondition and the types of the function's
arguments as the hole types.

%% In case, the argument is not a
%% hole, it simply looks for the next unfilled hole to synthesize a term
%% against. If there are no more holes, It applies the
%% weakest-precondition verification (rule BW\_Sub) to verify that the
%% synthesized expression is a correct solution.

%% The frame rule in forward-synthesis process although also applicable
%% in the backward reasoning, it is rarely useful in backward
%% reasoning/synthesis in practice. This is due to the fact that it is
%% rarely the case that we we have a R in the environment of the form
%% ($\Gamma$, R; $\Sigma$ $\vdash$ ...).  Rather, in most cases one first
%% needs to apply the consequence rule on the postcondition before
%% applying the frame rule.

The backward frame rule (\textsc{BW\_Frame}) identifies a frame $R$ using the
consequence judgments in the premise, applies frame rule checks on the
disjointness of variables and qualifiers, and establishes a synthesis
query on the framed pre- and postconditions ($P_1$ and $Q_1$).

\section{Synthesis Algorithm}
\label{sec:alg}

\begin{algorithm*}[t]
\begin{multicols}{2}
\SetAlgoNoLine
\SetNlSty{texttt}{(}{)}
%% \KwData { Input :: $\Sigma$, $\Gamma$, specification ($\Psi$) =  \{$\phi$\} v 
%% : t \{$\phi'$\}} 
%% \KwResult{ Output :: A $\lambda_{eff}$ expression $e$, such that $\Gamma$, 
%% $\Sigma$ $\vdash$ $e$ : $\Psi$ or $\bot$}
\SetKwFunction{synthesize}{\textsc{Synthesize}}
\Indm\synthesize{$\langle\Gamma$,$\Sigma$, $\Psi\rangle$, $\mathtt{F}$}\\
\Indp
\nl   $t$ := BW\_Rules ($\Gamma$, $\Sigma$ ,$\Psi$, 
$\mathtt{F}$) \\
\nl\eIf{$(t \neq\ (\bot, \_) )$}{
	\nl{\bf return} $t$\;
	}{
      \nl\If{$(t = (\bot, \langle\, H, e_b, \Psi' \rangle))$}{
	  \nl  $t'$ := \textsc{CDCL} ($\langle\Gamma$, $\Sigma$, $\Psi'\rangle$, $H$)\\
	  \nl\If{$(t' = e_f)$}{
	    \nl{\bf return} ($e_f$; $e_b$)\;
	  }{
	    \nl\ElseIf{$t'$ = $(\bot$, $\mathtt{F'})$}{
	    \nl \synthesize($\Gamma$, $\Sigma$, $\Psi$, ($\mathtt{F}$ 
	      $\cup$ $\mathtt{F'}$))\;
	    }
	  }
 	}    
}
\[\]
\[\]
\SetKwFunction{cdcl}{\textsc{CDCL}}
\Indm\cdcl{$\langle\Gamma$,$\Sigma$, $\Psi\rangle$, $H$}\\
\Indp

\nl D := $\forall$ $c_i$ $\in$ $\Sigma$. D ($c_i$) = ($\mathsf{true}, \mathsf{false}$) \\
\nl $\mathtt{F}$ := $\varnothing$, $p_i$ := $\bot$ \\

\nl\While{true}{
  \nl($\Gamma$, $D$, $c_i$) := $\mathcal{R}$\_Choice ($\Gamma$, $\Sigma$, D,  
$\Psi$, 
$H$, $p_i$)
     
  \nl\eIf{$c_i$ = $\bot$}{
    \nl\If{$(\mid$ $p_i$ $\mid\ > 0)$}{
      \nl$\mathtt{F}$ := $\mathtt{F}$ $\cup$ \{($p_i$)\} \; 
      \nl($p_i$, $D$) := $\mathcal{R}$\_Learn ($\Gamma$, $\Sigma$, $\langle$ D, $p_i$ 
$\rangle$)\; 
      %\nl$p_{i}$ $\leftarrow$ $c_1$, $c_2$, ..., $c_{i-1}$ \;
      }\nl\lElse{
      {\bf return} ($\bot$, $\mathtt{F}$)
      }
    }{
     \nl $e$ := FW\_SUB ($\Gamma$, $\Sigma$, 
($p_i$;$c_i$), $\Psi$)\;
      \nl\lIf{($e$ $\neq$ $\bot$)}{
	  {\bf return} $e$
	}\nl\lElse{
	$p_i$ := ($p_i$;$c_i$) 
      }
   }
 } 
\end{multicols}
\caption{The Synthesis Algorithm}
\label{fig:synthesis}
\end{algorithm*}

%% In this section we discuss our top-level synthesis algorithm and a 
%% conflict-driven learning-based enumeration algorithm. The enumeration is 
%% inspired from conflict-driven
%% learning based exploration used by modern SAT solvers~\cite{CDCL-sat}.
%% The components of our algorithm are similar to the vocabulary used in
%% CDCL-leaning algorithms.
Algorithm~\ref{fig:synthesis} outlines the top-level synthesis
algorithm and can be understood as pseudo-code for the overview of our
approach given in Figure~\ref{fig:overview}. The input to the
algorithm is a \name\ synthesis problem (a triple
$\langle\Gamma,\Sigma,\Psi\rangle$) along with a set of explored {\it
  stuck-paths} $\mathtt{F}$, initially empty.  The algorithm first
makes a call to the backward synthesis procedure using a function
\textsf{BW\_Rules}, a deterministic implementation of the backward
synthesis rules given in Figure ~\ref{fig:backward}). In case backward
synthesis does not succeed in producing a complete solution (line 4),
it returns a partial solution $e_b$, a hypothesis $H$, and a new
specification $\Psi'$, which is calculated by substituting the weakest
precondition for $\Psi$'s postcondition and the partial solution
$e_b$. The algorithm invokes the \textsf{CDCL} routine (line 5) with
the hypothesis $H$, the updated specification $\Psi'$.  The CDCL
routine if successful, returns a solution $e_f$ for $\Psi'$, which is
then sequenced (using a monadic-sequencing expression) with the
partial backward solution $e_b$, to give the required solution for the
original problem (line 8).  Otherwise, the synthesis routine is
recursively called (line 9) with an updated stuck-paths set
($\mathtt{F \cup F'}$).

\subsection{Conflict Driven Learning Based Enumeration}

The {\sf CDCL} routine takes as input a synthesis problem as well as a
hypothesis $H$ and returns either a $\lambda_{\mathit{eff}}$
expression satisfying $\Psi$ (line 20) or $\bot$ (line 18) if it
cannot find such an expression. It maintains three data-structures: 1)
a {\it Discriminating Propositions} map {\sf D} that maps components
{\sf $c_i$} to a pair of stuck-path and truncated-path propositions as
discussed in the last section; 2) a sequence of
components $p_i$ (a path) representing the partially synthesized
expression; 3) a set of already explored {\it stuck-paths}
$\mathtt{F}$.  The algorithm begins by initializing D by mapping each
component in $\Sigma$ with trivial propositions and the sequence $p_i$
as an empty sequence.  The search is performed by the main loop (lines
12-21) that iterates until it finds a correct expression (line 20) or
has exhausted path exploration (line 18), updating D and $p_i$ in each
iteration.

The algorithm makes a choice of the next component for a given $p_i$
and D, using a function $\mathcal{R}$\_Choice (line 13), a
deterministic implementation of the {\small\textsc{CDCL\_CHOICE}} rule
given in Figure~\ref{fig:cdcl}.  If the procedure is unable to find a
new component (line 14), it learns new discriminating propositions for
the stuck-node associated with $p_i$ and backtracks to the previous
path using a function $\mathcal{R}$\_Learn (line 17), a deterministic
implementation of the {\small\textsc CDCL\_LEARN} rule in
Figure~\ref{fig:cdcl}, or it has exhausted all paths and terminates
the loop (line 18).  If a candidate component has been found, a call
to the {\small\textsc FW\_Sub} function (corresponding to the rule
{\sf FW\_SUB} in Figure ~\ref{fig:forward}) is performed (line 19);
this call checks if the type for the expression corresponding to path
($p_i$;$c_i$) is a subtype of the original synthesis query $\Psi$, in
which case it returns this expression.  If not, the algorithm
continues with an updated path ($p_i$;$c_i$).

\paragraph{Learning Discriminating Propositions}
%% The forward synthesis rules are integrated with a conflict-driven
%% learning phase that learns/constructs a set of proposition called {\it
%%   discriminating propositions} for a k-bound-stuck-node, utlizing
%% these propositions while exploring newer components to prune away
%% equivalent-modulo-stuckness program terms.  Figure~\ref{fig:cdcl}
%% shows the two main rules governing this process.
We introduce discriminating propositions for a {\it
  k}-bound-stuck-node
%   \SJ{If we want to refer the reader to the formal
%   definition, then we should cite a technical report, not refer to
%   supplemental material.} 
$c_i$ using the {\small\textsc{CDCL\_LEARN}}
rule. A Detailed formal definitions for k-bound-stuck nodes can be found in the technical resort~\cite{cobalt-tech}. Given a stuck-path $p_i$, typing ($\Gamma$) and library
($\Sigma$) environments, and an incoming discriminating propositions
Map $D$, the rule generates a new set of discriminating propositions
for the stuck-node $c_i$, updating $D$ in the process, and returning a
smaller path to be explored next.  The learned proposition has two
components.  The first is a {\it stuck-path proposition} $\phi_s$ that
captures the strongest postcondition for $t_{p_i}$, the expression
corresponding to $p_i$ for the given goal precondition
$\phi$.\footnote{In the following, we abuse the use of $p_i$ to serve
  as both the path and the term $t_{p_i}$ it represents for
  perspicuity.}  The second component, a {\it truncated proposition}
$\phi_t$, is a disjunction over the preconditions of those components
$c_j$ that can in principle be invoked using the
{\small\textsf{FW\_call}} rule but which cannot due to the bound $k$
and which are thus prematurely truncated. This is ensured by the
implication ($\Sigma$, $\Gamma$ $\phi_s$ $\Rightarrow$ $\phi_{c_j}$).

\begin{figure}[htbp]
\begin{flushleft}

\fbox{\small
     $\begin{array}{c} H; D ; \Gamma; \Sigma  \vdash     
				    (\Psi, p_i) \hookrightarrow (p_i;c_i) \\
      \Gamma; \Sigma  \vdash (p_i, D) \hookrightarrow (p_i', D')
      \end{array}$
           
} \ \ {CDCL Rules}

\end{flushleft}
\small \begin{center}
\inference{\Psi \equiv \{ \phi \} v : t \{ \phi' \} \quad
  p_i \equiv c_1;c_2;...;c_i &  D(f_i) = \langle \phi_{s_{f_i}}, \phi_{t_{f_i}} \rangle &  \phi_s \equiv \mathbf{SP (\phi, p_i)} \\
	   \phi_{t} \equiv \{ \bigvee_{j}. \phi_{c_j} \mid 
	  (c_j : (\overline{x_i : \tau_i}) \rightarrow \{\phi_{c_j}\} v : t' \{\phi_{c_j}'\})  \in \Sigma \wedge (\Gamma, \phi_s => \phi_{c_j})\} \\
	  D' = D [c_i \mapsto \langle (\phi_{s_{f_i}} \wedge \phi_s), (\phi_{t_{f_i}} \vee \phi_t) \rangle]
	   }{\Gamma; \Sigma;  \vdash     
  (p_i, D) \hookrightarrow ((c_1;c_2;...c_{i-1}), D')}[{\tiny CDCL\_LEARN}]
\end{center}
\bigskip
\small\begin{center}
\inference{ \Psi \equiv \{ \phi \} v : t \{ \phi' \} & \Gamma; \Sigma \vdash \{\mathsf{SP}(\phi, p_i)\} \mathsf{v : t} \{ \phi' \} \twoheadrightarrow c_i \\
	    (p_i; c_i) \prec H \quad D(c_i) = \langle \phi_s, \phi_t \rangle  
	    \quad \mid (p_i; c_i) \mid \leq k  & X \equiv \{ \neg (\phi_s => \mathbf{SP (\phi, (p_i;c_i)})\} \vee  \\
    \{(\mathbf{SP (\phi, (p_i;c_i))} => \phi_t) \wedge \neg (\mathbf{SP (\phi, (p_i))} => 
\phi_t)\}	\\		   
				 ([\Gamma] \models X)}  
				  {H; D ; \Gamma; \Sigma;  \vdash     
				    (\Psi, p_i) \hookrightarrow (p_i;c_i) 
}[{\tiny CDCL\_CHOICE}]  
\end{center} 
\bigskip
\caption{Rules for constructing and using discriminating propositions.}
\label{fig:cdcl} 
\end{figure}

The {\small\textsc{CDCL\_CHOICE}} rule uses the discriminating
propositions introduced by the learning rule to prune away
equivalent-modulo-stuckness paths. It returns a new function
component $c_i$
%%SJ: not sure this adds much to the discussion
%%(\ADD{alternately, it can also return a conditional term 
%%which is handled as two new cases of CDCL-CHOICE, with appropriate 
%%environments})
that can be used to construct a bigger
$\lambda_{\mathit{eff}}$ expression, provided an existing path $p_i$,
a hypothesis $H$, typing and library environments, a discriminating
propositions map $D$, and the goal specification $\Psi$. The rule
first searches for a component $c_i$ using the forward synthesis
rules, and performs two additional checks for the new potential path
$p_{i+1}$ = ($p_i$;$c_i$): (1) that $p_{i+1}$ satisfies the shape
given by the hypothesis $H$, and (2)
that $p_{i+1}$ is not equivalent-modulo-stuckness to some earlier
visited stuck-path. The check first generates the strongest
postconditions for the expressions corresponding to paths $p_i$ and
$p_{i+1}$ respectively.  It extracts the discriminating proposition
pair ($\phi_{s}$, $\phi_t$) for the component $c_i$ and generates a
check with two disjuncts. Failure of the first disjunct intuitively
implies that any path (and hence the corresponding term) that can be
explored by choosing $c_i$ was explored earlier without leading to a
solution and hence the exploration of $c_i$ and all the following
paths can be safely skipped without effecting the completeness of the
search process. In such a case, we should choose $c_i$ only if we had
prematurely truncated some path earlier that can also be taken along
$p_{i+1}$ (checked using conjunct \#1 in the second disjunct) and
which cannot be explored without exploring $p_{i+1}$ (checked using
conjunct \#2 in the second disjunct).

\subsection{Soundness}

Programs synthesized by \name\ are correct with respect to the provided
query specification $\Psi$ assuming the validity of each library
function against their specifications.
Complete proofs for the theorems can be found in the technical report~\cite{cobalt-tech}.

\begin{theorem}[Soundness]
\label{thm:soundness}
Iff \textnormal{Synthesize} ({$\langle\Gamma$,$\Sigma$, $\Psi\rangle$, $\varnothing$}) = e then 
$\Gamma$;$\Sigma$ $\vdash$ e : $\Psi$.
\end{theorem}

%% \begin{proof}
%% The proof is by an induction on forward/backward synthesis rules and the type 
%% checking rules for $\lambda_{eff}$ expressions.
%% \end{proof}
% 
% We define a {\it small-subset} relation between two search-spaces $S_k$ and 
% ${S_k}^{\Psi}$ for programs of size upto $k$ and a given \name\ synthesis 
% problem $\Psi$, such that if ${S_k}^{\Psi}$ is a {\it small-subset} of $S_k$, 
% then a) it is a subset of $S_k$ and b) iff there exists a solution for the 
% synthesis problem $\Psi$ in $S_k$ then there must exists a solution to the 
% problem in ${S_k}^{\Psi}$. 
% 
% \begin{definition}[Small-Subset]
% A search space $S_k$ is a finite set of all possible expressions ${\mid e 
% \mid}_{k}$ of length upto {\it k}. A  { \it small-subset} ${S_{k}}^{\Psi}$ 
% $\subseteq$ $S_k$ is a search space such that if $\exists$ $e$ $\in$ $S_k$ , 
% such that $\Gamma$; $\Sigma$ $\vdash$ $e$ : $\Psi$, then $\exists$ $e'$ $\in$ 
% ${S_{k}}^{\Psi}$ with $e'$ possibly same as $e$ and $\Gamma$; $\Sigma$ $\vdash$ 
% $e'$ : $\Psi$. 
% \end{definition}

% \begin{lemma}
% \label{lem:small-subset}
% $\forall$ k. the CDCL-Search (Algorithm~\ref{alg:cdcl}) always maintains a {\it Small-subset} of the original search space $S_k$.
% \end{lemma}
% 
% \begin{proof}
% The proof is using the construction of the pruned search-space in CDCL-Search, 
% specifically in the CHOOSE and LEARN subroutines. 
% \end{proof}

\noindent Since the {\sf CDCL} routine (refer Algorithm~\ref{fig:synthesis}) can possibly discard a correct program if 
it can ensure that there exists another program satisfying the given query-spec of 
smaller size, the completeness argument is relative to a query spec.

\begin{theorem}[Completeness]
\label{thm:completeness}
$\forall$ k. If \textnormal{Synthesize} ({$\langle\Gamma$,$\Sigma$, $\Psi\rangle$, $\varnothing$}) =  $\bot$ 
then $\nexists$ e. $\mid$ e $\mid$ $\leq$ k and $\Gamma$; $\Sigma$ $\vdash$ e : $\Psi$. 
\end{theorem}

% 
% \begin{theorem}[Completeness]
% \label{thm:completeness}
% If $\exists$ e. $\Gamma$; $\Sigma$ $\vdash$ e : $\Psi$, then $\exists$ e'. 
% \textnormal{Synthesize} ({$\langle\Gamma$,$\Sigma$, $\Psi\rangle$, $\varnothing$}) =  e' and $\Gamma$; 
% $\Sigma$ $\vdash$ e' : $\Psi$.
% \end{theorem}
% 
% \begin{proof}
% The proof is by the exhaustiveness of the search over a given search space 
% $S_k$, iterative increment of $k$ and Lemma~\ref{lem:small-subset}.
% \end{proof}

%\input{implementation}
\section{Implementation and Evaluation}
\label{sec:impl}

\name\ is implemented in approximately 7300 lines of OCaml.
We rely on OCaml lexing and parsing libraries OCamllex~\cite{ocaml} for
handling the front end of our query specification language and use
Z3~\cite{z3} to discharge SMT queries.  The input to \name\ is a
specification file containing a library of functions and data
constructors, along with their specifications, followed by a goal
query specification.

% 
% \subsection{Exploration Order and Heuristics}

% 
% 
% 
% the following main questions:
% \begin{itemize}
%  \item Generality of the synthesis approach {\it vis-\`a-vis} the features in 
% the program and the domain of the effectful libraries.
% \item Effectiveness of the synthesis approach, in-terms of synthesis times.
% \item Improvements/advantages of our bidirectional-CDCL approach with respect 
% to a na\`ive base-line approaches, a unidirectional deductive enumeration and 
% an enumeration without CDCL-learning. 
% \end{itemize}
%\section{Evaluation}

We evaluate \name by synthesizing programs from several domains and
consider its effectiveness with respect to the following questions:
\begin{itemize}
 \item [$\textnormal{RQ}_1$] Is \name\ effective in synthesizing
   programs from available verified libraries?
 \item [$\textnormal{RQ}_2$] How does \name's integration of forward,
   backward, and CDCL search compare against each technique applied
   individually?
 \item [$\textnormal{RQ}_3$] How sensitive is \name\ synthesis to the
   complexity of library specifications and queries?
 \item [$\textnormal{RQ}_4$] How does \name\ compare against other
   state-of-the art component-based synthesis techniques when
   applied to specification-rich libraries?
 
\end{itemize}

\subsection{Benchmarks}
\begin{figure*}
\begin{subfigure}[b]{0.52\textwidth}
%\begin{minipage}{0.60\textwidth}
  %\vspace*{-.8em}
\centering
\advance\leftskip-1cm
\centering 
\includegraphics[width=1.0\textwidth]{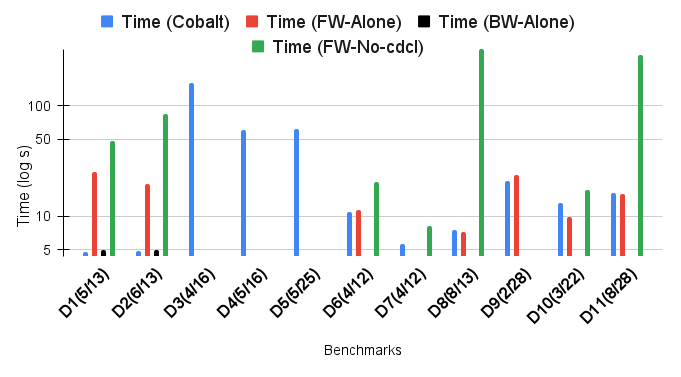}
 \caption{Database Benchmarks}
 \label{fig:chart_databases}
 %   \vspace*{-.8em}
 \end{subfigure}
\begin{subfigure}[b]{0.47\textwidth}
%\begin{minipage}{0.60\textwidth}
 % \vspace*{-.8em}
\centering
\advance\leftskip-1cm
\centering 
\includegraphics[width=1.0\textwidth]{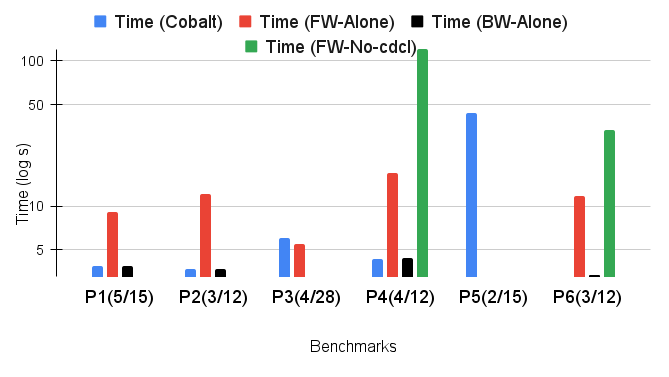}
 \caption{Parsing Benchmarks}
 \label{fig:chart_parsers}
  %  \vspace*{-.8em}
\end{subfigure}    
\begin{subfigure}{1.0\textwidth}
  %\begin{minipage}{0.60\textwidth}
 % \vspace*{-.8em}
\centering
\advance\leftskip-1cm
\centering 
\includegraphics[width=1.05\textwidth]{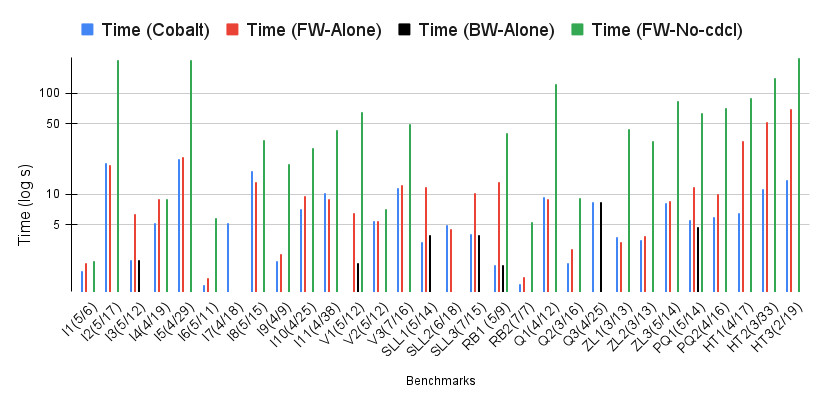}
\caption{Imperative Data Structure Benchmarks}
\label{fig:chart_imperatives}
\end{subfigure}
\label{fig:charts}
\caption{\small Synthesis time in seconds for \name\ (T \name) and
  uni-directional approaches (Time (\textsf{BW-alone})) and (Time
  (\textsf{FW-alone})) and a na\`ive forward synthesis (Time ({\sf
    no-cdcl})).  The horizontal axis enumerates different synthesis
  queries.
%Queries with single bar (like D3) represents a case where the the approach with missing bar did not 
%find a solution in a given time limit of 10 mins. 
% Labels on the top of the bar gives the size of synthesized
% \name\ expressions in terms of the number of AST nodes in the
% expression. 
Benchmarks for which a bar does not appear for a given approach
indicate that the synthesis problem was not solvable within a 10
minute time-bound.  Graphs are given in log-scale. The ratio of size of query specification, to 
the size of
synthesized expressions in terms of the number of AST nodes is given
 within parentheses on the labels of the x-axis. %\SJ{Can we make the labels a bit bigger?}
 }
    \vspace*{-.8em}
\end{figure*}

%explanation of the doamins
We consider a number of synthesis problems
for applications drawn from three different domains.
  A detailed characterization of the queries used is provided in the
  technical report~\cite{cobalt-tech}.
% \SJ{I think we should separate the charts by having D1-D11 and
%   P1-P6 in one (Database and Parsers) and the other chart having
%   all the imperative data structures.}
  The results of applying \name\ to these problem domains are shown in
  Figures~\ref{fig:chart_databases}, ~\ref{fig:chart_parsers} and
  ~\ref{fig:chart_imperatives}.  In these figures, synthesis problems
  for \emph{database} applications are prefixed with ``D'' and are
  adopted from~\cite{database-examples}.  These queries (D1-D11) are
  defined over two database applications.  The first is a {\it
    Newsletter} database with a single table NS with attributes
  {\it newsletter, user, subscribed, articles, code, etc.}  and
  effectful library functions such as {\it subscribe, unsubscribe,
    add, etc.}  An example query (D5) encodes the following problem:
  {\it given a newsletter n and a user u, return the list of articles
    available to u in n, and then unsubscribe u from n}; the solution
  must take care to first check that the user is subscribed to the
  newsletter before unsubscribing.  The second is a network firewall
  database that has two tables, a table of devices and a table storing
  sender-receiver links; its library functions include {\it
    add\_device, add\_connection, delete\_device etc.}  For example,
  query D7 encodes the following problem: {\it insert new devices d
    and x in the device table and create a connection between them}.
  Synthesizing programs from queries of this kind must take into
  account appropriate preconditions that reflect the effectful
  behavior of the library; e.g., to establish a connection to a device
  that is not currently in the device table requires that the device
  first be added.
%This application class has 13 library methods in total.

The second domain consists of \emph{parser} benchmarks prefixed
with ``P'' and include stateful combinator style parsers for
simplified grammars for a PNG image format and C-language declaration
syntax.  The libraries and specifications are constructed using the
grammars of stateful parsers ~\cite{yakker}.  The libraries include
subparsers and standard basic parsers for alphabet, identifier,
number, etc.  The synthesis queries describe the specification for
bigger parsers that can be constructed using these libraries.
E.g. benchmark P1 encodes the following data-dependent property: {\it
  synthesize a parser for a png-chunk using subparsers for length,
  typespec, content, etc. such that the combined size of typespec and
  content of the output chunk is always equal to the parsed length
  value.}  Synthesizing programs that satisfy these kinds of
properties must take into account the effects of upstream parsers on
the length value when considering parsing candidates downstream in a
parsing pipeline.

The third domain considers \emph{imperative data structure} libraries
that implement tables, queues written in OCaml; in the figures, these
benchmarks are prefixed with ``I''.  The \textsf{Table} library
described in Figure~\ref{fig:lib} and its specifications are adopted
from ~\cite{stateful-manifest-contract}, while libraries for Queue are
adopted from the development of mutable data structures given in
Software Foundations~\cite{AppelSF}. The queries we consider involve
multiple insertions, deletions, conditional insertions/deletions
etc. on tables and queues, maintaining library usage protocols.  For
instance, I4 encodes the following query : {\it Given a queue of
  unique integers, and an integer, synthesize a program which
  increments the size of the queue}.  The result must take into
account if the given integer is present in the queue or not and then
appropriately insert either the given or a new integer.
%These
%benchmarks operate over 15 library methods. 

This domain also includes other OCaml data structure libraries
imported from works attempting mechanical verification for OCaml
libraries~\cite{vocal}. These benchmarks are prefixed with the
appropriate data-structure name, for instance ``V'' for OCaml Vector
library, ``HT'' for Hash Tables, etc. The queries again include
standard textbook examples of the usages of these libraries.
E.g. benchmark HT3 encodes the following query: {\it given a hash
  table, and a key-value pair, add the pair in the table, create a new
  hash table and transfer the contents of the current table to the new
  table.}

%These libraries collectively have
%17 methods.

In total, these libraries span 48 files and contain a total of 251
functions, a size that makes memorization of their signatures and
specifications by clients impractical.

\DEL{The imperative data-structure Libraries contribute 105 functions,
  the parser library contributes 32, the database library contributes
  40, with the remaining functions include constructors
  (e.g. \textsf{Pair} and \textsf{Triple}, etc.) and pure functions
  from OCaml libraries like the OCaml Core~\cite{ocaml} (or functions
  translated from the Haskell Core libraries used by other purely
  functional component-based synthesis approach~\cite{hoogle,
    hoogleplus}).}

\DEL{Note that \name works on this complete library set; the alternative
could be to find the minimal set of functions required for each
domain. Finding such a library set apriori is not feasible as these
functions can be called across multiple domains. E.g. a database
domain benchmark may use a List library from the imperative
data-strcture domain; a Queue benchmark may use a pure pair creation
function, etc.  Indeed, we found that 28\% of our synthesized
solutions used at least one function from outside its domain.}

\subsection{Library Specification Annotations}

All the benchmarks in our evaluation were taken from verified
libraries whose specifications were provided by the library authors.
Fortunately there are multiple such projects currently available
across a number of different
domains~\cite{yakker,vocal,stateful-manifest-contract}.  We adopted
these specifications to the \name specification language, a
straightforward mechanical task for most of these benchmarks; four of
the libraries defined specifications that capture richer properties
than what \name\ currenty supports and their specifications had to be
slightly rewritten.  For example, the specification for the
\textsf{Vector} library in the \textsf{VOCal} suite leverages the
algebraic theory of lists which cannot be handled using our SMT driven
synthesis.  Here, we adopted these specification to use more abstract
notions like list membership, ordering, etc. These libraries are used
in benchmarks V1-V3, Q1-Q3, RB1, RB2, and ZL1-ZL3.
% \SJ{Which
%   benchmarks used these?}
Queries were chosen to ensure that every library method for each
application class would be used in at least one solution, that no two
solutions would be identical, and that each solution would entail some
combination of non-trivial control-flow (e.g., pattern-matching over
type constructors) with library calls, and non-trivial synthesis of
function call arguments.

\DEL{ To actually define queries over these annotated libraries, we
  adopted a mix of methodologies: For some benchmarks, we directly use
  the verification task defined by the authors and translate it to its
  synthesis dual.  For example, Figure 2 and benchmarks I10 and I11
  are direct verification queries given in
  ~\cite{stateful-manifest-contract}.  Similarly, the \emph{Firewall}
  example (D6) for deleting network devices is translated directly
  from the verification queries provided in~\cite{database-examples}.
  Additionally, we also manually defined queries using real-life
  scenarios and textbook examples, e.g. extracting read articles from
  a \emph{Newsletter}, while ensuring that the library protocol is followed
  (D5), replacing one device with another in a firewall as a central
  device (D8), etc. We also created several such real-life scenarios
  for databases and textbook examples over imperative data-structure
  libraries including inserting multiple elements in a hashtable,
  adding elements in a queue maintaining uniqueness, etc.  For parser
  examples, we relied directly upon specifications associated with
  known data-dependent grammars for parsers, e.g. a PNG chunk that
  must satisfy a length-payload dependence, is given as specification
  query P1.  }

%\AM{Say more about Query selection.}
% We also list a small set of synthetic benchmarks which highlight useful features of \name\ . These synthesis 
% problems have relatively smaller function specifications while many more paths in the programs to explore, 
% thus highlighting the effect of our CDCL based pruning strategy. 

%\AM{If required we can put Table and Chart side by side, in case Chart is not readable in the current form.}

% \begin{figure*}
%   %\begin{minipage}{0.60\textwidth}
%   \vspace*{-.8em}
% \centering
% \advance\leftskip-1cm
% \centering 
% \includegraphics[width=.8\textwidth]{fig/chart_log.png}
%     \caption{\small Synthesis time in seconds for \name\ (T \name) and
% uni-directional approaches (Time (\textsf{BW-alone})) and (Time (\textsf{FW-alone})). 
% The horizontal axis enumerates different synthesis queries.
% %Queries with single bar (like D3) represents a case where the the approach with missing bar did not 
% %find a solution in a given time limit of 10 mins. 
% Labels on the top of the bar gives the size of synthesized
% \name\ expressions in terms of the number of AST nodes in the
% expression. Benchmarks for which a bar does not appear for a given
% approach indicate that the synthesis problem was not solvable within
% a 10 minute time-bound. Time in log-scale.}
%     \label{fig:chart}
%     \vspace*{-.8em}
% \end{figure*}    

\subsection{Results}
The figures show synthesis times in seconds (the y-axis is in log
scale) executing on a standard Intel laptop with 16GB RAM.  All
queries were executed with a time-out limit of 10 mins and a bound {\it k}=5.  
The timings
are for four different instantiations: the \textcolor{blue}{blue} bar
shows timings for \name\ (with bidirectional synthesis +
CDCL-learning); the \textcolor{red}{red} bar (\textsf{FW-alone}) shows
times with backward synthesis disabled, but with CDCL-leaning enabled;
c) the \textcolor{black}{black} bar (\textsf{BW-alone}) shows times
for just backward synthesis, with forward synthesis and CDCL disabled;
and, d) the \textcolor{green}{green} bar shows the synthesis time for
a na\`ive forward alone synthesis without the CDCL learning component.
Benchmarks with no corresponding bar indicate that the particular
instantiation could not find a solution (i.e., it either timed-out or
got stuck). Each benchmark label along the horizontal axis has an associated numeric value in parenthesis indicating the 
size of the synthesized result for that query in terms of number of AST nodes, e.g. D1 (13).
%on a 2.7 GHz Intel Core i7 Lenovo Thinkpad 
%with 16GB RAM. 
%% We manually
%% verified the synthesized programs for each benchmark while also
%% verifying the program using the built-in verifier in
%% \name\.~\footnote{In future we plan to output weakest pre-condition
%%   formulas for the synthesized programs and verify them using a
%%   third-party imperative program verifier like
%%   $\mathtt{F}^{*}$~\cite{fstar}}
% \begin{figure}[!t]
% %\begin{subfigure}[t]{0.57\textwidth}
% \advance\leftskip-1cm
% \centering 
%     \includegraphics[width=.5\textwidth]{fig/chart_firstedit.png}
%     \caption{\small Synthesis time in seconds for \name\ (T \name) and
% unidirectional approaches (Time (BW-disabled)) and (Time (FW-disabled)). 
% The horizontal axis enumerates different synthesis queries.
% %Queries with single bar (like D3) represents a case where the the approach with missing bar did not 
% %find a solution in a given time limit of 10 mins. 
% Labels on the top of the bar gives the size of synthesized
% \name\ expressions in terms of the number of AST nodes in the
% expression. }
% \label{fig:chart}
%     %\end{subfigure}
% \end{figure}

%Description overall results, generality and usability
\subsubsection{$\textnormal{RQ}_1$ and $\textnormal{RQ}_2$: Effectiveness and impact of design decisions.}
% \SJ{We should indicate the size of the synthesized benchmarks in terms
%   of AST nodes in the graphs.}
Our results show that \name\ was successfully able to synthesize
component-based programs for all the benchmarks considered (\DEL{Note: the synthesis time using \name for V1 (Figure ~\ref{fig:chart_imperatives}) is 1.1 seconds, hence not visible on log scale}).  Overall
synthesis times for all benchmarks take less than one minute, with
approximately 32/47 completing in less than 10 seconds.  The variance
in synthesis times is primarily due to the number of quantified
variables that must be instantiated in queries supplied to Z3.  The
complexity of these generated formulae are in turn dependent on the
complexity of method specifications and synthesis queries, and the
specificity of the expected return type.  More significantly, the
chart also reveals that bi-directional synthesis can solve queries
that are not solvable using just the FW/BW-alone synthesis approaches.

Five of the benchmarks we consider (D3-D5, I7, P5) were unable to be
solved using either forward- or backward synthesis within the given
time bound.  Using just a backward synthesis (\textsf{BW-alone})
method fails to find solutions for 34/47 queries, while disabling
goal-directed search (\textsf{FW-alone}) fails to find a solution in
6/47 queries.  Finally, the na\`ive forward alone synthesis ({\sf
  FW-no-cdcl}) was unable to find a solution for 14/47 queries.  For
the 41 queries \textsf{FW-alone} is able to solve, \name\ is on
average {2x} faster, justifying the benefits of our
bi-directional synthesis strategy over a unidirectional synthesis with
CDCL.

There are 8 queries for which {\sf FW-alone} can find a solution but a
na\"{\i}ve synthesis without the learning component ({\sf FW-no-cdcl})
failed to find a solution. For the remaining queries where both
succeed in finding a solution, the {\sf FW-no-cdcl} is on-average {6x}
slower than the CDCL version, justifying the benefits of using a CDCL
mechanism as part of a forward search procedure.

\paragraph{Synthesized Programs}
The size of synthesized programs (given in parentheses along with the
benchmark name on the x-axis) range between 6 to 38 AST nodes. These
programs include function calls, conditional control-flow,
constructors applications, etc.  The number of
components (continuous chain lengths) across synthesized programs,
range from 2 to 7, comparable to other component-based-synthesis
systems [~\cite{sypet} (Fig. 8), ~\cite{rbsyn} (Table-1)].

As an example of the output \name\ generates,
Figure~\ref{fig:synthesized} presents the synthesis result for query
{\sf D11}, which asks to synthesize a program that, given a globally
shared {\it Firewall} database, and two devices: {\sf d}, a {\it
  central} device, and {\sf x}, deletes {\sf d} and makes {\sf x} as
the central device.
%% Note that each of the components used
%% in the program has an effectful specification.
%% for instance, the function
%% {\sf delete\_device} specification has the following pre-conditions:
%% {\small \begin{lstlisting}[escapechar=\@,basicstyle=\linespread{0.9}\small\sf,numbers=right,
%%       breaklines=true,language=ML] delete_device : (d:device) ->
%%     (x:device) -> {device(D, d)=true $\wedge$ device(D, x)=true
%%       $\wedge$ central(CS, x)=true} $\nu$:unit {...}
%% \end{lstlisting}}
%% It takes the device {\sf d} to delete and requires another device {\sf x} to be {\it central}.
The conditional branches
(lines~\ref{line:outerif},~\ref{line:outerelse}) distinguish cases
when we need to add {\sf x} to the database before deleting {\sf
  d}. Similarly, the nested conditional
(\ref{line:innerif},~\ref{line:innerelse}) distinguishes cases when we
can directly delete {\sf d} (if {\sf x} is {\it central}) or when we
need to first make {\sf x} a {\it central} device.

%% Moreover, unlike other systems
%% ~\cite{sypet}, \name can synthesize programs with conditionals.
% 
% \SJ{This is not a very interesting program; other than the initial check
%   against b, the two branches are identical.  We should try to find a
%   benchmark that exhibits more interesting difference between the two
%   branches.}
\begin{wrapfigure}{l}{.35\textwidth}
\vspace*{-.8em}
\begin{lstlisting}[escapechar=\@,basicstyle=\linespread{0.9}\small\sf,numbers=left, breaklines=true,language=ML]
$\lambda$ (d : device) (x : device).
  b1 <- is_device x;
  if (b1) @\label{line:outerif}@
    b2 <- is_central x; 
    if (b2) then @\label{line:innerif}@
      _ <- delete_device d x;
      ret () 
    else @\label{line:innerelse}@	
      _ <- make_central x;
      _ <- delete_device d x; 
      ret ()
  else @\label{line:outerelse}@
    _ <- add_device x; 
    _ <- make_central x;
    _ <- delete_device d x; 
    ret ()
\end{lstlisting}
\caption{Synthesized Program for query the D11}
\label{fig:synthesized}
\end{wrapfigure}

\paragraph{Utility and Specification Efforts.}
\DEL{Each benchmark label along the horizontal axis also has an
  associated ratio ($p$/$q$) in parenthesis, where $p$ is the size of
  the query specifications in terms of the number of conjuncts in the
  specification, and $q$ is the size of the synthesized result for
  that query in terms of the number of AST nodes. E.g., the label D1
  (5/13) implies that \name\ given query D1, a table insertion query
  whose specification has five conjuncts, produces a synthesized
  program with 13 AST nodes.}

\DEL{These ratios highlight that for simple programs, the size of the
  synthesized programs is comparable to the size of the
  specifications. However, for programs with intricate control flows
  found in some of the database queries (e.g., D9 and D10) or
  conditional queries found in some of the imperative data structure
  benchmarks (e.g., Q3 and HT2), queries are simpler because their
  preconditions are weaker.  At the same time, the synthesized
  programs generated are more complex, especially highlighting the
  power of \name\'s efficient enumerative search with limited
  specification.  }

\DEL{Although \name performs well on this traditional metric, we
  found that writing such programs from scratch (even in OCaml),
  without the use of libraries would typically involve non-trivial
  complexity with intricate control flows, loops, recursion, etc. For
  instance, in the absence of component-based synthesis support,
  synthesizing a program for the query in Figure 2 would need to
  synthesize code/auxiliary functions for tasks like table insertion,
  checking membership, taking the average etc. This makes it
  challenging to apply state-of-the-art deductive synthesis techniques
  directly to our queries, given that synthesizing auxiliary functions
  with these complex features in an effectful setting remains very
  much an open problem~\cite{suslik}.  Thus, a more reasonable and
  precise assessment of \name's capabilities would involve comparing
  the complexity of defining queries with the complexity of the
  overall function synthesized, i.e. the combined size of the
  synthesized code plus the size of each library function used in the
  code.}

In summary, these results support our two main claims: (1) a
bi-directional synthesis strategy is beneficial to reason over
effectful libraries - unlike \name, neither \textsf{FW-alone} nor
\textsf{BW-alone} could successfully discharge all the synthesis
problems in our benchmark suite; note that at least one benchmark in
each application class failed to be solved by either uni-directional
method, indicating that our technique is not specialized to a
particular application class.  And, (2) CDCL learning in this setting
is demonstrably useful - since \textsf{FW-alone} is also equipped with
CDCL, its execution times are competitive with \name\ for the
benchmarks it completes.  We note that disabling CDCL in
\textsf{FW-alone} causes at least an order of magnitude increase in
synthesis times while more than doubling the number of failing
benchmarks.
\vspace{.01em}

%% \paragraph{Other Observations.}
%% \SJ{I don't think these observations add much to the discussion.}
%% For some benchmarks (e.g. P1, P2, SLL1, I9) forward synthesis
%% \textsf{FW-no-cdcl} enters repetitive cycles, e.g.  SLL1, has a function {\sf
%%   create} to create new cells and {\sf clear} to clear new cells, but
%% \textsf{FW-no-cdcl} gets stuck creating chains of programs given by the
%% regular expression: $\textnormal{(create (); clear ())}^{*}$ and thus
%% fails to find a solution; CDCL learning identifies such repetitive
%% paths, prunes their repetitive synthesis, and enables a solution to be
%% found in reasonable time.

%% For several benchmarks we found a \textsf{BW-alone} strategy is
%% sufficient to find a solution (e.g. D1, D2, D2, P2, P1, etc.), while
%% in several others, backward synthesis does not contribute to the
%% overall solution (e.g. D6, D9, RB2, etc.).  We examined these cases
%% and found that backward synthesis is effective when a) libraries have
%% precise (stronger) specifications and/or b) when the synthesis query
%% has strong pre-condition.  These two features ensure that the function
%% specifications and the query provides sufficient information in the
%% environment to apply the \textsf{BW-call} and \textsf{BW-frame} rules,
%% thus guiding the backward synthesis effectively.  This observation
%% again underlines the {\it effectiveness of our goal-oriented backward
%%   search technique} for libraries with rich specifications.

\subsubsection{$\textnormal{RQ}_3$ : Sensitivity to specification complexity and library size.}

Synthesis complexity (and hence  synthesis times) is dominated by
the complexity of the queries discharged to Z3. Synthesis time
increases as function specifications and queries become more complex,
where complexity of specifications is directly correlated with the
number of uninterpreted functions and variables in the query and
number of conjuncts in propositional formulas.

\paragraph{Case Study}

\begin{figure}[htb]
\begin{minipage}{0.45\textwidth}\begin{lstlisting}[escapechar=\@,basicstyle=\linespread{0.9}\small\sf,breaklines=true,language=ML]
subscribe : (n : nl)-> (u :user) -> 
    { nlmem (D , n , u) = true $\wedge$ 
      confirmed (D, n, u) = true $\wedge$ 
      subscribed (D, n, u) = false
    } v :  unit   
    { 
      nlmem (D', n, u) = true $\wedge$ 
      subscribed (D', n, u) = true $\wedge$
      confirmed (D', n, u) = false $\wedge$
     @\texttt{\textcolor{gray}{     
      subsize (D', u) == subsize (D, u) + 1 $\wedge$ \\	
      nlreach (D', n) == nlreach (D, n) + 1}}@
    }	
\end{lstlisting}
\end{minipage}
\begin{minipage}{0.45\textwidth}
\begin{lstlisting}[escapechar=\@,basicstyle=\linespread{0.9}\small\sf,breaklines=true,language=ML]
goal : (n : nl)-> (u :user) ->
    { 
      nlmem (D , n , u) = true $\wedge$ 
      subscribed (D, n, u) = true $\wedge$
      confirmed (D, n, u) = false $\wedge$
      @\texttt{\textcolor{gray}{activenl (D, n) = true $\wedge$
      activeuser (D, u) = true $\wedge$
      subsize (D, u) > 0 $\wedge$
      nlreach (D, n) > 0}}@
    } v : [string]  
    { 
      v = articles (D') $\wedge$ nlmem (D', n, u) = false $\wedge$
      @\texttt{\textcolor{gray}{activenl (D, n) = true $\wedge$
      subsize (D', u) == subsize (D, u) - 1  $\wedge$
      nlreach (D, n) == nlreach (D, n) - 1)}}@
    }    
\end{lstlisting}
\end{minipage}
\caption{Effectful specifications for a \textsf{Newsletter} library
  function and a synthesis query {\sf goal}.  Shaded specifications
  are additional properties that were added to the original to assess
  the \name's sensitivity to specification complexity and size.}
\label{fig:newspec}
\end{figure}

To understand the impact of specification complexity on synthesis
capability, we compared the synthesis times for queries D1-D11 using
its provided specifications, comparing it against the synthesis times
taken when additional qualifiers are added to these specifications.
For instance, the {\it Newsletter} benchmark has three qualifiers in
its original specification {\it viz.}  {\sf nlmem} (a membership
qualifier), {\sf subscribed} (a Boolean-valued subscription function)
and {\sf confirmed} (a Boolean-valued function, indicating if the user
has confirmed an action).  To these, we additionally include the
following four new qualifiers in a new variant of the benchmark: {\sf
  activenl} (a Boolean-valued function that is true if a newsletter
has at least one active subscription), {\sf activeuser} (a
Boolean-valued function capturing if a user has at least one active
subscription), {\sf subsize} (an integer-valued function that gives
the number of newsletters a user is subscribed to) and {\sf nlreach}
(the number of users which are subscribed to a newsletter).

Figure~\ref{fig:newspec} shows the specification for a
library function {\sf subscribe}, which takes a newsletter {\sf n} and
a user {\sf u} and sets the subscription of the user for the
newsletter to true and a synthesis query ({\sf goal}) to synthesize a
program which returns the list of articles read by {\sf u} in {\sf n}
and then unsubscribes the user from the newsletter.  The original
specification and the query is shown in {\bf black}; the modified
variant includes the original formulas plus the new conjuncts (shown
in \textcolor{gray}{gray}).
In a similar fashion we also define revised specifications for the
{\it Firewall} libraries and its associated queries (D6-D11).

\begin{wrapfigure}{l}{.48\textwidth}
  %\begin{minipage}{0.60\textwidth}
  \vspace*{-.8em}
\centering
  \advance\leftskip-1cm
\centering 
\includegraphics[width=.5\textwidth]{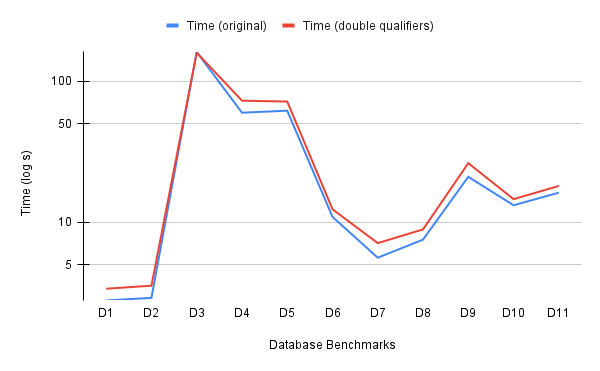}
    \caption{\small Running time comparison between original Database benchmarks against doubling the number of qualifiers in specifications.}
    \label{fig:stability}
    \vspace*{-.8em}
\end{wrapfigure}

Figure ~\ref{fig:stability} shows two line graphs comparing the time
for the original run (\textcolor{blue}{Time (original)}) compared to
the time taken to synthesize a result when these new qualifiers are
added to specifications and queries (\textcolor{red}{Time (double
  qualifiers)}).  Synthesis times increase from {\sf 0\%} to a maximum
of 26\% (case D7); for most other cases, the increase is less than
20\%, an indication that \name's synthesis strategy scales reasonably
well against specification complexity.

%% The results highlight the relative stability of \name against
%% the complexity of the specifications.

\subsubsection{$\textnormal{RQ}_4$ : Comparison to other enumerative
  and deductive component-based synthesis techniques.}

\paragraph{Comparison with other type-directed, component-based enumerative synthesis approaches.}
To address how \name compares against other systems, we
consider the effectiveness of type and example-based synthesis
approaches ~\cite{sypet, table-synthesis, hoogleplus} in solving
effectful queries, using example demonstrations instead of
specifications to guide the synthesis procedure.  For instance, 
the query,
\begin{lstlisting}[basicstyle=\linespread{0.9}\small\sf,breaklines=true,language=ML]
(l : int list) $\rightarrow$ (i : int) $\rightarrow$ { true } (v : int list) { size (v') = size (l) +  1 }
\end{lstlisting}
can be synthesized using type-and-examples by giving the type,
{\sf (int list $\rightarrow$ int list)}, and a set of input-output
examples, e.g., i) {\sf (l=[1; 2], i=3 , output=[1;2;3])} ii) {\sf (l=[1], i=1, output=[1,1])}.

TYGAR~\cite{tygus} is a type-directed component-based tool that
operates over polymorphic Haskell data-types and components.  We
conducted an experiment on an extension of TYGAR, named
Hoogle+~\cite{hoogleplus} that allows using examples to further guide
the TYGAR synthesis process.  To perform our comparison, we modeled
the \textsf{Table} datatype used in our running example as a
functional list over an abstract type, erasing effect annotations from
each of the libraries, and making sure to include suitable libraries
that were available in Hoogle+.

%% In order to make the comparison somewhat interesting and meaningful,
%% we can extend this comparative evaluation by using both types and
%% examples as done in a follow-up work for TYGAR~\cite{hoogleplus}. The
%% idea in this work is to first synthesize a possibly unsound program
%% using TYGAR and then refine the search process using input/output
%% examples there by rejecting unsound programs.

%% We perform an interesting case-study of performing this hybrid
%% synthesis using TYGAR and our introductory {\sf Table} example and
%% present our findings.  {\it Firstly}, wew

% \begin{lstlisting}[basicstyle=\linespread{0.9}\small\sf,breaklines=true,language=ML]
% type pair = Pair of float * int
% type table = [a] 
% 
% add_tbl : (tbl : table) -> (s : a) -> table 
% 
% mem_tbl : (tbl : table) -> (s : a) ->  bool
% ...
% \end{lstlisting}
To simplify things further, we also modified the original query to
just return a new table (rather than the original \textsf{Pair} value)
as follows:
\begin{lstlisting}[basicstyle=\linespread{0.9}\small\sf,breaklines=true,language=ML]
goal: (tbl : table) ->  (s : a) -> {True} v : table 
     {sel (h, tbl) = Tbl $\wedge$ sel (h', tbl) = Tbl' $\wedge$ mem (Tbl', s) $\wedge$ size (Tbl') = size (Tbl) + 1};
\end{lstlisting}
We translated the \name\ query above to the following Hoogle+ query:
\begin{lstlisting}[basicstyle=\linespread{0.9}\small\sf,breaklines=true,language=ML]
goal : (tbl : table) -> (s : a) -> table
\end{lstlisting}
\noindent Running Hoogle+ on this query returns the following synthesized term:
\begin{lstlisting}[basicstyle=\linespread{0.9}\small\sf,breaklines=true,language=ML]
goal = $\lambda$ (tbl : table) (s : a) . (add_tbl s)
\end{lstlisting}
This program is unsound given the original interface of
the {\sf add\_tbl} function since it can violate the uniqueness
invariant of the table, a property enforced by the library via the
precondition ({\sf not (mem (Tbl, s)}) of the {\sf add\_tbl} function.

To refine this result, we next supplied input-output examples to
Hoogle+ to help guide it to find the required (sound) solution.  Some
of the examples provided included:\\
\hspace*{.25in}\begin{minipage}{0.45\textwidth}
\begin{lstlisting}[escapechar=\@,basicstyle=\linespread{0.9}\small\sf, numbers=left, breaklines=true,language=ML]
Input : tbl = [],   s = 'b' ; Output  : ['b']
Input : tbl = ['b'],   s = 'b' ; Output : ['b';'c']
Input : tbl = ['a'],   s = 'b' ; Output : ['a';'b']
\end{lstlisting}
\hfill
\end{minipage}
\begin{minipage}{0.45\textwidth}
\begin{lstlisting}[escapechar=\@,basicstyle=\linespread{0.9}\small\sf, numbers=left, firstnumber=4, breaklines=true,language=ML]
Input : tbl = ['a';'c'],   s = 'b' ; Output : ['a';'c';'b']
Input : tbl = ['a';'b'],   s = 'b' ; Output : ['a';'b';'d']
\end{lstlisting}
\end{minipage}

% \begin{lstlisting}[basicstyle=\linespread{0.9}\small\sf,breaklines=true, numbers=left,language=ML]
% \end{lstlisting}
Unfortunately, these examples were ineffective in helping Hoogle+ to
find a solution.  This is because, although examples are effective at capturing
structural properties like ordering, size or reformatting of inputs,
they are not very useful in defining logical cumulative properties
like membership or its negation.  The input-output pairs at lines 2-5
above try to capture such a property, but fail to do so as the
synthesizer has no way of knowing that the new elements inserted
(i.e. \textsf{'c'}, \textsf{'d'}) are related to the input table by
the property {\it not a member} or are intended to be just another
character.  This example illustrates the difficulty in relating the
shape and contents of an input-output example to a provided logical
specification, especially when these specifications capture effectful
behavior.

\paragraph{Comparison with specification-guided heap manipulating program synthesis.}

\DEL{A direct comparison with other heap- and
  effect-aware synthesis tools~\cite{IPP+21} like Suslik~\cite{suslik}
  or Cypress~\cite{cypress} is not feasible because of fundamental
  differences in approaches and goals.  For example, Suslik supports
  queries over separation-logic formulas with limited support for
  component-based synthesis, and limited expressiveness to specify
  effectful but non-separation specifications.  Conversely, Cobalt
  defines a specification language for reasoning over components with
  non-trivial effectful semantics and rich qualifiers, but does not
  support separation logic formulas that capture fine-grained sharing
  and aliasing properties of the heap.}

\DEL{These differing capabilities are in service of differing goals:
  Suslik aims to synthesize recursive, pointer-manipulating programs
  from inductive specifications using the shape properties expressed
  in these specifications. Cobalt, on the other hand, uses pre/post
  specifications of effectful libraries to guide a component-based
  synthesis procedure for synthesizing non-recursive (albeit
  conditional) programs for complex, effectful (albeit non-separation)
  specifications that do not appeal to sophisticated shape
  properties.}

\DEL{These differences pose major technical challenges in running
  such tools on our benchmarks.  For example, in theory, our queries
  correspond to non-spatial specifications in Suslik. However, both
  our queries and specifications allow rich formulas with
  qualifiers/method-predicates like \textsf{mem}, \textsf{size}, etc. Unfortunately,
  such specifications are beyond what is currently supported for pure
  (non-spatial) formulas in Suslik, which only supports qualifiers
  over a simple theory of linear arithmetic. This limitation is
  discussed by the authors in follow-up work
  ~\cite{IPP+21}, Sec 4.2.}

\DEL{To attempt to better quantify these differences, we translated
  the Cobalt synthesis problem (Sec. 3.1) to separation-logic
  (spatial) formulas in Suslik because spatial formulas do allow
  method predicates; each of these, however, must be given a logical
  interpretation. We then ran Suslik on this translated problem.}

\DEL{ For example, we translated the Cobalt synthesis problem given in
  Figure~\ref{fig:solution} to a Suslik query as follows.  We first
  define a unique list (\textsf{ulist}) that is a singly-linked list with
  unique elements to model the table data structure.}

\begin{lstlisting}[escapechar=\@,basicstyle=\linespread{0.9}\small\sf, breaklines=true,language=ML]
predicate ulist(loc x, set s) {
    | x == 0 => { s == {} ; emp }
    | not (x == 0) => { s == {v} ++ s1 $\wedge$ not (v in s1); [x, 2] ** x :-> v ** (x + 1) :-> nxt ** ulist(nxt, s1) }
}
\end{lstlisting}
\DEL{We then define two qualifiers \textsf{sll\_mem} and \textsf{sll\_len}
  over the table as inductive separation-logic formulas:}
\begin{minipage}[t]{.42\textwidth}
\begin{lstlisting}[escapechar=\@,basicstyle=\linespread{0.9}\small\sf, breaklines=true,language=ML]
predicate sll_len(loc x, int len) {
    | x == 0 => { len == 0 ; emp }
    | not (x == 0) =>
       { len == len1 + 1; [x, 2] ** x :-> v **
         (x + 1) :-> nxt ** sll_len(nxt, len1) }
}
\end{lstlisting}
\end{minipage}
\hfill
\begin{minipage}[t]{.52\textwidth}
\begin{lstlisting}[escapechar=\@,basicstyle=\linespread{0.9}\small\sf, breaklines=true,language=ML]
predicate sll_mem(loc x, int str, set s, bool mem){
    | x == 0 => { s == {} && false; emp }
    | not (x == 0) =>
       { s == {v} ++ s1 && (str == v || mem1);
        [x, 2] ** x :-> v ** (x + 1) :-> nxt
        ** sll_mem(nxt, str, s1, mem1) }
}
\end{lstlisting}
\end{minipage}
\DEL{We next defined a library of functions using \textsf{ulist} and these
qualifiers in terms of separation formulas.
Finally, we took \name queries and translated these to Suslik
  queries; for example, the functional query specification shown in
  Figure~\ref{fig:solution} can be written as follows:}
\begin{lstlisting}[escapechar=\@,basicstyle=\linespread{0.9}\small\sf,  breaklines=true,language=ML]
void goal (loc r, loc ret)
{ r :-> x ** ret :-> val ** ulist (x, s) ** sll_len (x, n) }
{(mem == true) $\wedge$ n1 == n + 1; r :-> y ** ulist (y, s1) ** sll_mem (y, val, s1, mem) ** sll_len (y, n1)}
\end{lstlisting}
\DEL{A formula in Suslik has two components \{ $\phi$; $P$\}; a non-separation formula (authors call it a \textit{pure} formula) $\phi$ and an \textit{impure} component $P$ possibly containing separation formulas. $P$ contains {\it points-to} specification (written as {\sf x} {:->} {\sf y}) and separating conjuncts (written as ($P1$ ** $P2$)). 
The above query's post-condition thus requires that the size of the output {\sf ulist} (represented by {\sf y}), which is pointed-to by {\sf r}, has size one greater than the input {\sf ulist} (represented by {\sf x}), and the returned value (represented by {\sf val}) is present in {\sf y}.}

\begin{wrapfigure}{l}{.35\textwidth}
\begin{tabular}{| l | c | c |}
 \hline %inserts double horizontal lines
 Benchmark & Cobalt & Suslik \\ 
 \hline
    I1 & $\checkmark$ & err \\
    I4 & $\checkmark$ & t/o \\
    I5 & $\checkmark$ & t/o \\
    I6 & $\checkmark$ & err \\
    I9 & $\checkmark$ & t/o \\
    I10 & $\checkmark$ & t/o \\
    I11 & $\checkmark$ & t/o \\
    V1 & $\checkmark$ & err \\
\hline
 \end{tabular}
 \caption{Results of running selected Cobalt queries on Suslik.} %
 \label{fig:suslik}
\end{wrapfigure}
\DEL{On this query, the tool timed out with a timeout of 30000
  seconds\footnote{This is the timeout value set in the online version
    of the tool.}  (approx.  8 hours 20 minutes).  We ran a similar
  experiment on several other \name benchmarks wherever it was
  possible for us to define an inductive separation-logic predicate
  corresponding to the qualifiers in our specifications for the
  data-structure used in the benchmarks. Figure~\ref{fig:suslik} shows
  the results for these benchmakrs. For each of these experiments,
  Suslik either timed-out (t/o) or generated a program with an {\sf
    error} expression, a result indicating an inconsistent internal
  state encountered while solving the query.}

%% Unfortunately, it was not feasible for us to
%% define appropriate inductive separation predicates for qualifiers
%% required in the data-base domain as well as parser domain.

\DEL{These experiments give anecdotal evidence about our claim of
  differing capabilities and goals.  We conjecture the timeout happens
  because Suslik hides complexities of the function call rule into a
  call-abduction routine. This routine prepares the current heap (say
  \textsf{H}) to take it to another heap (say \textsf{H'}) such that
  the precondition of a function \textsf{f} holds in
  \textsf{H'}. Thus, Suslik uses an abduction (forward search)
  procedure for resolving multiple function calls.  While this may be
  sufficient in the case of a single recursive function call, its
  generalization to handle multiple library functions requires a
  multi-abduction decision procedure~\cite{Albarghouthi2016} to
  resolve, a challenging problem in the presence of expressive
  data structures of the kind used in our benchmarks~\cite{ZD+22}.}

  %% (true for
  %% most of their benchmarks) but in the presence of multiple library
  %% functions, this is likely to be insufficient and will require a
  %% muti-abduction decision procedure to resolve. For example, the
  %% "then" branch of the synthesized function in Fig 2, has 3 function
  %% calls and will require multi-abductive
  %% inference. Devising multi-abduction decision
  %% procedures over formulas with rich data structures of the kind used
  %% in our work remains a challenging problem, however. Suslik's
  %% single-abduction procedure is analogous to the FW-NO-CDCL variant of
  %% our synthesis in theory.}

\subsection{Limitations}
\DEL{\name relies on automated verification of the programs for its
  forward, backward as well as the CDCL search procedures.  This
  reliance requires \name to make fundamental assumptions about
  library behavior that (a) each heap object (OCaml reference) is
  always referenced by a unique path (e.g, a variable \textsf{x} or a
  field access \textsf{x.f.y}) and, (b) that there is no sharing of
  heap objects. This allows us to reason locally and automatically
  about effectful behavior of library functions without the need for
  reasoning over separation-logic formulas.  In practice, this
  restriction prevents us from synthesizing programs fron libraries
  that do require heap sharing, for example, those that implement
  cyclic data structures, graphs, etc.  We leave incorporating
  approaches such as~\cite{grasshopper,natural} that enable some
  degree of automated verification for programs specified using
  separation formulas into our synthesis pipeline as a topic for
  future work.}

\section{Related Work}
\label{sec:rel}
%\vspace{-.8em}
%\TODO{Make Changes based on the Reviews}
\noindent \paragraph{Deductive Program Synthesis.}  Closely related
deductive synthesis approaches to ours include Suslik~\cite{suslik}
and its follow-up work Cypress~\cite{cypress}, both of which take
Hoare-triple style specifications and have synthesis rules for
function calls.  An important difference between \name and Suslik
specifications stems from the form of Suslik pre- and post-conditions,
\{$\phi$; P\} $\rightarrow$ \{$\phi'$ ; Q\}, that are expressed using
two components - a non-separation part $\phi$ defining constraints on
the logical data structure associated with the actual mutable data
structure, and a {\it spatial} part defining assertions related to the
shape of the heap.  The pre- and post-formulas in \name specifications
are analogous to the non-separation part of Suslik specifications; our
specification framework has no corresponding analog to Suslik's
spatial component.  For non-spatial properties, however, our queries,
and consequently our specifications, enable rich formulas with
qualifiers, expressivity that is beyond what is currently supported
for pure formulas in Suslik, which only supports a simple theory of
linear arithmetic. 
These significant differences lead to fundamentally different
synthesis strategies making any kind of direct comparison infeasible.
We note that Suslik does allows libraries of functions through their ({\sf Abduce Call}) rule
which is a na\`ive call-abduction routine that is equivalent to \name's {\it no-cdcl}
approach in theory.

Other deductive synthesis~\cite{fiat,synquid,leon,narcissus} and
proof-search guided synthesis efforts~\cite{myth,type-and-example} use
deductive proof rules for synthesizing pure terms.  In contrast, our
synthesis rules operate over heap manipulating expressions ({\it viz.}
effectful function calls and sequencing).  Synquid~\cite{synquid} uses
a bi-directional typing calculus to synthesize functional
programs. However their notion of bi-directionality is related to
bi-directional typing~\cite{bidirectional}, which is unrelated to the
notion of bi-directionality used in \name\ that is defined with
respect to forward and backward proof search over programs with
effectful specifications~\cite{htt,ynot,fstar}.  Viser~\cite{viser}
uses a light-weight bidirectional abstract interpretation coupled
tightly to table transformation tasks that can reason specifically
about simple table inclusion constraints; however, it cannot handle
general effectful specifications of the kind intended to be used in
\name.

%% In
%% fact, we can use Synquid or one of these systems as a purely
%% functional component in our work.

\noindent \paragraph{Component-based Synthesis and Learning.}  There
is a long line of work on the use of component-based synthesis in the
context of domain-specific languages~\cite{table-synthesis,
  oracle-guided-synthesis} as well as general-purpose programming
domains~\cite{sypet,tygus,cdcl-synthesis,rbsyn,frangel,viser}.
\name\ is distinguished from these other systems in the form of the
query specifications that we consider (formal query specifications
expressed as Hoare triples vs.  input-output or informal
specifications as found in~\cite{sypet,rbsyn,cdcl-synthesis,frangel});
and, in our expectation that the libraries from which synthesized
programs are constructed are effectful, in contrast to the assumptions
made in~\cite{tygus,cdcl-synthesis}. Although Sypet~\cite{sypet} does
support effectful libraries, it does not use library protocol
specifications to guide synthesis and consequently cannot enforce
associated library protocols or programs that require conditional
control-flow.

%% We differ from these
%% efforts in the following ways : a) query specifications foramt {\it viz.}
%% input-out/informal specifications
%% vs formal query specifications in \name\ b) Pure, functional libraries
%% and specifications  vs possibly
%% effectful libraries in \name\ c) Using fine-grained effect
%% specifications in Hoare
 
Our CDCL-learning based enumeration is similar in spirit to the
conflict-driven learning based synthesis of pure
components~\cite{cdcl-synthesis}, and is inspired by conflict-driven
learning based enumeration techniques found in modern SAT
solvers~\cite{cdcl-sat}.  The discriminating propositions learned in
\name, however, must include the path-sensitive, effectful semantics
of failed programs in terms of the strongest postconditions associated
with these programs, in contrast to the simpler, global set of Boolean
propositional formulas associated with the partial failed programs
found in ~\cite{cdcl-synthesis}, such a global formula in presence of effectful libraries
with path-sensitive information will grow too large and overwhelm the solver.
Further, we must also account for
program failures due to bounded exploration of an unbounded search
space without losing completeness, a challenge ~\cite{cdcl-synthesis}
does not face.

\section{Conclusions}
\label{sec:conc}

We present a new specification-guided synthesis framework capable of
synthesizing effectful programs from a library of effectful
components.  We capture the behavior of these components using a rich
specification language that capture effectful behavior in terms of
Hoare-style pre- and post-conditions.  The synthesis procedure itself
combines forward and backward proof search with respect to these
specifications, integrating a CDCL-style learning framework to enable
scalability.  Experimental results on a tool (\name) that integrates
these ideas are promising, demonstrating \name's ability to
efficiently synthesize programs over complex effectful queries,
guaranteed to be consistent with component specifications.

%% There are a few important directions to take \name\ in Future: Several
%% effectful programing tasks rely on a global inductive invariants to
%% hold e.g. data-base applications with invariants, Document object
%% model based programming, etc. Since \name\ specifications currently
%% does not allow such global invariants, we cannot synthesize such
%% programs. Extending \name\ to be cognizant of global inductive
%% invariants is one inetersting direction for future work.

%% More generally, component-based synthesis primarily focuses on straight line codes, however some 
%% programs may require recursive calls/loops similar to other deductive synthesis approaches, 
%% thus extending 
%% \name\ to support interplay between component-based synthesis with recursive procedure calls can be a 
%% direction for future work.

%% Acknowledgments
\begin{acks}                            %% acks environment is optional
                                        %% contents suppressed with 'anonymous'
  %% Commands \grantsponsor{<sponsorID>}{<name>}{<url>} and
  %% \grantnum[<url>]{<sponsorID>}{<number>} should be used to
  %% acknowledge financial support and will be used by metadata
  %% extraction tools.

  We thank the anonymous reviewers for their detailed comments and
  suggestions.  Funding for this work material is supported in part by
  DARPA, under the Safe Documents (SafeDocs) program.
\end{acks}

%% Bibliography
\bibliography{paper}

%% Appendix
% \appendix
% \section{Appendix}
% 
% Text of appendix \ldots
\section{Supplemental Material for the Main Paper}
\label{sec:supplementary}

\subsection{Extended Forward Synthesis Rules}
\begin{figure}[h]
\begin{flushleft}
\fbox{\small
     $\Gamma  \ottsym{;}  \Sigma  
\vdash  \tau  \twoheadrightarrow  \ottnt{e}$          
} \ \ {\ottcom{Forward Synthesis}}
\end{flushleft}
\small \begin{center}
\inference[FW\_ConsApp]{\Gamma; \Sigma \vdash \tau_j \twoheadrightarrow y_j \\
		      D_i \overline{(x_j: \tau_j)} \rightarrow \{\nu : TN | \phi_i\} \in \Sigma \\
		      {\phi_i}[y_j/x_j] => \phi_i'
		    } {\Gamma;\Sigma \vdash \{\nu : TN | \phi_i'\} \twoheadrightarrow D_i (\overline{y_j})}
\end{center}
\bigskip
\small \begin{center}
\inference[FW\_Ref]{  l : t \ ref \in \Gamma} 
		  {\Gamma;\Sigma \vdash t \ ref  \twoheadrightarrow l}
\end{center}

% \begin{minipage}{0.5\textwidth}
% \end{minipage}

\bigskip

\small \begin{center}
\inference[FW\_Lambda]{\Gamma, \overline{x_i : \tau_i}; \Sigma \vdash \tau \twoheadrightarrow e} 
	      {\Gamma;\Sigma \vdash (\overline{x_i : \tau_i}) \rightarrow \tau \twoheadrightarrow \lambda 
(\overline{x_i :\tau_i}). e}
\end{center}
\bigskip

\caption{Extra FW\_Synthesis Rules}
\label{fig:fw-full}
\end{figure}

\subsection{Typing Rules for $\lambda_{\mathit{eff}}$}
Fig.~\ref{fig:typing-eff} presents typing semantics for $L$-expressions. Each typing judgment is of the 
form 
$\Gamma;\Sigma \vdash$ {\sf e}  : $\tau$, saying, in a environment $\Gamma;\Sigma$, an expression $e$ has a 
type $\tau$.
\begin{figure}[h]

\small \begin{center}
\inference[T-var]{{\sf x} : \tau \in \Gamma} 
				{
				\Gamma; \Sigma \vdash {\sf x} : \tau}
\end{center}
\bigskip

\small \begin{center}
\inference[T-consapp]{ D_i \overline{(x_j: \tau_j)} \rightarrow \{\nu : TN | \phi_i\} \in \Sigma \\
		      \Gamma; \Sigma \vdash \overline{x_i : \tau_i}
		      } {\Gamma; \Sigma \vdash D_i (\overline{y_j}) : \{\nu : TN | \{\phi_i [y_i/x_i]\} }
\end{center}
\bigskip

\small \begin{center}
\inference[T-call]{ \Gamma; \Sigma \vdash f : (\overline{x_i : \tau_i}) \rightarrow \{P\} {\sf v : t} \{Q\} \\ 
		    \Gamma; \Sigma \vdash \overline{y_i : \tau_i} 
		    }
		      {\Gamma ; \Sigma \vdash f (\overline{y_i}) : \{P[y_i/x_i]\} {\sf v : t} \{ Q[y_i/x_i]\}}
\end{center}
\bigskip

\small \begin{center}
\inference[T-match]{
		      D_i \overline{(x_j: \tau_j)} \rightarrow \{\nu : TN | \phi\} \in \Sigma \\
		      \Gamma;\Sigma \vdash e : \{\nu : TN | \phi\} \\
		      \Gamma, \overline{x_j : \tau_j}; \Sigma \vdash e_i : \tau}
		      {\Gamma ; \Sigma \vdash {\bf match} \ {\sf e} \ {\bf with} \  D_i (x_j) -> e_i : \tau}
\end{center}
\bigskip

\small \begin{center}
\inference[T-if]{ \Gamma;\Sigma \vdash e : \{ \nu : bool | \phi_t \wedge \phi_f\} \\
		   \Gamma, \phi_t[true/\nu];\Sigma \vdash e_t : \tau \\
		   \Gamma, \phi_f[false/\nu];\Sigma \vdash e_f : \tau}
		      {\Gamma ; \Sigma \vdash {\bf if} \ {\sf e} \ {\bf then} \  e_t \ {\bf else} \ e_f : \tau}
\end{center}
\bigskip

\small \begin{center}
\inference[T-frame]{ \Gamma;\Sigma \vdash e : \{P\} \nu : {\sf t} \{Q\}\\
		    \ottsym{(}  \ottsym{(}  \ottkw{Vars} \, \ottsym{(}  \ottmv{R}  \,  
\ottsym{)}  \ottsym{)} \, \cap \, \ottsym{(}  \ottkw{EVars} \, \ottsym{(}  
\ottmv{P}  \,  \ottsym{,}  \ottmv{Q}  \,  \ottsym{)}  \ottsym{)}  \ottsym{)}  
\!\! = \!\! \, \varnothing  %
\ottsym{(}  \ottsym{(}  \ottkw{Qual} \, \ottsym{(}  \ottmv{R}  \,  
\ottsym{)}  \ottsym{)} \, \cap \, \\ %NEWLINE%
\ottsym{(}  \ottsym{(}  \ottkw{Qual} \, 
\ottsym{(}  \ottmv{P}  \,  \ottsym{)}  \ottsym{)} \, \cup \, \ottsym{(}  
\ottkw{Qual} \, \ottsym{(}  \ottmv{Q}  \,  \ottsym{)}  \ottsym{)}  \ottsym{)}  
\ottsym{)}  \!\! = \!\!  \ottsym{(}  \varnothing  \ottsym{)} 
		    }
		      {\Gamma  \ottsym{;}  \Sigma  \vdash e :  \ottsym{\{}  \ottmv{P}  \,  \wedge  
\ottsym{(}  \ottmv{R}  \,  \ottsym{)}  \ottsym{\}}  \nu  \ottsym{:}  \ottnt{t}  
\ottsym{\{}  \ottmv{Q}  \,  \wedge  \ottsym{(}  \ottmv{R}  \,  \ottsym{)}  
\ottsym{\}} }
\end{center}
\bigskip

\small \begin{center}
\inference[T-seq]{ \Gamma; \Sigma \vdash e_1 : \{P1 \} \nu : t_1 \{ Q1\} \\
		  \Gamma, x : t_2 \vdash e_2 : \{P2 \} \nu : t_2 \{Q2\} 
		    }
		      {\Gamma; \Sigma \vdash \{P1 \wedge Q1 => P2\} \nu : t2 \{Q1 \wedge Q2 \} }
\end{center}
\bigskip

\small \begin{center}
\inference[T-sub]{\Gamma; \Sigma \vdash e : \{P1\} \nu : t \{Q1\} \\
		   \Gamma; \Sigma \vdash  P => P1  & \Gamma; \Sigma \vdash  Q1 => Q
		    }
		      {\Gamma; \Sigma \vdash e : \{P\} \nu : t \{Q\} }
\end{center}
\bigskip

\small \begin{center}
\inference[T-hole]{
		    }
		      {\Gamma; \Sigma \vdash ((??) : \tau) : \tau }
\end{center}
\bigskip

\small \begin{center}
\inference[T-skip]{
		    }
		      {\Gamma; \Sigma \vdash skip : \{True\} v : unit \{True\} }
\end{center}
\bigskip
\caption{Typing Rules for $\lambda_{\mathit{eff}}$ expressions}
\label{fig:typing-eff}
\end{figure}

% 
% 
% \subsection{Complete Backward Synthesis Rules}
% \begin{figure}[htbp]
% 
% \small \begin{center}
% \inference[T-fun]{\Sigma, {\sf x} : \tau \vdash e : \textnormal{state} \ \{ \phi \} \ t1  \{ \phi' \}} 
% 				{
% 				\Sigma \vdash \textnormal{$\lambda$ x : t}. {\sf e} : \textnormal{state} \ \{ 
% \phi \} \ t1  \{ \phi' \} }
% \end{center}
% \bigskip
% 
% 
% \caption{Typing Semantics for $\lambda_{\mathit{eff}}$ expressions}
% \label{fig:bw-full}
% \end{figure}
% 

% \AM{These semantics are not interesting in themselves, so we can put them in the Appendix}

\clearpage
\section{Details of Soundness Theorems and Proofs}

\subsection{Soundness}
\begin{lemma}[BW-Soundness]{\label{lemma:bw-sound}}
If $\Gamma  ;  \Sigma  \vdash  \{ \phi_{{\mathrm{1}}} \}  \diamondsuit  \{ 
\phi_{{\mathrm{2}}}  \} \twoheadleftarrow \ottnt{e}$ or
$\Gamma  \ottsym{;}  \Sigma  \vdash  \tau  \twoheadleftarrow  \ottnt{e}$ then 
$\Gamma$;$\Sigma$ $\vdash$ e : $\{ \phi_{{\mathrm{1}}} \}  v : t  \{ 
\phi_{{\mathrm{2}}}  \}$ or $\Gamma$;$\Sigma$ $\vdash$ e : $\tau$ respectively.
\end{lemma}

\begin{proof}
The proof is via induction over BW\_Synthesis rules (Figure 8 in the main paper):
First, the bases cases:
\begin{itemize}
 \item Case BW\_FW : Using Lemma ~\ref{lemma:fw-sound}
 \item Case BW\_Hole : Using Typing rule for hole (T-hole) (refer Figure~\ref{fig:typing-eff}) and typing rule 
for sequencing (T-seq) for the term y $\leftarrow$ (??) : $\tau$ ; skip.  
\end{itemize}
Next the inductive cases:
\begin{itemize}
 \item Case BW\_Call : We give a proof for a function call with 1 argument, the general case will be an 
inductive argument over number of arguments i:
\begin{enumerate}
 \item Using IH $\Gamma; \Sigma$ $\vdash$ $y_i$ : $\tau_i$ and $\Gamma; \Sigma$ $\vdash$ $e_i$ : \{ P\} (??) : 
$\tau_i$ \{P'\}.
\item Using Typing rule for function application (T-call), the type for the function 
call f ($y_i$) for the given post condition Q is given by \{P1 $\wedge$ (Q1 $\implies$ Q)\} v : t' \{Q\}.
Which is equivalent to \{ P' \} v:t' \{Q\} using the definition of P'.
\item Using the typing rule for sequencing for the sequence term ($y_i$ $\leftarrow$ $e_i$; f ($y_i$)) we get 
\{P\} v : t' \{P' $\wedge$ Q\}.
\item Using 1-3 and the subtyping rule for effectful computations, T-sub, we get $\Gamma; \Sigma$ $\vdash$ e 
: \{P\} v : t' \{Q\}.  
\end{enumerate}
\item Case BW\_Frame : 
  \begin{enumerate}
   \item Using IH over the premise we get $\Gamma; \Sigma$ $\vdash$ e : \{$P_1$\} (??) : t \{$Q_1$\}.
   \item Using the T-frame rule and the constraints in the premise {\it viz.} (P $\vdash$ $P_1$ $\wedge$ $R$), 
($Q_1$ $\wedge$ $R$ $\vdash$ Q), we get the goal condition.
  \end{enumerate}
\item Case BW\_Sub : Using IH and subtyping rule for computations (T-sub). 
\end{itemize} 
\end{proof}

\begin{lemma}[FW-Soundness]{\label{lemma:fw-sound}}
If $\Gamma  \ottsym{;}  \Sigma  \vdash  \tau  \twoheadrightarrow  \ottnt{e}$ then 
$\Gamma$;$\Sigma$ $\vdash$ e : $\tau$.
\end{lemma}

The proof methodology is similar to the BW-Soundness, but is defined over the FW\_Synthesis rules 
\begin{proof}
Applying Induction on the Forward Synthesis rules:
(Figure 7 in the main-paper and Figure~\ref{fig:fw-full} in this Supplemental material for extended rules) and 
the typing rules for $\lambda_{\mathit{eff}}$ expressions (Figure~\ref{fig:typing-eff}).
First the base cases : 
\begin{itemize}
 \item Case FW\_Var : Directly using the T-var rule.
 \item Case FW\_Ref : Using the T-ref rule.
 \end{itemize}
Next the inductive Cases:
\begin{itemize}
 \item Case FW\_ConsApp : Using IH, $\Gamma;\Sigma$ $\vdash$ $y_i$ : $\tau_i$ and using the T-consapp rule, we 
get $\Gamma;\Sigma$ $\vdash$ $D_i$ ($\overline{y_i}$) : \{v : TN | $\phi_i'$\}.
 \item Case FW\_Lambda : Using IH and T-lambda rule.
 \item Case FW\_Match : Using IH and T-match rule.
 \item Case FW\_If : Using IH on the two branches and combining the results using the T-if rule.
 \item Case FW\_Call : The proof again is for function call with 1 argument, the general case will be an 
inductive argument over the number of arguments i:
\begin{enumerate}
 \item Using IH $\Gamma; \Sigma$ $\vdash$ $y_i$ : $\tau_i$.
\item Using Typing rule for function application (T-call), the type for the function 
call f ($y_i$) for the given post condition Q is given by \{P  $\wedge$ (P $\implies$ P1)\} v : t' \{P 
$\wedge$ Q1\}.
Which is equivalent to \{ P \} v:t' \{Q'\} using the definition of Q'.
\item Using IH, $\Gamma; \Sigma$ $\vdash$ $e$ : \{ Q'\} v : t {Q}.
\item Using the typing rule for sequencing for the sequence term (f ($y_i$);$z$ $\leftarrow$ $e$; ) we get 
\{P\} v : t' \{Q' $\wedge$ Q\}.
\item Using 1-3 and the subtyping rule for effectful computations, T-sub, we get $\Gamma; \Sigma$ $\vdash$ e 
: \{P\} v : t' \{Q\}.
\end{enumerate}
\item Case FW\_Frame : Using IH over the premise we get $\Gamma; \Sigma$ $\vdash$ \{$P$\} v:t \{$Q$\}. Using 
the T-frame rule we get the goal condition.
\item Case FW\_Sub : Using IH and the T-Sub rule for computation Subtyping.  
\end{itemize}
\end{proof}

\begin{theorem}[Soundness]
\label{thm:soundness}
Iff \textnormal{Synthesize} ({$\langle\Gamma$,$\Sigma$, $\Psi\rangle$, $\varnothing$}) = e then 
$\Gamma$;$\Sigma$ $\vdash$ e : $\tau$.
\end{theorem}

The proof of the soundness theorem is by a case analysis over the {\sf Synthesize} routine followed by an 
induction on forward/backward synthesis rules, while using $\lambda_{\mathit{eff}}$ typing semantics.

\begin{proof}

From the premise of the Soundness statements : Given \textnormal{Synthesize} ({$\langle\Gamma$,$\Sigma$, 
$\Psi\rangle$, $\varnothing$}) = e. 
Performing a case-split over , when {\sf Synthesize} return an e $\neq$ $\bot$ we get the following cases: We 
first take the simpler of the two cases:
\begin{enumerate}
 \item The {\bf if} case, (line 3, Algorithm 1 in main-paper): e = BW\_Rules ($\Gamma, \Sigma, \Psi$, F) 
 We get this directly using the Lemma~\ref{lemma:bw-sound}.
 
 \item The {\bf else} case, (line 7 Algorithm 1):  e = $e_f$ ; $e_b$, where 
 $e_f$ = CDCL ($\langle\Gamma$,$\Sigma$, $\Psi\rangle$, $\mathtt{H}$) and $e_b$ = BW\_Rules 
($\langle\Gamma$,$\Sigma$, $\Psi\rangle$, $\mathtt{F}$).
\begin{enumerate}
 \item Using the definition of CDCL, iff e = CDCL ($\langle\Gamma$,$\Sigma$, $\Psi\rangle$, $\mathtt{H}$) then 
  e = FW\_SUB ($\Gamma$,$\Sigma$, ($p_i$),  $\Psi\rangle$) (line 19, Algorithm 1).
  \item Further, using the soundness result for forward synthesis rules (Lemma~\ref{lemma:fw-sound}, we get 
$\Gamma;\Sigma$ $\vdash$ $e_f$ : $\Psi$.
\item  $\Psi'$ is defined as \{P\} v : \_ \{WP ($e_b$, Q)\}
\item Using a-c and T-seq, the type for e = $e_f$; $e_b$ is \{P\} v : t \{Q\}. 
\end{enumerate}
\end{enumerate}
\end{proof}

\subsection{Completeness}
Since the {\sf CDCL} routine (refer Algorithm 1) can possibly discard a correct program if 
it can ensure that there exists another program satisfying the given query-spec, the completeness 
argument is relative to a query spec.

\begin{theorem}[Completeness]
\label{thm:completeness}
$\forall$ k. If \textnormal{Synthesize} ({$\langle\Gamma$,$\Sigma$, $\Psi\rangle$, $\varnothing$}) =  $\bot$ 
then $\nexists$ e. $\mid$ e $\mid$ $\leq$ k and $\Gamma$; $\Sigma$ $\vdash$ e : $\Psi$. 
\end{theorem}

For a given value of k, let us call the original search space of programs of size less than or equal to k 
as $S_k$. The CDCL search strategy is exhaustive over the search-space it maintains while filtering out 
k-equivalent-modulo-stuckness program terms, let us call this filtered search space as ${S_{k}}^{\Psi}$ thus, 
the only source of incompleteness leading to the failure of the above theorem can occur iff there exists a term 
e in the original unfiltered search space (e $\in$ $S_{k}$) such that $\Gamma; \Sigma$ $\vdash$ e : $\Psi$, 
but e $\notin$ ${S_{k}}^{\Psi}$ and $\nexists$ e' $\in$  ${S_{k}}^{\Psi}$, such that $\Gamma; \Sigma$ $\vdash$ 
e' : $\Psi$. Our completeness proof argues that this condition can never occur in CDCL.

We define a {\it small-subset} relation between two search-spaces $S_k$ and 
${S_k}^{\Psi}$ for programs of size upto $k$ and a given \name\ synthesis 
problem $\Psi$, such that if ${S_k}^{\Psi}$ is a {\it small-subset} of $S_k$, 
then a) it is a subset of $S_k$ and b) iff there exists a solution for the 
synthesis problem $\Psi$ in $S_k$ then there must exists a solution to the 
problem in ${S_k}^{\Psi}$. 

\begin{definition}[Small-Subset]
A search space $S_k$ is a finite set of all possible expressions ${\mid e 
\mid}_{k}$ of length upto {\it k}. A  { \it small-subset} ${S_{k}}^{\Psi}$ 
$\subseteq$ $S_k$ is a search space such that if $\exists$ $e$ $\in$ $S_k$ , 
such that $\Gamma$; $\Sigma$ $\vdash$ $e$ : $\Psi$, then $\exists$ $e'$ $\in$ 
${S_{k}}^{\Psi}$ with $e'$ possibly same as $e$ and $\Gamma$; $\Sigma$ $\vdash$ 
$e'$ : $\Psi$. 
\end{definition}

\begin{definition}[k-bound-stuck-path]{\label{def:k-stuck}}
For a given k and $\Psi$ $\equiv$ \{$\phi$\} v : t \{ $\phi'$\}, a path $p_i$ $\equiv$ $c_1;c_2;...c_i$ is a 
{\it k}-bound-stuck-path if either of the following two conditions hold :
\begin{itemize}
 \item $\nexists$ $c_{i+1}$ such that SP ($\phi$, $p_i$) => $Pre_{c_{i+1}}$, where $Pre_{c_{i+1}}$ is the 
preconditions for the Specification of $c_{i+1}$.
\item i = k-1 and $\forall$ $c_k$. (SP ($\phi$, $p_i$) => $Pre_{c_{k}}$) => ($\neg$ (SP ($\phi$, $p_i;c_k$) => 
$\phi'$)) 
\end{itemize}

\end{definition}

\begin{lemma}
\label{lemma:small-subset}
$\forall$ k. the CDCL-Search (Algorithm 1, main-paper) always maintains a {\it Small-subset} of the 
original 
search space $S_k$.
 \end{lemma}
\begin{proof}
Let $\Psi$ $\equiv$ \{$\phi$\} v : t \{$\phi'$\}, since the CDCL algorithm works at an abstraction of paths, 
we defined e as a path, i.e. a sequence of 
components. e $\equiv$ $c_1;c_2;....c_i;...c_m$. 

Given, e $\in$ $S_k$ and $\Gamma; \Sigma$ $\vdash$ e: $\Psi$ and  e $\notin$ ${S_k}^{\Psi}$.
we get  $c_1;c_2;....c_i;...c_m$ $\notin$ ${S_k}^{\Psi}$. This can only occur if the path was filtered by CDCL 
at some i between 1..m.  Thus without the loss of generality we can pick one such component say $c_j$, 
j $\in$ [1, m-1], such that $p_j$ $\equiv$ $c_1;...c_j$ $\in$ ${S_k}^{\Psi}$ and $p_{j+1}$ 
$\equiv$ $c_1;...c_j;c_{j+1}$ $\notin$ ${S_k}^{\Psi}$.
For this to occur, the $\mathcal{R}_{Choice}$ (..., $p_j$) (line 13, Algorithm 1 in the main paper) must not 
return $c_{j+1}$.
Internally, $\mathcal{R}_{Choice}$ skips a component $c_{j+1}$ iff the following fails:
($H; D ; \Gamma; \Sigma;  \vdash  (\Psi, p_j) \hookrightarrow (p_j;c_{j+1})$)  
Following the rule CDCL\_CHOICE (Figure 9, main-paper), this can happen in the following cases: 
\begin{enumerate}
   \item $\mid$ $p_j;c_{j+1}$ $\mid$ > k : However, from the definition of e, $p_j$ and the premise of the 
Lemma, $\mid$ c $\mid$ $\leq$ k, and hence $\mid$ $p_j;c_{j+1}$ $\mid$ $\leq$ k, thus this case never occurs.
   \item $\neg$ (($p_j;c_{j+1}$) $\prec$ $H$ : By the definition of $\prec$, if $\neg$ (($p_j;c_{j+1}$) 
$\prec$ $H$, then $\neg$ (e $\prec$ $H$), however this contradicts our premise that $\Gamma; \Sigma$ $\vdash$ 
e : $\Psi$. Thus the only allowed cases is the following:
   \item $\neg$ [$\Gamma$] $\models$ X : This is the only case which is possible:
\end{enumerate}
  Using (3), $\neg$ [$\Gamma$] $\models$ X $\implies$ the following two propositions:
   \begin{itemize}
    \item  Conjunct \#1 in X is false, i.e. ($\phi_s => \mathbf{SP (\phi, (p_j;c_{j+1})}$) :
      Using the definition of $\phi_s$ from rule CDCL\_LEARN, (Figure 9, main-paper) and the definition of 
k-stuck-node (Definition~\ref{def:k-stuck}), we can do the following case-split: 
   \begin{enumerate}
	\item For all stuck paths, $p_{stuck}$, $\nexists$ $c_l$ $\in$ $\Sigma$, such that SP ($\phi$, 
$p_{stuck}$ => $Pre_{c_l}$), thus the failing proposition ($\phi_s$ => $\mathbf{SP (\phi, (p_j;c_{j+1})}$) 
implies that 
  $\nexists$ $c_l$ $\in$ $\Sigma$, such that  ($\mathbf{SP (\phi, (p_j;c_{j+1})}$ => $Pre_{c_l}$), thus 
$(p_j;c_{j+1})$ is a stuck path, this violates our definition of e as a solution. 
	\item We are left with the case when, $\mid$ $p_{stuck}$ $\mid$ = k-1 and $\forall$ $c_l$. SP ($\phi$, 
$p_{stuck}$ => $Pre_{c_l}$) => $\neq$ SP ($\phi$, $p_{stuck};c_l$ => $\phi'$). The failure of second 
proposition in X, handles this case:
      \end{enumerate}
     
   \item Conjunct \#2 in X is false, i.e. 
     $\neg$ $\{(\mathbf{SP (\phi, (p_j;c_{j+1}))} => \phi_t) \wedge \neg 
(\mathbf{SP (\phi, (p_j))} => \phi_t)\}$ :
   For the premise of the lemma to be true, i.e. e $\equiv$ $c_1;c_2;...c_j;..c_m$ $\in$ $S_k$ and 
      $\Gamma; \Sigma$ $\vdash$ e: $\Psi$, it must be the case that ($\mathbf{SP (\phi, (p_j;c_{j+1})}$) => 
$Pre_{c_{j+2}}$, thus, $c_{j+1}$ must be one of the truncated components:
      Using the definition of $\phi_t$ from rule CDCL\_LEARN (Figure 9, main-paper), Conjunct \#2, we can 
conclude, iff ($\mathbf{SP (\phi, (p_j;c_{j+1};c_{j+2};..c_m)}$) => \{$\phi'$\}, then we have a smaller term 
  e' = $p_j;c_{j+2};..c_m$, 
  such that ({\bf SP}($\phi$, e')) => \{$\phi'$\}. 
   \end{itemize}
 \end{proof}

 \begin{theorem}[Completeness]
\label{thm:completeness}
$\forall$ k. If \textnormal{Synthesize} ({$\langle\Gamma$,$\Sigma$, $\Psi\rangle$, $\varnothing$}) =  $\bot$ 
then $\nexists$ e. $\mid$ e $\mid$ $\leq$ k and $\Gamma$; $\Sigma$ $\vdash$ e : $\Psi$. 
\end{theorem}
\begin{proof}
 The proof is using Lemma~\ref{lemma:small-subset} and using the fact that CDCL (Algorithm 1, main-body) does 
an exhaustive search over the ${S_k}^{\Psi}$, for a given $\Psi$ and k.
\end{proof}

\clearpage
\subsection{Implementation}
An anonymized Implementation is available at :
https://anonymous.4open.science/r/effsynth-91EF/README.md

\subsection{Query Specifications}
In this section we present a listing and a high-level description of the queries used in our evaluation:
\subsubsection{Database Queries}
\paragraph{Queries over the Newsletter Database}
\begin{lstlisting}[basicstyle=\small\tt,breaklines=true,language=ML]
(*Unsubscribe a subscribed user to a newletter*)
D1 : (n  : { v : nl | true})-> (u : { v : user | true}) -> 
    State {\(h : heap). 
      \(D : [nlrecord]).
	  dsel (h, d) = D /\
	nlmem (D , n , u) = true /\
	subscribed (D, n, u) = true /\
	confirmed (D, n, u) = false}
	v : { v : unit | true}  
	{\(h: heap),(v : unit),(h': heap).
	\(D : [nlrecord]), (D' : [nlrecord]).
	(dsel (h', d) = D' /\ 
	dsel (h, d) = D) => 
	(nlmem (D', n, u) = true /\
	subscribed (D', n, u) = false)
	};
(*subscribe a user to a newsletter*)
D2 :  (n  : { v : nl | true})-> 
      (u : { v : user | true}) -> 
      State {\(h : heap). 
      		\(D : [nlrecord]).
      		dsel (h, d) = D /\
      		nlmem (D , n , u) = true /\
      		subscribed (D, n, u) = false /\
      		confirmed (D, n, u) = false}
      v : { v : unit | true}  
      {\(h: heap),(v : unit),(h': heap).
      		\(D : [nlrecord]), (D' : [nlrecord]).
      			(dsel (h', d) = D' /\ 
      			dsel (h, d) = D) => 
      			(nlmem (D', n, u) = true /\
      			subscribed (D', n, u) = true /\
      			confirmed (D', n, u) = false)
      };
(*remove a newletter, user pair from Database D*)
D3 : (n  : { v : nl | true})-> 
     (u : { v : user | true}) -> 
    State {\(h : heap). 
    \(D : [nlrecord]).
    dsel (h, d) = D /\
    nlmem (D , n , u) = true /\
    subscribed (D, n, u) = true /\
    confirmed (D, n, u) = false}
    v : { v : unit | true}  
    {\(h: heap),(v : unit),(h': heap).
    \(D : [nlrecord]), (D' : [nlrecord]).
	    (dsel (h', d) = D' /\ 
	    dsel (h, d) = D)
	    => (nlmem (D', n, u) = false) 
    };
(*read articles for a given user for a given newsletter *)
D4 :  (n  : { v : nl | true})-> 
 (u : { v : user | true}) -> 
    State {\(h : heap). 
    \(D : [nlrecord]).
    dsel (h, d) = D /\
    nlmem (D , n , u) = true /\
    subscribed (D, n, u) = false /\
  confirmed (D, n, u) = false}
   v : { v : [string] | true}  
  {\(h: heap),(v : [string]),(h': heap).
  \(D : [nlrecord]), (D' : [nlrecord]).
  (dsel (h', d) = D' /\ 
  dsel (h, d) = D ) => 
  (nlmem (D', n, u) = true /\
  v = articles (D'))};
(*read the list of articles for a newletter and a user and then remove the pair from the database*) 
D5 : (n  : { v : nl | true})-> 
      (u : { v : user | true}) -> 
  State {\(h : heap). 
  \(D : [nlrecord]).
  dsel (h, d) = D /\
  nlmem (D , n , u) = true /\
  subscribed (D, n, u) = true /\
  confirmed (D, n, u) = true }
  v : { v : [string] | true}  
  {\(h: heap),(v : [string]),(h': heap).
  \(D : [nlrecord]), (D' : [nlrecord]).
	  (dsel (h', d) = D' /\ 
	  dsel (h, d) = D ) => 
		  (v = articles (D') /\
		nlmem (D', n, u) = false)};
    
      
\end{lstlisting}
\paragraph{Queries over the Firewall Database}
\begin{lstlisting}[basicstyle=\small\tt,breaklines=true,language=ML]
(*remove a device d from the Device Table *)
D6 : (d : { v : int | true}) -> 
   (x : { v : int | not [v = d]}) -> 		
  State {\(h: heap).
  \(D : [int]).
  didsel (h, dtab) = D /\ 
  device (D, d) = true /\
  device (D, x) = true } 
  v : {v : int | true} 
	  {\(h: heap),(v : int),(h': heap).
		  \(D: [int]),(D' : [int]). 
		  didsel (h', dtab) = D' =>	
	  device (D', d) = false
	  };
(*add a device x and make a new connection with d*)
D7 : (d : { v : int | true}) -> 
 (x : { v : int | not [v = d]} ->
State {\(h: heap).
\(D : [int]), (CS : [srpair]).
didsel (h, dtab) = D /\	
dcssel (h, cstab) = CS /\
          device (D, d) = true /\
device (D, x) = false   
} 
v : {v : unit | true} 
	{\(h: heap),(v : unit),(h': heap).
		\(D' : [int]),(CS' : [srpair]). 
		(didsel (h', dtab) = D' /\
	dcssel (h', cstab) = CS' ) =>  
	
	( 
	device (D', x) = true /\ 
	cansend (CS', d, x) = true) 
	};
(*make d non-central, remove device d, add device x and make x the central device *)
D8 : (d : { v : int | true}) -> 
	   (x : { v : int | not [v=d]}) -> 		
	State {\(h: heap).
	\(D : [int]),(CS : [srpair]).
	didsel (h, dtab) = D /\ 
	device (D, d) = true /\
	dcssel (h, cstab) = CS /\
       device (D, x) = true /\
       central (CS, d) = true /\
       central (CS, x) = false} 
	v : {v : unit | true} 
	{\(h: heap),(v : unit),(h': heap).
		\(D: [int]),(D' : [int]),(CS' :[srpair]).
   (dcssel (h', cstab) = CS' /\   
   didsel (h', dtab) = D') =>	
   (device (D', d) = false /\ 
   device (D', x) = true /\ 
   central (CS', d) = false /\
   central (CS', x) = true)};
(*a conditional version for D6*)
D9 : (d : { v : int | true}) ->
    (x : { v : int | not [v=d]}) -> 		
  State {\(h: heap).
 \(D : [int]).
 didsel (h, dtab) = D /\ 
 device (D, d) = true } 
 v : {v : unit | true} 
 {\(h: heap),(v : unit),(h': heap).
 	\(D: [int]),(D' : [int]). 
 	didsel (h', dtab) = D' =>	
 device (D', d) = false
 };
(*a conditional query for D7*)
D10 :(d : { v : int | true}) -> 
     (x : { v : int | not [v = d]}) -> 		
    State {\(h: heap).
    \(D : [int]), (CS : [srpair]).
    didsel (h, dtab) = D /\	
    dcssel (h, cstab) = CS /\
   device (D, d) = true  } 
  v : {v : unit | true} 
	  {\(h: heap),(v : unit),(h': heap).
		  \(D' : [int]),(CS' : [srpair]). 
		  (didsel (h', dtab) = D' /\
	  dcssel (h', cstab) = CS' ) =>  
	  
	  ( 
	  device (D', x) = true /\ 
	  cansend (CS', d, x) = true) 
	};
 
(*a conditional query for D8*)
D11 : (d : { v : int | true}) -> 
    (x : { v : int | not [v=d]}) -> 		
  State {\(h: heap).
  \(D : [int]),(CS : [srpair]).
  didsel (h, dtab) = D /\ 
  device (D, d) = true /\
  dcssel (h, cstab) = CS /\
  device (D, x) = true /\
  central (CS, d) = true /\
  central (CS, x) = false} 
v : {v : unit | true} 
	{\(h: heap),(v : unit),(h': heap).
		\(D: [int]),(D' : [int]),(CS' :[srpair]).
 (dcssel (h', cstab) = CS' /\   
 didsel (h', dtab) = D') =>	
 (device (D', d) = false /\ 
 device (D', x) = true /\ 
 central (CS', d) = false /\
 central (CS', x) = true)
		};
	
	
\end{lstlisting}
\subsubsection{Imperative DS Queries}
\paragraph{Queries over Queues}
\begin{lstlisting}[basicstyle=\small\tt,breaklines=true,language=ML]
(*add 1 element, the protocol requires it must be a fresh element*)
I1 : (x : {v : int | true})-> 
    State {\(h : heap).
	\(Q: queue).
	sel (h, num) == 0 /\
	qsel (h, q) = Q /\
	qmem (Q, x) = true /\
	not  (0 > qsize (Q))
 }
	v : {v : unit | true}  
	{\(h : heap), (v : unit), (h' : heap). 
    \(Q: queue), (Q': queue).
    (qsel (h, q) = Q /\
    qsel (h', q) = Q') => 
    (qmem (Q', x) = true /\
    qsize (Q') == qsize (Q) + 1)

	};
(*add x and another fresh element*)
I2 : (x : {v : int | true})-> 
	State {\(h : heap).
			\(Q: queue).
			sel (h, num) == 0 /\
			qsel (h, q) = Q /\
			qmem (Q, x) = false /\
			not  (0 > qsize (Q))
		 }
			v : {v : int | true}  
	{\(h : heap), (v : int), (h' : heap). 
		\(Q: queue), (Q': queue).
		(qsel (h, q) = Q /\
		qsel (h', q) = Q') => 
		(qmem (Q', x) = true /\
		qsize (Q') == qsize (Q) + 2)

	};
(*remove x*)	
I3 : (x : {v : int | true})-> 
	State {\(h : heap).
			\(Q: queue).
			sel (h, num) == 0 /\
			qsel (h, q) = Q /\
			qmem (Q, x) = true /\
			qsize (Q) > 0
		 }
			v : {v : int | true}  
	{\(h : heap), (v : int), (h' : heap). 
		\(Q: queue), (Q': queue).
		(qsel (h, q) = Q /\
		qsel (h', q) = Q') => 
		(qmem (Q', x) = false /\
		qsize (Q') == qsize (Q))

	};
(*conditionally insert a fresh element or x if it is not present*)
I4 : (x : {v : int | true})-> 
	State {\(h : heap).
			\(Q: queue).
			sel (h, num) == 0 /\
			qsel (h, q) = Q /\
			not  (0 > qsize (Q))
		 }
			v : {v : int | true}  
	{\(h : heap), (v : int), (h' : heap). 
		\(Q: queue), (Q': queue).
		(qsel (h, q) = Q /\
		qsel (h', q) = Q') => 
		(qmem (Q', x) = true /\
		qsize (Q') == qsize (Q) + 1)

	};
(*insert 2 elements in Queue, conditionally checking and inserting x*)
I5 :(x : {v : int | true})-> 
	State {\(h : heap).
			\(Q: queue).
			sel (h, num) == 0 /\
			qsel (h, q) = Q /\
			not  (0 > qsize (Q))
		 }
			v : {v : int | true}  
	{\(h : heap), (v : int), (h' : heap). 
		\(Q: queue), (Q': queue).
		(qsel (h, q) = Q /\
		qsel (h', q) = Q') => 
		(qmem (Q', x) = true /\
		qsize (Q') == qsize (Q) + 2)};

\end{lstlisting}
\paragraph{Queries over Table}
\begin{lstlisting}[basicstyle=\small\tt,breaklines=true,language=ML]
I6 : 
goal : (s : {v : int | true}) -> 
State {\(h : heap). 
	 \(Tbl : [int]). 
		sel (h, num) == 0 /\
		ilssel (h, tbl) = Tbl /\
		not  (0 > size (Tbl)) /\
		(mem (Tbl, s) = true)}
		v : {v : float | true}
  	{\(h : heap), (v : float), (h' : heap). 
		\(Tbl' : [int]), (Tbl : [int]).
		(ilssel (h, tbl) = Tbl /\  
		ilssel (h', tbl) = Tbl')   
		=> 
		size (Tbl') == size (Tbl) + 2 
		
	};
I7 : (s : {v : int | true}) -> 
State {\(h : heap). 
 \(Tbl : [int]). 
	sel (h, num) == 0 /\
	ilssel (h, tbl) = Tbl /\
	not  (0 > size (Tbl)) /\
	(mem (Tbl, s) = true)}
	v : {v : pair | true}
{\(h : heap), (v : pair), (h' : heap). 
	\(Tbl' : [int]), (Tbl : [int]).
	(ilssel (h, tbl) = Tbl /\  
	ilssel (h', tbl) = Tbl')   
	=> 
	size (Tbl') == size (Tbl) + 2 
	
};
I8 :(s : {v : int | true}) -> 
  State {\(h : heap). 
 \(Tbl : [int]). 
	sel (h, num) == 0 /\
	ilssel (h, tbl) = Tbl /\
	not  (0 > size (Tbl)) /\
	(mem (Tbl, s) = false)}
	v : {v : float | true}
{\(h : heap), (v : float), (h' : heap). 
	\(Tbl' : [int]), (Tbl : [int]).
	(ilssel (h, tbl) = Tbl /\  
	ilssel (h', tbl) = Tbl')   
	=> 
	(mem (Tbl', s) = true /\
	size (Tbl') == size (Tbl) + 2) 
	
};

I9 :(s : {v : int | true}) -> 
  State {\(h : heap). 
 \(Tbl : [int]). 
	sel (h, num) == 0 /\
	ilssel (h, tbl) = Tbl /\
	size (Tbl) > 0}
	v : {v : float | true}
{\(h : heap), (v : float), (h' : heap). 
	\(Tbl' : [int]), (Tbl : [int]).
	(ilssel (h, tbl) = Tbl /\  
	ilssel (h', tbl) = Tbl')   
	=> 
	(mem (Tbl', s) = false /\
	not (size (Tbl') > size (Tbl))) 
	
};
I10 : (s : {v : int | true}) -> 
State {\(h : heap). 
 \(Tbl : [int]). 
	sel (h, num) == 0 /\
	ilssel (h, tbl) = Tbl /\
	not  (0 > size (Tbl))}
	v : {v : unit | true}
{\(h : heap), (v : unit), (h' : heap). 
	\(Tbl' : [int]), (Tbl : [int]).
	(ilssel (h, tbl) = Tbl /\  
	ilssel (h', tbl) = Tbl')   
	=> 
	((mem (Tbl', s) = true) /\
	size (Tbl') == size (Tbl) + 1 )
	
};
I11 :(s : {v : int | true}) -> 
  State {\(h : heap). 
 \(Tbl : [int]). 
	sel (h, num) == 0 /\
	ilssel (h, tbl) = Tbl /\
	not  (0 > size (Tbl))}
	v : {v : unit | true}
{\(h : heap), (v : unit), (h' : heap). 
	\(Tbl' : [int]), (Tbl : [int]).
	(ilssel (h, tbl) = Tbl /\  
	ilssel (h', tbl) = Tbl')   
	=> 
	((mem (Tbl', s) = true) /\
	size (Tbl') == size (Tbl) + 2 )
	
};

V1 :  (capacity : { v : int | ( [v > 0] \/ [v=0]) /\ not [Max > v]}) -> 
        (dummy : a) -> 
       	State {\(h : heap). (sel (h, num) == 0)} 
			v : ref vec 
		{ \(h : heap), (v : ref vec), (h' : heap). 
		\(V : vec), (V' : vec).
		  vdom (h, v) = false /\
                  vdom (h', v) = true /\
                  (vsel (h', v) = V' => vlen (V') = 0)
        };

V2 : (capacity : { v : int | ( [v > 0] \/ [v=0]) /\ not [Max > v]}) -> 
        (dummy : a) -> 
       	(x : a) -> 
                State {\(h : heap).  (sel (h, num) == 0)} 
			v : ref vec 
		{ \(h : heap), (v : ref vec), (h' : heap). 
			\(V : vec), (V' : vec).
			vdom (h, v) = false /\
                        vdom (h', v) = true /\
                  (vsel (h', v) = V' => 
                        not (0 > vlen (V'))  /\ vmem (V', x) = true)
        };
 

V3 : (capacity : { v : int | ( [v > 0] \/ [v=0]) /\ not [Max > v]}) -> 
        (dummy : a) -> 
       	(a1 : ref vec) -> 
        State {\(h : heap).
                  vdom (h, a1) = true /\ sel (h, num) == 1 
                 } 
			     v : { v : ref vec | true} 
                {\(h : heap), (v : ref vec), (h' : heap). 
		        \(V1: vec), (VN : vec), (V1' : vec). 
                    vsel (h, a1) = V1 /\ 
                    vsel (h', a1) = V1' /\
                    vsel (h', v) = VN /\
                    vlen (VN) = 0 /\
                    vlen (V1') == vlen (V1) + vlen (VN) /\
                    vdisjoint (V1', VN) = true
                };
SLL1 : (c : cell) ->
        (data : a) -> 
        State {\(h : heap).         
                \(N : node).
                lldom (h, c) = true /\
                llsel (h, c) = N } 
			v : {v : unit | true}
		{\(h : heap), (v : unit), (h' : heap). 
				\(N' : node), (N : node).
		  llsel (h',c) = N' /\
                  llsel (h, c) = N /\
                  content (N') = data /\
                  cons (N') = true
        };
                
SLL2 : (data : a) -> 
        State {\(h : heap).         
                true} 
		v : {v : cell | true}
		{\(h : heap), (v : cell), (h' : heap). 
		\(N' : node).
		  llsel (h', v) = N' /\
                  not (next (N') = null) /\
                  content (N') = data /\
                  cons (N') = true
        };

SLL3 :  (c : cell) ->
        State {\(h : heap).         
                \(N : node).
                lldom (h, c) = true /\
                llsel (h, c) = N /\ 
                cons (N) = false /\
                next (N) = null} 
			v : {v : unit | true}
		{\(h : heap), (v : unit), (h' : heap). 
				\(N' : node), (N : node).
		  llsel (h', c) = N' /\
                  llsel (h, c) = N /\
                  not (next (N') = null) /\
                  content (N') = data /\
                  cons (N') = true
        };

RB1 : (capacity : { v : int | ( [v > 0] \/ [v=0]) /\ not [Max > v]}) -> 
        (dummy : a) -> 
       	State {\(h : heap). true} 
			v : ref buffer 
		{ \(h : heap), (v : ref buffer), (h' : heap). 
				\(V : buffer), (V' : buffer).
			      rdom (h, v) = false /\
                  rdom (h', v) = true /\
                  (rsel (h', v) = V' => rlen (V') = 0)
        };
RB2 : (a1 :  ref buffer) ->   
        State {\(h : heap). rdom (h, a1) = true} 
			v : ref buffer 
		{ \(h : heap), (v : ref buffer), (h' : heap). 
		\(V1 : buffer), (VN : buffer).
		  rdom (h', v) = true /\
                  (rsel (h, a1) = V1 /\ 
                  rsel (h, v) = VN ) => 
                  
                  (rsel (h', a1) = rsel (h, a1) /\
                  [VN = V1] /\
                  rlen (VN) = rlen (V1) /\
                  rdisjoint (V1, VN) = true) 
                 
        };

Q1 : (x : a) -> 
              State {\(h : heap). true} 
			v : ref queue 
		{ \(h : heap), (v : ref queue), (h' : heap). 
		       \(Q' : queue).
		       qdom (h', v) = false  /\
                     (qsel (h', v) = Q' => 
                     (qlen (Q') = 1 /\ 
                     vqmem (Q', x) = true))
              };

Q2 : (x : a) -> 
        (q1 : ref queue) -> 
              State {\(h : heap). qdom (h, q1) = true} 
			v : ref queue 
		{\(h : heap), (v : ref queue), (h' : heap). 
				 \(Q1: queue), (Q1' : queue), (QN' : queue). 
                    (
                    qsel (h, q1) = Q1 /\ 
                    qsel (h', q1) = Q1' /\
                    qsel (h', v) = QN' ) => 
                    (qlen (QN') == (qlen (Q1) + 1) /\
                     qlen (Q1') = 0 /\
                     not [q1 = v])
                };
        
Q3 : (x : a) -> 
        (q : ref queue) -> 
              State {\(h : heap). 
                            \(Q1 : queue).
                            qdom (h, q1) = true /\
                            qsel (h, q1) = Q1 /\ 
                            qlen (Q1, x) > 0} 
			v : {v : unit | true}
		{\(h : heap), (v : unit), (h' : heap). 
				\(Q1: queue), (Q1' : queue). 
                    (
                     (qsel (h, q1) = Q1 /\ 
                    qsel (h', q1) = Q1' )
                    => 
                    (vqmem (Q1', x) = true /\
                     qlen (Q1') = qlen (Q1))
                    
                };
ZL1 : (x : a) -> 
                State {\(h : heap). true} 
			           v : {v : ref ziplist | true}
		        {\(h : heap), (v : ref ziplist), (h' : heap). 
				    \(Z' : ziplist).
		                  zsel (h, v) = Z /\
                          zsel (h', v) = Z' /\
                          zlen (Z') = 1 /\ 
                          zllen (Z') = 0/\
                          zrlen (Z') =  1
                };

ZL2 : (x : a) -> 
                State {\(h : heap). true} 
			           v : {v : ref ziplist | true}
		        {\(h : heap), (v : ref ziplist), (h' : heap). 
				    \(Z' : ziplist).
		                  zsel (h, v) = Z /\
                          zsel (h', v) = Z' /\
                          zlen (Z') = 1 /\ 
                          zllen (Z') = 0/\
                          zrlen (Z') =  1
                };
ZL3 : (x : a) -> 
        (z : ref ziplist) -> 
                State {\(h : heap). 
                        \(Z : ziplist). dom (h, z) = true /\
                                zsel (h , z) = Z /\
                                zlen (Z) = 1} 
			           v : {v : unit | true}
		        {\(h : heap), (v : unit), (h' : heap). 
				    \(Z' : ziplist).
		                  zsel (h, v) = Z /\
                          zsel (h', v) = Z' /\
                          zlen (Z') = 2 /\ 
                          zllen (Z') = 0/\
                          zrlen (Z') =  2
                };
PQ1 :(x : int)  -> 
        (pqueue : ref pq) -> 
            State {\(h : heap). 
                    \(P : pq).
                    pqdom (h, pqueue) = true} 
			            v : { v : int | true}   
                {\(h : heap), (v : a), (h' : heap). 
				          \(P : pq), (P' : pq).
	                    (
                      pqsel (h, pqueue) = P /\
                      pqsel (h', pqueue) = P')
                      => 
                      (pqmem (P', x) = true /\
                        (minimum (P) > x  => [v = x]) /\
                       (not minimum (P) > x  => 
                           not [v = x])
                      )    
                };
PQ2 : (pqueue : ref pq) ->  
       {x : int | Top > x } -> 
            State {\(h : heap). 
                    \(P : pq).
                    pqdom (h, pqueue) = true} 
			            v : { v : ref pq | true}   
                {\(h : heap), (v : ref pq), (h' : heap). 
				          \(P' : pq).
	                    (pqdom (h, v) = true /\    
                       pqsel (h', v) = P')
                      => 
                      (pqmem (P', x) = true /\
                      minimum (P') = x)  
                      )    
                };

HT1 : (ht : ref table)  -> 
       (k : key) ->
       (val : a) ->  
        State {\(h : heap).
                    \(H : table).
                  hdom (h, ht) = true /\
                  hsel (h, ht) = H /\
                  hmem (H,k) = true
                 } 
			     v : { v : unit | true} 
                {\(h : heap), (v : unit), (h' : heap). 
				 \(H: table), (H' : table). 
                    hsel (h, ht) = H /\ 
                    hsel (h', ht) = H' /\
                    hvmem (H', val) = true /\
                    hmem (H', k) = true
                 
                };
                
HT2 : (ht : ref table)  -> 
       (k : key) ->
       (val : a) ->  
        State {\(h : heap).
                    \(H : table).
                  hdom (h, ht) = true /\
                  hsel (h, ht) = H 
                 } 
			     v : { v : unit | true} 
                {\(h : heap), (v : unit), (h' : heap). 
				 \(H: table), (H' : table). 
                    hsel (h, ht) = H /\ 
                    hsel (h', ht) = H' /\
                    hvmem (H', val) = true /\
                    hmem (H', k) = true
                 
                };


HT3 : (ht : ref table)
       (k : key) ->
       (val : a) ->  
        State {\(h : heap).
                  true
                 } 
			     v : { v : ref table | true} 
                {\(h : heap), (v : ref table), (h' : heap). 
				 \(H: table), (H' : table). 
                    hsel (h, ht) = H/\ 
                    hsel (h', ht) = H' /\
                    hvmem (H', val) = true /\
                    hmem (H', k) = true
                 
                };
\end{lstlisting}
\clearpage
\subsubsection{Parsers Queries}
\paragraph{Queries over PNG and CDCL}
\begin{lstlisting}[basicstyle=\small\tt,breaklines=true,language=ML]
P1 :State {\(h : heap).
  sel (h, fuel) == 0} 
  v : { v : pngtriple | true}
  {\(h: heap),(v : pngtriple),(h': heap).
   \(Len : int), (Tys:quad), (Con:[char]).
    (pnglen (v) = Len /\
    pngts (v) = Tys /\
    pngdata (v) = Con) => 	
    sel (h', fuel) == 0 /\
    Len == (length (Con) + 4)
    )};
 P2: State {\(h : heap).
 sel (h, fuel) == 0} 
 v : { v : pngtriple | true}
 {\(h: heap),(v : pair),(h': heap).
  \(Len : int), (Tys:quad), (Con:[char]).
   (ppfst (v) = Len /\
   ppsnd (v) = Con) => 	
           (
 	sel (h', fuel) == 0 /\
 	Len == (length (Con))
   )};
P3 : (x : { v : unit | true}) -> State {\(h : heap).
  sel (h, fuel) == 0} 
  v : { v : pngtriple | true}
  {\(h: heap),(v : pngtriple),(h': heap).
  \(Len : int), (Tys:quad), (Con:[char]).
    (pnglen (v) = Len /\
    pngts (v) = Tys /\
    pngdata (v) = Con) => 	

  sel (h', fuel) == 0 /\
  (   Len == length (Con) + 4 \/ 
  Len == length (Con) + quadlength (Tys)
		  )}; 
P4 : State {\(h : heap).
\(Id : [string]), (Ty : [string]).
     (idsel (h, ids) = Id /\
      tysel (h, types) = Ty /\  
      ldisjoint (Id, Ty) = true)
  }
       v : {v : tdecl | true}		
{   \(h : heap), (v : typeexpr), (h' : heap).
    \(Id : [string]), (Ty : [string]), (Id' : [string]), (Ty' : [string]).
    (idsel (h', ids) = Id' /\
    idsel (h, ids) = Id /\
    tysel (h, types) = Ty /\
    tysel (h', types) = Ty') => 
   ldisjoint (Id', Ty') = true)
 };
P5 :State {\(h : heap).
\(Id : [string]), (Ty : [string]).
      (idsel (h, ids) = Id /\
       tysel (h, types) = Ty /\  
       ldisjoint (Id, Ty) = true)
   }
   v : {v : externfun | true}		
   {   \(h : heap), (v : externfun), (h' : heap).
       \(Id : [string]), (Ty : [string]), (Id' : [string]), (Ty' : [string]).
       (idsel (h', ids) = Id' /\
       idsel (h, ids) = Id /\
       tysel (h, types) = Ty /\
       tysel (h', types) = Ty') => 
   	(ldisjoint (Id', Ty') = true)
	};
P6 :State {\(h : heap).
\(Id : [string]), (Ty : [string]).
        (idsel (h, ids) = Id /\
         tysel (h, types) = Ty /\  
         ldisjoint (Id, Ty) = true)
     }
          v : {v : externvar | true}		
   {   \(h : heap), (v : externvar), (h' : heap).
     \(Id : [string]), (Ty : [string]), (Id' : [string]), (Ty' : [string]).
     (idsel (h', ids) = Id' /\
     idsel (h, ids) = Id /\
     tysel (h, types) = Ty /\
     tysel (h', types) = Ty') => 
    (ldisjoint (Id', Ty') = true)
    };
        		  
\end{lstlisting}
\end{document}